\documentclass[aps,notitlepage,nofootinbib,longbibliography,
twocolumn,
superscriptaddress,10pt]{revtex4-2}

\usepackage{bbold}
\usepackage{simplewick}

\usepackage[usenames,dvipsnames]{color}

\usepackage{amsmath}
\usepackage{amsthm, amssymb}
\usepackage{enumitem}
\usepackage{slashed}
\usepackage{graphicx}
\usepackage{xcolor}
\usepackage{tikz}
\usetikzlibrary{calc,fadings,decorations.pathreplacing,shapes,shapes.multipart,arrows,shapes.misc,intersections,positioning,patterns}
\usepackage{bm}
\usepackage[colorlinks=true,citecolor=blue,linkcolor=blue,urlcolor=blue]{hyperref}
\usepackage{dsfont}
\usepackage{changepage}
\usepackage{array}
\usepackage{soul}
\usepackage[capitalise]{cleveref}
\crefname{section}{Sec.}{Secs.}
\Crefname{section}{Sec.}{Secs.}
\usepackage{xifthen}
\usepackage{xargs}
\usepackage{ulem}

\newcommand{\ie}{\textit{i.e.} }

\newcommand{\I}{\mathbb{I}}
\newcommand\eq[1]{\begin{align}#1\end{align}}

\usepackage{bm}
\renewcommand{\vec}[1]{\boldsymbol{\mathbf{#1}}}

\graphicspath{{./}{./images/}}

\newcommand{\bit}{\begin{itemize}}
\newcommand{\eit}{\end{itemize}}

\newcommand{\f}{\frac}
\renewcommand{\>}{\right\rangle}
\newcommand{\<}{\left\langle}
\newcommand{\ba}{\begin{align}}
\newcommand{\ea}{\end{align}}
\newcommand{\be}{\begin{equation}}
\newcommand{\ee}{\end{equation}}
\newcommand{\bi}{\begin{itemize}}
\newcommand{\ei}{\end{itemize}}
\newcommand{\lf}{\left(}
\newcommand{\ri}{\right)}

\newcommand{\ii}{\item}

\newcommand{\Tr}{\operatorname{Tr}}
\newcommand{\tr}{\operatorname{tr}}

\newcommand{\seq}{s_\text{eq}}
\newcommand{\Grms}{G_\text{rms}}
\newcommand{\Gtyp}{G_\text{typ}}
\newcommand{\UU}{\mathbf{V}}

\newcommandx{\fineq}[4][1=-.8ex,2=1,3=1]{
  \begin{tikzpicture}[baseline={([yshift=#1]current  bounding  box.center)}, scale = #2, every node/.style={scale = #3}]
    #4
  \end{tikzpicture}
}

\newcommand{\fullstate}{
\begin{tikzpicture}
\fill[black] (0,0) circle (0.1cm); 
\end{tikzpicture}
}

\newcommand{\emptystate}{
\begin{tikzpicture}
\draw (0,0) circle (0.1cm); 
\end{tikzpicture}
}

\newcommand{\emem}{
  \begin{tikzpicture}
\draw (0,0) circle (0.1cm);
\draw  (0.3,0) circle (0.1cm);
 \end{tikzpicture} 
}

\newcommand{\emfu}{
  \begin{tikzpicture}
\draw (0,0) circle (0.1cm);
\fill[black]  (0.3,0) circle (0.1cm);
 \end{tikzpicture} 
}
\newcommand{\fuem}{
  \begin{tikzpicture}
\fill[black]  (0,0) circle (0.1cm);
\draw  (0.3,0) circle (0.1cm);
 \end{tikzpicture} 
}
\newcommand{\fufu}{
  \begin{tikzpicture}
\fill[black]  (0,0) circle (0.1cm);
\fill[black]   (0.3,0) circle (0.1cm);
 \end{tikzpicture} 
}

\newcommandx{\wickpair}[5][1=0,2=0,3=0.3,4=0.2,5=0.8]{
  \begin{scope}[shift={(#1,#2)}]
    \draw[line width = 0.5pt] (0,0) to[out=90,in=180] (#3/2,#4) to[out=0,in=90] (#3,0);
    \begin{scope}[shift={(0,#5)}]
      \draw[line width = 0.5pt] (0,0) to[out=-90,in=180] (#3/2,-#4) to[out=0,in=-90] (#3,0);
    \end{scope}
  \end{scope}
}

\newcommandx{\wickswitch}[4][1=0,2=0,3=0.3,4=0.8]{
  \begin{scope}[shift={(#1,#2)}]
    \pgfmathsetmacro{\w}{#3};
    \pgfmathsetmacro{\h}{#4};
    \draw[line width = 0.5pt] (0,0) to[out=90,in=-135] (\w/2,\h/2) to[out=45,in=-90] (\w,\h);
    \draw[line width = 0.5pt] (\w,0) to[out=90,in=-45] (\w/2,\h/2) to[out=135,in=-90] (0,\h);
  \end{scope}
}

\newcommandx{\bcswitch}[1][1=0.3]
{
  \fineq[-0.4ex]{
    \pgfmathsetmacro{\w}{#1};
    \pgfmathsetmacro{\h}{0.8};
    \wickswitch[0][0][\w][\h];
    \node () at (0,\h/2) {};
    \node () at (\w,\h/2) {};
  }
}

\newcommandx{\bcid}[4][1=0,2=0,3=0.3,4=0.8]
{
  \begin{scope}[shift={(#1,#2)}]
    \draw (0,0)--++(0,#4);
    \draw (#3,0)--++(0,#4);
  \end{scope}
}

\newcommandx{\bcwick}[1][1=0]
{
  \fineq[-0.4ex]{
    \pgfmathsetmacro{\h}{0.8};
    \pgfmathsetmacro{\w}{0.3};
    \ifthenelse{\equal{#1}{0}}{
      \wickpair[0][0];
    }{}
    \ifthenelse{\equal{#1}{1}}{
      \wickpair[0][0][\w*2];
    }{}
    \ifthenelse{\equal{#1}{2}}{
      \wickpair[0][0][\w*3];
    }{}
    \node () at (0,\h/2) {};
    \node () at (\w,\h/2) {};
  }
}

\newcommandx{\idw}[0]
{
  \draw[line width = 0.5pt] (0,0) to[out=90,in=180] (0.15,0.2) to[out=0,in=90] (0.3,0);
  \draw[line width = 0.5pt] (0.6,0) to[out=90,in=180] (0.6+0.15,0.2) to[out=0,in=90] (0.9,0);
}

\newcommandx{\swapw}[0]
{
  \draw[line width = 0.5pt] (0,0) to[out=90,in=180] (0.45,0.25) to[out=0,in=90] (0.9,0);
  \draw[line width = 0.5pt] (0.3,0) to[out=90,in=180] (0.45,0.15) to[out=0,in=90] (0.6,0);
}

\newcommandx{\idweq}[0]
{
  \fineq{
    \idw;
    \node () at (0,0.1) {};
    \node () at (0.9,0.1) {};
  }
}

\newcommandx{\swapweq}[0]
{
  \fineq{
    \swapw;
    \node () at (0,0.1) {};
    \node () at (0.9,0.1) {};
  }
}

\newcommandx{\topcont}[3][1=0,2=0,3=0]
{
  \begin{scope}[shift={(#1,#2)}]
    \ifthenelse{\equal{#3}{0}}{
      \idw;
    }{}
    \ifthenelse{\equal{#3}{1}}{
      \swapw;
    }{}
  \end{scope}
}

\newcommandx{\bcwickfouru}[2][1=0,2=]
{
  \fineq[-0.6ex]{
    \pgfmathsetmacro{\h}{0.8};
    \pgfmathsetmacro{\w}{0.3};
    \ifthenelse{\equal{#1}{0}}{
      \wickpair[0][0];
      \bcid[2*\w][0];
    }{}
    \ifthenelse{\equal{#1}{1}}{
      \wickpair[\w][0];
      \bcid[0][0][3*\w];
    }{}
    \ifthenelse{\equal{#1}{2}}{
      \wickpair[0][0][\w*3];
      \bcid[\w][0]
    }{}
    \ifthenelse{\equal{#1}{3}}{
      \wickpair[2*\w][0];
      \bcid[0][0]
    }{}
    \node () at (0,\h/2) {};
    \node () at (3*\w,\h/2) {};
    \ifthenelse{\equal{#2}{}}{}{
      \topcont[0][\h+0.1][#2];
    }
  }
}

\newcommandx{\bcidfouru}[2][1=,2=]
{
  \fineq[-0.6ex]{
    \pgfmathsetmacro{\h}{0.8};
    \pgfmathsetmacro{\w}{0.3};
    \bcid;
    \bcid[2*\w];
    \node () at (0,\h/2) {};
    \node () at (3*\w,\h/2) {};
    \ifthenelse{\equal{#2}{}}{}{
      \topcont[0][\h+0.1][#2];
    }
  }
}

\newcommandx{\bcswitchfouru}[2][1=0,2=]
{
  \fineq[-0.6ex]{
    \pgfmathsetmacro{\h}{0.8};
    \pgfmathsetmacro{\w}{0.3};
    \ifthenelse{\equal{#1}{0}}{
      \wickswitch[0][0][2*\w][\h]
      \bcid[\w][0][2*\w];
    }{}
    \ifthenelse{\equal{#1}{1}}{
      \wickswitch[\w][0][2*\w][\h]
      \bcid[0][0][2*\w];
    }{}
    \node () at (0,\h/2) {};
    \node () at (3*\w,\h/2) {};
    \ifthenelse{\equal{#2}{}}{}{
      \topcont[0][\h+0.1][#2];
    }
  }
}

\newcommandx{\linearrow}[4][1=0,2=0,3=0.8,4=0]
{
  \pgfmathsetmacro{\h}{#3};
  \begin{scope}[shift={(#1,#2)}]
    \ifthenelse{\equal{#4}{0}}{
      \draw[-stealth] (0,0)--(0,\h*0.6);
      \draw (0,\h*0.6)--(0,\h);
    }{}
    \ifthenelse{\equal{#4}{1}}{
      \draw[-stealth] (0,\h)--(0,\h*0.45);
      \draw (0,\h*0.45)--(0,0);
    }{}
  \end{scope}
 
}

\newcommandx{\hket}[2][1=01,2=]
{
  \ifthenelse{\equal{#2}{}}{
    |
  }{}
  \fineq[-0.7ex]{
    \ifthenelse{\equal{#1}{0}}{
      \draw (0,0) circle (0.08);
    }{}
    \ifthenelse{\equal{#1}{1}}{
      \fill (0,0) circle (0.08);
    }{}
    \ifthenelse{\equal{#1}{00}}{
      \draw (0,0) circle (0.08);
      \draw (0.3,0) circle (0.08);
    }{}
    \ifthenelse{\equal{#1}{01}}{
      \draw (0,0) circle (0.08);
      \fill (0.3,0) circle (0.08);
    }{}
    \ifthenelse{\equal{#1}{10}}{
      \fill (0,0) circle (0.08);
      \draw (0.3,0) circle (0.08);
    }{}
    \ifthenelse{\equal{#1}{11}}{
      \fill (0,0) circle (0.08);
      \fill (0.3,0) circle (0.08);
    }{}
  }
  \rangle    
}

\newcommandx{\sumbasis}[4][1=0,2=0,3=0,4=0]
{
  \fineq[-0.4ex]{
    \pgfmathsetmacro{\h}{0.8};
    \pgfmathsetmacro{\w}{0.3};
    \begin{scope}[shift={(#1,#2)}]
      \linearrow[0][0][\h][#3];
      \linearrow[\w][0][\h][#4];
      \node () at (0,\h/2) {};
      \node () at (\w,\h/2) {};
    \end{scope}
  }
}

\newcommandx{\dhcontraction}[0]
{
  \fineq[-0.4ex]{
    \pgfmathsetmacro{\h}{0.8};
    \pgfmathsetmacro{\w}{0.3};
    \linearrow[0][0][\h][0];
    \linearrow[\w][0][\h][1];
    \linearrow[2*\w][0][\h][0];
    \linearrow[3*\w][0][\h][1];
    \node () at (0,\h/2) {};
    \node () at (\w,\h/2) {};
  }
}
\newcommandx{\idket}[0]{
  |
  \fineq[-0.6ex]{
    \idst[0][0][r]
  } \rangle 
}

\newcommandx{\swapket}[0]{
  |
  \fineq[-0.4ex]{
    \swapst[0][0][r]
  } \rangle 
}

\newcommandx{\idbra}[0]{
  \langle 
  \fineq{
    \idst[0][0][]
  } |
}

\newcommandx{\swapbra}[0]{
  \langle 
  \fineq{
    \swapst[0][0][]
  } |
}

\newcommandx{\idst}[3][1=0,2=0,3=]{
  \ifthenelse{\equal{#3}{r}}{
    \pgfmathsetmacro{\flag}{-1};
  }{
    \pgfmathsetmacro{\flag}{1};
  }
  \begin{scope}[shift={(#1,#2)}]
    \draw[line width = 0.5pt] (0,0) to[out=\flag*90,in=180] (0.15,\flag*0.2) to[out=0,in=\flag*90] (0.3,0);
    \draw[line width = 0.5pt] (0.4,0) to[out=\flag*90,in=180] (0.4+0.15,\flag*0.2) to[out=0,in=\flag*90] (0.7,0);
  \end{scope}
}

\newcommandx{\swapst}[3][1=0,2=0,3=]{
  \ifthenelse{\equal{#3}{r}}{
    \pgfmathsetmacro{\flag}{-1};
  }{
    \pgfmathsetmacro{\flag}{1};
  }
  \begin{scope}[shift={(#1,#2)}]
      \draw[line width = 0.5pt] (0,0) to[out=\flag*90,in=180] (0.4,\flag*0.25) to[out=0,in=\flag*90] (0.8,0);
      \draw[line width = 0.5pt] (0.3,0) to[out=\flag*90,in=180] (0.4,\flag*0.15) to[out=0,in=\flag*90] (0.5,0);
  \end{scope}
}

\newcommandx{\lowarc}[2][1=,2=]{
  \ifthenelse{\equal{#2}{r}}{
    \pgfmathsetmacro{\flag}{-1};
  }{
    \pgfmathsetmacro{\flag}{1};
  }
  \ifthenelse{\equal{#1}{}}{
    \draw[line width = 0.5pt] (0,0) to[out=\flag*90,in=180] (0.3,\flag*0.3) to[out=0,in=\flag*90] (0.6,0);
  }{
    \draw[line width = 0.5pt] (0,0) to[out=\flag*90,in=190] (0.3-0.2,\flag*0.28);
    \draw[line width = 0.5pt] (0.3+0.2,\flag*0.28) to[out=-10,in=\flag*90] (0.6,0);
    \node () at (0.3,\flag*0.28) {#1};
  }
}

\newcommandx{\higharc}[2][1=,2=]{
  \ifthenelse{\equal{#2}{r}}{
    \pgfmathsetmacro{\flag}{-1};
  }{
    \pgfmathsetmacro{\flag}{1};
  }
  \ifthenelse{\equal{#1}{}}{
    \draw[line width = 0.5pt] (0,0) to[out=\flag*90,in=180] (1,\flag*0.5) to[out=0,in=\flag*90] (2,0);
  }{
    \draw[line width = 0.5pt] (0,0) to[out=\flag*90,in=180] (1-0.2,\flag*0.55);
    \draw[line width = 0.5pt] (1+0.2,\flag*0.55) to[out=0,in=\flag*90] (2,0);
    \node () at (1,\flag*0.55) {#1};
  }
}

\newcommandx{\smallarc}[2][1=,2=]{
  \ifthenelse{\equal{#2}{r}}{
    \pgfmathsetmacro{\flag}{-1};
  }{
    \pgfmathsetmacro{\flag}{1};
  }
  \ifthenelse{\equal{#1}{}}{
    \draw[line width = 0.5pt] (0,0) to[out=\flag*90,in=180] (0.5,\flag*0.5) to[out=0,in=\flag*90] (1,0);
  }{
    \draw[line width = 0.5pt] (0,0) to[out=\flag*90,in=190] (0.5-0.2,\flag*0.45);
    \draw[line width = 0.5pt] (0.5+0.2,\flag*0.45) to[out=-10,in=\flag*90] (1,0);
    \node () at (0.5,\flag*0.45) {#1};
  }
}

\newcommandx{\oidst}[5][1=0,2=0,3=,4=,5=]{
  \begin{scope}[shift={(#1,#2)}]
    \smallarc[#3][#5]
    \begin{scope}[shift={(1.5,0)}]
      \smallarc[#4][#5]
    \end{scope}
  \end{scope}
}

\newcommandx{\oswapst}[5][1=0,2=0,3=,4=,5=]{
  \begin{scope}[shift={(#1,#2)}]
    \higharc[#3][#5]
    \begin{scope}[shift={(0.75,0)}]
      \lowarc[#4][#5]
    \end{scope}
  \end{scope}
}

\begin{document}
\date{\today}

\newcommand{\bra}[1]{\< #1 \right|}
\newcommand{\ket}[1]{\left| #1 \>}
\newcommand{\bbra}[1]{\<\< #1 \right|\right|}
\newcommand{\kket}[1]{\left|\left| #1 \>\>}

\title{Real-time correlators in chaotic quantum many-body systems}

\author{Adam Nahum}
\affiliation{Laboratoire de Physique de l'\'Ecole Normale Sup\'erieure, ENS, Universit\'e PSL, CNRS, Sorbonne Universit\'e, Universit\'e Paris-Diderot, Sorbonne Paris Cit\'e, Paris, France.}
\affiliation{Rudolf Peierls Centre for Theoretical Physics, Clarendon Laboratory, Oxford University, Parks Road, Oxford OX1 3PU, United Kingdom}
\author{Sthitadhi Roy}
\affiliation{International Centre for Theoretical Sciences, Tata Institute of Fundamental Research, Bengaluru 560089, India}
\affiliation{Rudolf Peierls Centre for Theoretical Physics, Clarendon Laboratory, Oxford University, Parks Road, Oxford OX1 3PU, United Kingdom}
\author{Sagar Vijay}
\affiliation{Department of Physics, University of California, Santa Barbara, CA 93106, USA}
\author{Tianci Zhou}
\affiliation{Kavli Institute for Theoretical Physics, University of California, Santa Barbara, CA 93106, USA}
\affiliation{Center for Theoretical Physics, Massachusetts Institute of Technology, Cambridge, MA 02139, USA}

\begin{abstract}
We study real-time local correlators $\langle\mathcal{O}(\mathbf{x},t)\mathcal{O}(0,0)\rangle$ in chaotic quantum many-body systems. These correlators show universal structure at late times, determined by the dominant operator-space Feynman trajectories for the evolving operator $\mathcal{O}(\mathbf{x},t)$. The relevant trajectories involve the operator contracting to a point at both the initial and final time and so are structurally different from those dominating the out-of-time-order correlator. In the absence of conservation laws, correlations decay exponentially: $\langle\mathcal{O}(\mathbf{x},t)\mathcal{O}(0,0)\rangle\sim\exp(-s_\mathrm{eq} r(\mathbf{v}) t)$, where $\mathbf{v}= \mathbf{x}/ t$ defines a spacetime ray, and  $r(\mathbf{v})$ is an associated decay rate. We express $r(\mathbf{v})$ in terms of cost functions for various spacetime structures. In 1+1D, operator histories can show a phase transition at a critical ray velocity $v_c$, where  $r(\mathbf{v})$ is nonanalytic. At low $v$, the dominant Feynman histories are "fat":  the operator grows to a size of order $t^\alpha\gg 1$ before contracting to a point again. At high $v$ the trajectories are "thin": the operator always remains of order-one size. In a Haar-random unitary circuit, this transition maps to a simple binding transition for a pair of random walks (the two spatial boundaries of the operator). In higher dimensions, thin trajectories always dominate. We discuss ways to extract the butterfly velocity $v_B$ from the time-ordered correlator, rather than the OTOC. Correlators in the random circuit may alternatively be computed with an effective Ising-like model: a special feature of the Ising weights for the Haar brickwork circuit gives $v_c=v_B$. This work addresses lattice models, but also suggests the possibility of morphological phase transitions for real-time Feynman diagrams in quantum field theories.
\end{abstract}

\maketitle

\tableofcontents

\section{Introduction}

This paper is about dynamical correlation functions in chaotic quantum many-body systems.
Our aim is to  characterise the  spacetime processes that contribute to correlation functions of the form
\begin{equation}
G(\vec x, t) = \< \mathcal{O}({\bf x},t)\mathcal{O}(\vec 0,0)\>.
\end{equation}
Here, $\mathcal{O}$ is a local operator in (for concreteness) a lattice spin model.  Throughout most of the paper we will consider infinite temperature, so that the expectation value is  $G(\vec x, t) = \Tr \lf \mathcal{O}({\bf x},t)\mathcal{O}(\vec 0,0) \ri / \Tr \mathbb{1}.$

In a system that  is able to equilibrate, $G(\bf x,t)$ will decay to zero at late times, but the nature of this decay is universal.
In systems with slow hydrodynamic modes there are tails whose basic features can be understood from classical  hydrodynamics \cite{kadanoff_hydrodynamic_1963,chaikin1995principles}. 
However our starting point will instead be systems with no slow hydrodynamic modes, where the late time relaxation of correlations is exponential, because even in this simplest case there is universal structure in the relaxation dynamics.

\begin{figure}[t]
\centering
\includegraphics[width=\columnwidth]{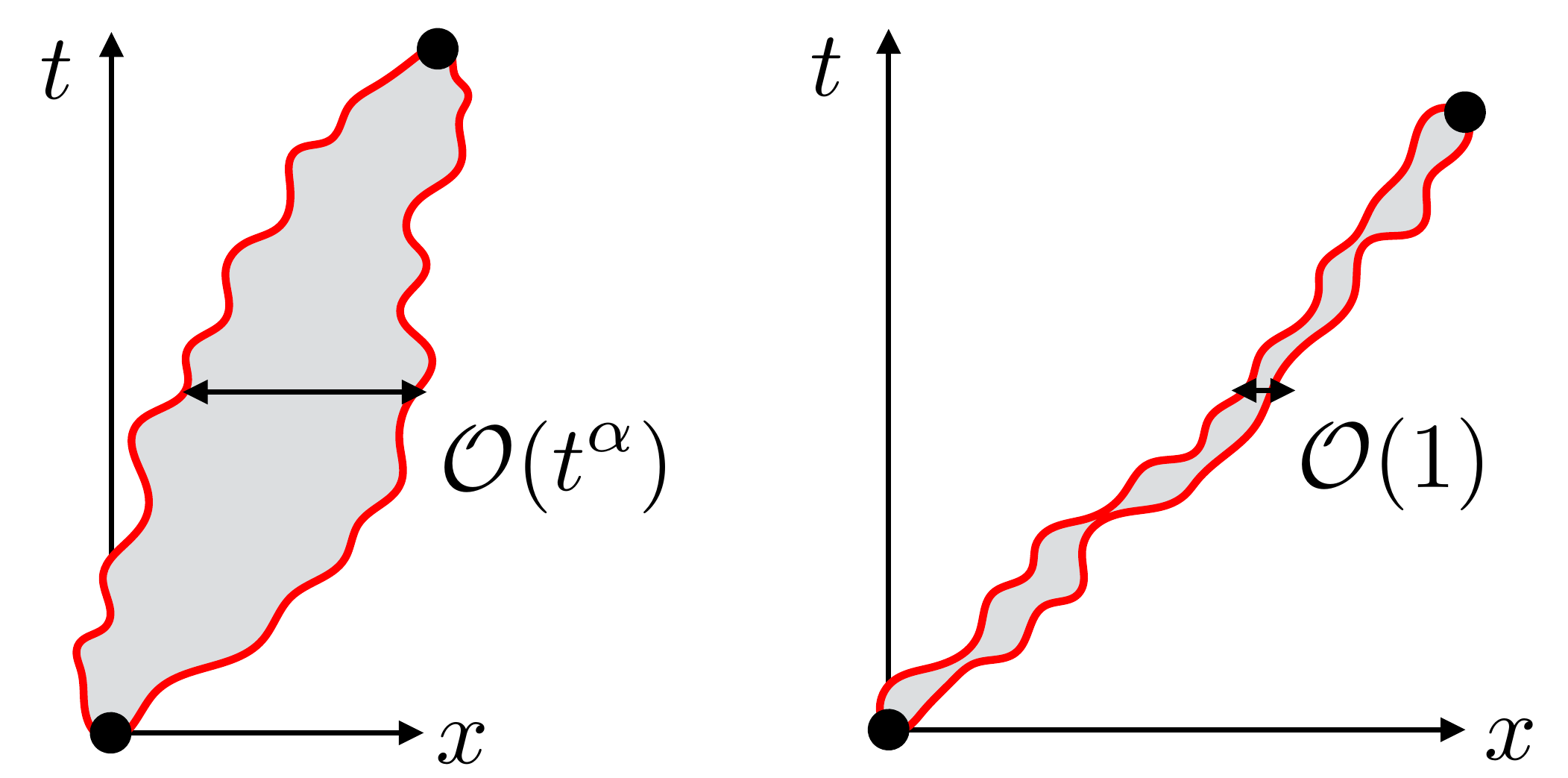}
\caption{Two types of operator Feynman history that may contribute to the local correlator. 
In both cases an operator string (product of local basis operators) propagates from $(0,0)$ to $(x,t)$.
Left: The support of the operator string grows parametrically large in $t$ before shrinking to a point ($\alpha=1/2$ in the simplest case).
Right: The typical size of the string remains of order 1 throughout the trajectory.}
\label{fig:path_cartoon}
\end{figure}

A correlator such as $G(\vec x,t)$ may be written as a sum over Feynman histories in operator space, describing the evolution of the Heisenberg-picture operator $\mathcal{O}(\vec x,t')$ from $t'=0$ to $t'=t$.
The late time decay of the correlator along a given ray in spacetime is generically exponential, so we define a rate function $r(\vec v)$ for each velocity $\vec v$:
\begin{align}
\label{eq:Cdecayintro1}
G(\vec x, t) &  \asymp \exp \lf - s_\text{eq}\, r(\vec v)\, t \ri,
&
\vec v & = \vec x / t
\end{align}
($s_\text{eq}$ is the equilibrium entropy density of the system and is included for later convenience).
We will see that the rate function $r(\vec v)$ may be understood in terms of ``costs''  associated with different types of Feynman histories. 
Fig.~\ref{fig:path_cartoon} is a cartoon of two kinds of history that can contribute in 1+1D. 
We will use random circuits as analytically tractable examples,  building on discussions of operator spreading and the out-of-time-order correlator (OTOC) in random circuits in Refs.~\cite{nahum_operator_2018,von_keyserlingk_operator_2018,nahum2018dynamics}.

Our first task is to determine what kinds of history dominate. 
We find that, for a class of 1+1D systems, there is a phase transition between the two types of trajectory in Fig.~\ref{fig:path_cartoon} as a function of the ray velocity $v$.  
At large velocities, the trajectories resemble the one shown on the right: the support of the operator remains of order one size during its evolution.
At small velocities, on the other hand, the  operator becomes parametrically large before contracting again, 
as in the left panel.  
This transition leads to a nonanalyticity in the decay rate $r(\vec v)$ in Eq.~\ref{eq:Cdecayintro1}. 
The phase transition,
which occurs at a critical velocity $v_c$,
maps to an unbinding transition for a pair of random walks.

Interestingly, in the simplest 1+1D random circuit, made of Haar-random unitary gates,
the location $v_c$ of this ``morphological transition'' coincides with the butterfly velocity $v_B$ that appears in the out-of-time-order correlator. 
However, this is not generic, and in other 1+1D circuits $v_c$ and $v_B$ can differ; the ``fat'' phase can even be eliminated entirely, for example if interactions have a large enough range.
The transition between the two phases can also be crossed by tuning a model parameter.

In higher dimensions, in contrast to 1+1D, we find that ``thin'' trajectories always dominate. (This is due to a cost for a fat trajectory  that scales with the size of the perimeter of the operator's footprint.) 
This means that in a sense the two-point function $G(\vec 0, t)$ in a generic higher-dimensional circuit can in fact be simpler than in 1+1D.

We emphasize that the operator trajectories contributing to $G(\vec x, t)$ are very different from those contributing to the OTOC.
The OTOC probes ``typical'' Feynman histories for $\mathcal{O}({\bf x},t')$, in which the operator grows ballistically and has size proportional to $v_B t$ at the final time
(rate functions may be defined for the OTOC too \cite{xu2020accessing,khemani2018velocity}).
By contrast, the  trajectories contributing to the two-point function are ones in which $\mathcal{O}(\vec x,t')$ becomes ``small'' at $t'=t$, in order to overlap with the other operator $\mathcal{O}(\vec 0,0)$ in the trace that defines $G$.
These trajectories are highly atypical, and this gives rise to the exponential suppression of $G(\vec x,t)$. 
In the random circuit, operator averages can be mapped to a Markov process \cite{oliveira_generic_2007,dahlsten_emergence_2007,znidaric_exact_2008,harrow_random_2009,nahum_operator_2018,von_keyserlingk_operator_2018,khemani_operator_2018}, and this allows ``atypicality'' to be understood via an exact correspondence with rare events in the Markov process.\footnote{To be more precise, this mapping holds for the simplest quantity, which is the average of $G^2$ over the random unitaries. We will comment below on how to take fluctuations into account.}
(Despite the fundamental difference between the OTOC and the two point function, we will argue that in a class of 1+1D models it is possible to deduce $v_B$ from a two-point function.)

One of our basic results is that the rate function $r(\vec v)$ can be expressed in terms of ``line tensions'' of certain types of  path in spacetime. 
This is reminiscent of various lattice statistical mechanics models, 
in which correlations can be given a sum-over-paths formulation, 
and the  decay constant, characterising exponential decay of correlations in a disordered phase,
 may be interpreted as a renormalized free energy for a path.\footnote{For example, the high-temperature expansion of the classical Ising model relates the correlator of two spins to an effective partition function for a path, connecting the two spins. The exponential decay length of the correlator is equal to the line tension (free energy per unit length) of this path \cite{kardar2007statistical}.}
Here, however, the meaning of the relevant paths depends on the nature of the dominant operator histories. In cases like the left panel of Fig.~\ref{fig:path_cartoon}, the relevant paths through spacetime are the spacetime trajectories of the \textit{boundaries} (left and right)  of the operator string, marked in red in the figure.  But in cases like the right panel of Fig.~\ref{fig:path_cartoon}, the relevant paths are simply paths of the \textit{operator} itself. 
In  this situation we can say  that the left and right boundaries have formed a ``bound state'', and it is the line tension of this bound state that we need to compute. 

These concepts, for example the assignment of rate functions (line tensions) to different types of spacetime trajectory, can be applied to general ``realistic'' many-body Hamiltonians that may have no randomness, as we discuss. However, random circuits are a useful testing ground where explicit calculations are possible. 

Our plan is as follows.
We will first clarify the type of Feynman trajectories that are of interest to us (Sec.~\ref{sec:operatorhistories}),
define the basic rate functions (Sec.~\ref{sec:classesoftrajectory}), and note some constraints that they satisfy.
We then do the simplest calculation of the rate functions, for a 1+1D Haar-random circuit. This illustrates many of our basic points
(Sec.~\ref{sec:Haartransition}). We also discuss some subtleties relating to disorder.

We discuss higher dimensions in 
Sec.~\ref{sec:higherdims}.

In  Sec.~\ref{sec:Ising} 
we revisit the calculation for the 1+1D Haar-random circuit in a  different language,
that of an effective Ising model \cite{nahum_operator_2018,zhou_emergent_2019,von_keyserlingk_operator_2018,hunter2019unitary}. 
This clarifies which features of the Haar circuit are general. 
The relation ${v_c=v_B}$  is obeyed by the simplest Haar brickwork circuit, but not for more general circuits: we explain this in terms of the  vanishing of a particular vertex weight in the effective Ising model for the Haar circuit. 
This feature may be relevant for other applications of the effective Ising model for the Haar circuit.

Sec.~\ref{sec:numericalcasestudy} is a numerical case study of noisy 1D Hamiltonians whose rate functions have some differences to the Haar case. 

In Sec.~\ref{sec:instances} we discuss applications to more general systems, in particular systems without randomness.
We also make connections with recent works on efficient numerical methods for computing correlators \cite{von2021operator}
 and on dual unitary circuits  
\cite{kos_correlations_2021}. 
Finally in Sec.~\ref{sec:outlook}  we discuss various  questions for the future.

\section{Operator Feynman histories}
\label{sec:operatorhistories}

\subsection{Basic setup}

We will consider chaotic unitary dynamics, in discrete time, for a lattice of $q$-state spins (for example $q=2$ for spin-1/2).
For concreteness we may imagine a 1+1D quantum circuit with the brickwork layout shown in Fig.~\ref{fig:circuit_walk}, though the basic definitions in this section are independent of the structure of evolution operator (for example the presence or absence of translation invariance) or the spatial dimensionality. 
We often refer to the case  where the circuit is made up of Haar-random two site [i.e. $\mathrm{U}(q^2)$] unitaries in a brickwork pattern, see  Fig.~\ref{fig:circuit_walk}.

The time evolution operator from time 0 to time $t$, written as $U(t;0)$, is a product of unitaries $U_\tau$ for individual timesteps:
\begin{equation}
U(t;0) = U_t U_{t-1} \ldots U_1.
\end{equation}
In the brickwork circuit example, we take a single timestep to be a single layer of the circuit. 

We assume for now that the model has no conserved quantities, and (in the thermodynamic limit) a unique local equilibrium state given by the infinite temperature Gibbs state. We will always compute quantum expectation values in this ensemble:
\begin{equation}
\label{eq:definegibbsstate}
\<\cdots\>\equiv \f{\Tr(\cdots)}{\Tr \mathbb{1}}.
\end{equation}
We will study time-ordered two-point functions of the form
(we place the earlier operator at the origin for notational simplicity)
\begin{equation}
\label{eq:twoptfnbasic}
G(\vec x,t) = \< \mathcal{O}(\vec x,t) \mathcal{O}'(\vec 0,0) \>,
\end{equation}
where $\mathcal{O}(\vec x,0)$  and $\mathcal{O}'(\vec 0,0)$  are local Hermitian operators at sites $\vec x$ and $\vec 0$ respectively, and the time-evolved operator is 
\begin{equation}
 \mathcal{O}(\vec x,t) = U^\dag_t  \mathcal{O}(\vec x,0) U_t. 
\end{equation}
We may take the local operators to be traceless without loss of generality. Then, assuming that in the thermodynamic limit the dynamics under consideration is able to reach local equilibrium, 
 $G(\vec x,t)$ will relax to zero in the limit of large $t$, and our focus will be on the nature of this relaxation.

In the simple chaotic models we will  consider, which have a unique local equilibrium state, this relaxation is exponential and so we may characterise it by a rate function $r(\vec v)$, which is independent of the choice of local operators $\mathcal{O}$ and $\mathcal{O}'$:\footnote{Here $A\asymp B$ means that ${(\ln A)/(\ln B) \rightarrow 1}$ at large $t$.}
\begin{align}\label{eq:Gratefn}
|G(\vec x,t)| & \asymp \exp \lf - \seq \, r(\vec v) \, t \ri, 
& 
\vec v & = \vec x / t.
\end{align}
In the case where the circuit is made up of random unitaries it will be necessary to define the meaning of the left hand side more precisely: for example the rate function for the typical value of $|G(\vec x,t)|$ will differ from that for the average. We defer this point until later, see Sec.~\ref{sec:Haarsimplification}.

\subsection{Evolution in Pauli string basis}
\label{sec:evolutionstring}

Let us make a slight change to our notational convention which will make the boundary conditions more natural in later calculations.
Eq.~\ref{eq:twoptfnbasic} is written in the standard Heisenberg picture,  where ${\mathcal{O}({\bf x},t) = U_t^\dag \mathcal{O}({\bf x},0) U_t}$. 
We may also write
\begin{equation}
\label{eq:twoptfnreverse}
G(\vec x,t) = \< \mathcal{O}(\vec x,0) \mathcal{O}'_r(\vec 0,t) \>,
\end{equation}
where we define $\mathcal{O}'_r(\vec 0,t)$ by 
${\mathcal{O}'_r(\vec 0,t) = U_t  \mathcal{O}^\prime(\vec 0,0) U_t^\dag}$ (the subscript $r$ indicates that we have reversed the convention). This (equivalent) rewriting  of the correlator will be slightly more convenient below.

It is convenient to express the evolving operator $\mathcal{O}^\prime_r(\vec 0, t)$ as a superposition of ``strings'' $\mathcal{S}$, i.e. products of local Hermitian basis operators \cite{roberts_localized_2015-1,ho_entanglement_2017-1}.
 To simplify notation, let us restrict for now to the $q=2$ case. 
The strings may then be taken to be products of Pauli matrices at distinct lattice sites:
\begin{equation}
\mathcal{O}^\prime_r(\vec 0, t) = \sum_\mathcal{S} a_\mathcal{S}(t) \mathcal{S}.
\end{equation}
The strings are orthonormal, i.e. $\< \mathcal{S}\mathcal{S}'\> = \delta_{\mathcal{S},\mathcal{S}'}$.  It is convenient to normalize $\mathcal{O}$ so that  $\< \mathcal{O}^2\> = 1$; then the weights are normalized as
\begin{equation}
\sum_\mathcal{S} a_\mathcal{S}(t)^2 = 1.
\end{equation}
Making a formal analogy between the quantum operator and a quantum state, the coefficients $a_\mathcal{S}$ are the wavefunction amplitudes in the basis of product operators $\mathcal{S}$, and the weights $a_\mathcal{S}^2$ are the quantum probabilities associated with these basis operators.

The amplitudes $a_\mathcal{S}$ evolve in a given time step with a unitary matrix $V^{(t)}_{\mathcal{S}, \mathcal{S'}}$, \be
a_\mathcal{S}(t) = \sum_{\mathcal{S}'} \UU^{(t)}_{\mathcal{S}, \mathcal{S}'} \, a_{\mathcal{S}'}(t-1), 
\end{equation}
which is just a rewriting of the evolution for a single time-step,
 $\mathcal{O}^\prime_r( \vec 0,t ) = U_t \mathcal{O}^\prime_r( \vec 0, t-1 ) U_t^\dag$, 
 in the string basis. 
Explicitly, $\UU^{(t)}_{\mathcal{S}, \mathcal{S}'}  = \big\langle \mathcal{S} U_t \mathcal{S}' U_t^\dag \big\rangle$.
Formally, $\UU$ is the unitary evolution operator $U_t \otimes U_t^*$ that acts on the Hilbert space of operators, written out in a particular choice of basis given by Pauli strings. 
Thinking of operators as vectors in a doubled Hilbert space --- sometimes referred to as ``superspace'' --- is useful in many areas~\cite{mori1965transport,mermin1966absence,auerbach2019equilibrium}.

The matrix $\UU^{(t)}$ describes allowed transitions between operator strings. 
In a finite chain of $L$ sites, $\UU^{(t)}$ is a ${4^L\times 4^L}$ matrix ($q^{2L}\times q^{2L}$ in the general case). 
It necessarily has a $1\times 1$ block for the trivial string ${\mathcal{S}=\mathbb{1}}$, which is invariant under any unitary dynamics.
The action  of $\UU^{(t)}$ is constrained by locality: for example in a circuit model a local string can only grow within the lightcone.

For concreteness, let us take our local operators $\mathcal{O}'(\vec 0,0)$ and $\mathcal{O}(\vec x, 0)$ to be the local Pauli-Z  operators $Z_{\vec 0}$ and $Z_{\vec x}$.
Since these will set the initial ($I$) and final ($F$) conditions for the strings in the Feynman path expansion below, we write from now on
\begin{align}
\mathcal{S}_I & \equiv Z_{\vec 0}, 
& 
\mathcal{S}_F & \equiv Z_{\vec x}. 
\end{align}
Then the correlator (\ref{eq:twoptfnbasic}) just extracts   
the amplitude of the final (target) string $\mathcal{S}_F$ in the time-evolved $\mathcal{S}_I$:
\begin{align}
\label{eq:Gasamplitude}
G(\vec x, t)  = a_{\mathcal{S}_F}(t), \,\,\, \text{with $a_{\mathcal{S'}}(0) = \delta_{\mathcal{S}, \mathcal{S}_I}$}.
\end{align}
We may write the desired amplitude $a_{\mathcal{S}_F}(t)$ at time $t$ as a sum over histories $(\mathcal{S}_F, \mathcal{S}_{t-1}, \ldots, S_1,\mathcal{S}_I)$ of the operator. 
To avoid clutter, we suppress the arguments of $G$ in Eq.~\ref{eq:Gasamplitude}:
\begin{equation}\label{eq:GVVVV}
G =  \sum_{\mathcal{S}_{t-1}, \ldots \mathcal{S}_{1}} 
\UU^{(t)}_{\mathcal{S}_F, \mathcal{S}_{t-1}}
\ldots
\UU^{(2)}_{\mathcal{S}_2, \mathcal{S}_{1}}
\UU^{(1)}_{\mathcal{S}_1, \mathcal{S}_{I}}. 
\end{equation}
This product of elements of $\UU$ gives the amplitude for a given Feynman trajectory ${(\mathcal{S}_F, \mathcal{S}_{t-1}, \ldots, S_1,\mathcal{S}_I)}$.
In Eq.~\ref{eq:GVVVV} time is discrete but analogous expressions can of course be written for continuous time evolution.

Note that the operator string propagates from the spacetime point ${({\bf 0},0)}$ to the point ${({\bf x},t)}$: this was the reason for using the nonstandard convention in Eq.~\ref{eq:twoptfnreverse}, where the evolution operators were grouped with  $\mathcal{O}'$ rather than with $\mathcal{O}$ .

\subsection{Simplifications in Haar circuits}
\label{sec:Haarsimplification}

This kind of expansion can be made (in either discrete or continuous time) for almost any model. 
In the Haar random circuit there is a significant simplification when we consider the average of $a_{\mathcal{S}} (t)^2$ over circuit realisations, or equivalently the root-mean-square (rms) value of the correlation function: 
\begin{equation}\label{eq:rmsdefn}
\Grms(\vec x,t)^2 = \overline{G( \vec x,t)^2}.
\end{equation}
The overline represents the average over random unitaries.
This object can be written as the probability of a rare event in a classical stochastic cluster growth process \cite{nahum_operator_2018,von_keyserlingk_operator_2018, nahum2018dynamics}.
This mapping makes the analysis of different types of operator Feynman history very intuitive.

In the Haar circuit the average $\overline{G}$ vanishes by trivial phase cancellation reasons, so $\Grms$ is the simplest nontrivial average.
We should be aware that $\Grms$ will in general not be numerically close to the typical value of $|G|$ defined by 
\begin{equation}
\label{eq:Gtypdefn}
 \Gtyp(\vec x,t) = \exp \, \overline{\ln |G( \vec x,t) |}.
\end{equation}
We will discuss $\Gtyp$ later in 
Sec.~\ref{sec:avversustypbrief}.
For now, studying $\Grms$ is sufficient to understand some basic features of operator Feynman histories that we argue are more general.

In general $a_{\mathcal{S}} (t)^2$ is a double sum, over two trajectories, 
$(\mathcal{S}_F, \mathcal{S}_{t-1}, \ldots, S_1,\mathcal{S}_I)$  and 
$(\mathcal{S}_F, \tilde{\mathcal{S}}_{t-1}, \ldots, \tilde{\mathcal{S}}_1,\mathcal{S}_I)$. 
But when we average, this double sum collapses to a single sum, because of the result  \cite{oliveira_generic_2007,dahlsten_emergence_2007,znidaric_exact_2008,harrow_random_2009}  (we restrict to nontrivial strings, $\mathcal{S}\neq \mathbb{1}$) 
\begin{equation}
\overline{\UU^{(t)}_{\mathcal{S}, \mathcal{S}'} \UU^{(t)}_{\tilde{\mathcal{S}}, \tilde{\mathcal{S}}'}} 
= 
\delta_{\mathcal{S}, \tilde{\mathcal{S}}}
\delta_{\mathcal{S}', \tilde{\mathcal{S}}'}
T^{(t)}_{\mathcal{S}, \mathcal{S}'}.
\end{equation}
We will specify $T^{(t)}$ below.
The Kronecker deltas force the two histories to be the same, and we have 
\begin{equation}
\label{eq:G2Ttrajectories}
\overline{G^2} =  \sum_{\mathcal{S}_{t-1}, \ldots \mathcal{S}_{1}} 
T^{(t)}_{\mathcal{S}_F, \mathcal{S}_{t-1}}
\ldots
T^{(2)}_{\mathcal{S}_2, \mathcal{S}_{1}}
T^{(1)}_{\mathcal{S}_1, \mathcal{S}_{I}}.
\end{equation}
Furthermore, the matrix $T^{(t)}_{\mathcal{S}, \mathcal{S}'}$ is the transition matrix for a classical stochastic process \cite{oliveira_generic_2007}, in which the basis string is randomly updated, in a local fashion, each time a unitary is applied to a pair of sites.
The transition probabilities of this stochastic process are simple and are reviewed in Appendix.~\ref{app:transitionrates}.
(The explicit $t$-dependence  of $T^{(t)}_{\mathcal{S}, \mathcal{S}'}$ is trivial and  arises only from the even-odd structure of the circuit illustrated in Fig.~\ref{fig:circuit_walk}, or analogous higher-dimensional geometries.) 

For a local operator, this process simplifies to a stochastic dynamics of the boundary of the operator cluster  \cite{nahum_operator_2018, von_keyserlingk_operator_2018}.
Let the ``occupation numbers'' $n_{\bf x}$ of the sites be ${n_{\bf x}= 1}$ (represented  $\begin{tikzpicture}
\fill[black] (0,0) circle (0.1cm); \end{tikzpicture}$) if  the site is in the string and ${n_{\bf x}=0}$ (represented  as  $\begin{tikzpicture}
\draw (0,0) circle (0.1cm); \end{tikzpicture}$) if it is not.
These occupation numbers undergo a simple stochastic dynamics. 
We will refer to the occupied sites (the support of the string) as the cluster.  In 1+1D, let $x_L$ and $x_R$ be the left and right endpoints of the operator string ${(x_L \leq x_R}$). 
The two points $x_L$, $x_R$ obey their own, autonomous, stochastic dynamics.
That is, the dynamics of the boundary points of the cluster are independent of its internal structure. 
When they are separated, each point does a simple random walk, with a net drift velocity that is equal to $-v_B$ for $x_L$, and $+v_B$ for $x_R$, 
reflecting the tendency of the operator support to grow with ``butterfly speed'' $v_B$~\cite{roberts_lieb-robinson_2016,roberts_localized_2015-1}. The two walkers have a contact interaction when they collide (Sec.~\ref{sec:Haartransition}).

With these simplifications, the dynamics of the operator string reduces to random classical motion of two points $x_L(t)$, $x_R(t)$.
This reduction is exact in the Haar circuit, but we will argue later for similar structures in a much broader range of systems. Loosely speaking, the idea is that the two boundaries of the cluster are the relevant ``slow'' degrees of freedom, because the interior of the string $\mathcal{S}$ rapidly reaches a simple local equilibrium.

This equilibrium is very simple: a given site is equally likely to be any of the basis operators; e.g. equally likely to be $\mathbb{1}$, $X$, $Y$, $Z$ in the $q=2$ case. As a consequence of unitarity, the stochastic process preserves this equilibrium state.

\section{Classes of trajectory in 1+1D}
\label{sec:classesoftrajectory}

\subsection{Defining line tensions}\label{sec:defininglinetensions}

We now characterise various types of trajectory in 1+1D, assuming for now that we have a mapping to an effective stochastic process for a string $\mathcal{S}$ like that described above.
In this Section we jump ahead to a coarse-grained picture, anticipating the microscopic calculation of Sec.~\ref{sec:Haartransition}.

We will define three line tensions: 
one associated with the spacetime trajectory of the right endpoint of the string, $r_R({\bf v})$, 
one associated with the left endpoint, $r_L({\bf v})$,
and one associated with their ``bound state'', $r_B({\bf v})$.
These determine the exponential costs associated with different kinds of spacetime trajectory.
In the Markov picture, these costs are ``rate functions'' setting the probabilities  of various kinds of rare event.

To define $r_R(v)$,  imagine a semi-infinite cluster, 
with $x_L\rightarrow -\infty$ and $x_R(0)=0$,
that is initially equilibrated in its interior. 
At a large time $t$, the probability that $x_R(t)$ has travelled a distance ${v}t$ 
(which can be positive, negative, or zero) 
scales as 
\begin{equation}
\label{eq:rRdefn}
P\big[x_R = v t\big] \asymp \exp \lf {- \seq \, r_R(v)\, t} \ri.
\end{equation}
$r_L(v)$ is defined analogously by considering a left endpoint of an operator that is semi-infinite in the opposite direction.
The rate functions $r_{R,L}(v)$ are non-negative, 
convex functions of $v$. Each vanishes at a single velocity, which is the  butterfly velocity $v_{R,L}$ for the corresponding endpoint. This is because if $v_R$ is the typical velocity for the right endpoint, then $r_R(v_R)$ must be zero.

Next, consider a cluster of finite size supported initially between $[x_L(0), x_R(0)]$, and consider the probability that at time $t$ the endpoints have advanced by a distance $vt$ to $[x_L(0) + vt, x_R(0) + vt]$. Taking the limit of large $t$ with ${x_R(0)-x_L(0)}$ \textit{fixed}, there are two possibilities for the nature of the trajectories.
Either the typical separation ${x_R-x_L}$ remains of order 1 throughout the trajectories, or the typical separation during the trajectory [defined say as ${x_R(t/2)-x_L(t/2)}$] diverges  in the large $t$ limit.
In the former case we say that the endpoints are bound together, for the given velocity $v$, and we define the bound state line tension using the above probability:
\begin{equation}
P\Big[ (x_L,x_R)_t =(x_L,x_R)_0 + v t\Big] \asymp \exp \lf {- \seq \, r_B(v)\, t} \ri.
\end{equation}
$r_B(v)$ is well-defined, for a given $v$, only if bound trajectories dominate over trajectories in which $x_L$ and $x_R$ wander parametrically far apart.
The cost of trajectories of the latter kind can be computed using the line tensions $r_{R,L}$ for isolated endpoints, and is ${\sim\exp (- \seq [r_R(v) + r_L(v)]t)}$. 
Therefore, 
\begin{align}
\notag
& \text{if $x_L$, $x_R$ are bound at velocity $v$,} \\ 
& \text{then:} \, \, \, \, r_B(v) \leq r_R(v) + r_L(v).\label{eq:rBrRrLinequality}
\end{align}
While $r_{R,L}(v)$ are well-defined for any $v$ (if we allow infinity as a value), 
$r_B(v)$ is only well-defined in a range where (\ref{eq:rBrRrLinequality}) holds. This range could be empty.

\begin{figure}[t]
\centering
\includegraphics[width=\columnwidth]{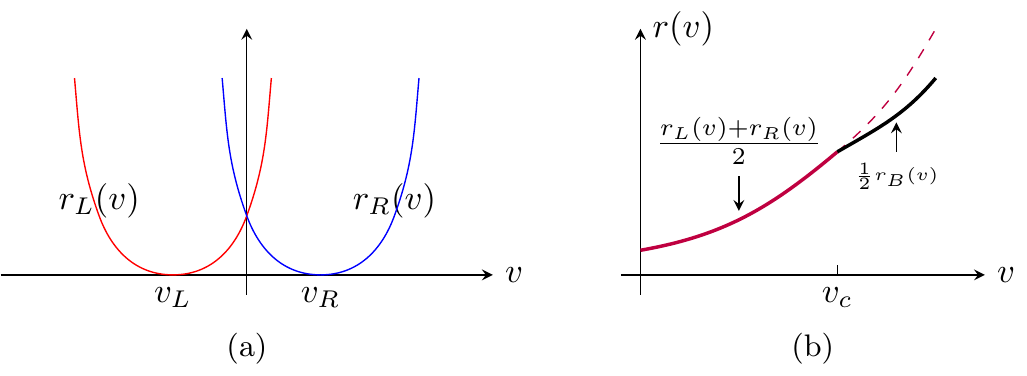}
\caption{Schematic of (a) $r_L(v)$ and $r_R(v)$ and (b) $r(v)$.
We show the case where a bound state exists for ${v>v_c}$ but not for ${v<v_c}$ (as in the Haar circuit). See Eq.~\ref{eq:rvalternative}.}
\label{fig:rL_rR}
\end{figure}

Fig.~\ref{fig:rL_rR} illustrates a scenario which holds in various 1+1D random circuits: the bound state exists for large enough speed, but is absent below a critical speed $v_c$ at which  (\ref{eq:rBrRrLinequality}) becomes an equality.

\subsection{Some symmetry relations}
\label{sec:symmrels}

We have already mentioned that
\begin{align}
r_R(v_{R}) &= 0,  &
r_L(v_{L}) &= 0.
\end{align}
Additional constraints follow
from symmetries. For simplicity, 
we  continue to assume the effective Markovian picture which holds for $G_\text{rms}$ in the random circuit 
(we defer a  discussion of symmetry constraints in the more general setting to the future).

Time-reversal symmetry, which is present in the random circuit after averaging, leads to  detailed balance in the effective Markov process of the \textit{string} $\mathcal{S}$, which gives the constraints:
\begin{align}
\label{eq:detailedbalancerels}
r_R(v) & = r_R(-v) - 2 v, \\
r_L(v) &=  r_L(-v)  + 2 v, \\
r_B(v) & = r_B(-v).
\end{align}
For example, the first of these comes from the relating the probability of  the process 
$x_R \rightarrow  x_R+d$ 
(for a semi-infinite cluster)
to the probability of the 
reverse process $x_R+ d\rightarrow x_R$.
The ratio of these probabilities is the inverse of the ratio of equilibrium probabilities for the two possible situations, and this is simply computed because the  equilibrium state of a string is trivial --- see App.~\ref{app:transitionrates}.

Spatial reflection (parity) symmetry, which is also present in the  the random circuit after averaging, gives  the symmetry relations
\ba 
r_L(v) & = r_R(-v), & 
r_B(v) & = r_B(-v).
\end{align}

Finally, in the  mapping of $G_\text{rms}$ in the random circuit to an effective Ising model that is discussed in Sec.~\ref{sec:Ising},
the replicalike symmetry of the Ising model implies
\be\label{eq:relnfromreplica}
r_R(v) = r_L(v) - 2v.
\ee
Eq.~\ref{eq:relnfromreplica} is also implied by the combination of time reversal and parity, but the Ising picture implies that it remains  valid if these symmetries are broken.\footnote{Ref.~\cite{stahl2018asymmetric} gives an example of a random circuit that breaks time reversal and parity, but in that specific case the combination of time reversal and parity  is preserved.}
Eq.~\ref{eq:relnfromreplica} means that in the random circuit $r_R$ and $r_L$ can be written in terms of a single function as
\begin{align}
r_R(v) &= \mathcal{E}_2(v) -v,
&
r_L(v) & = \mathcal{E}_2(v) +v.
\end{align}
The quantity $\mathcal{E}_2(v)$ is  an ``entanglement line tension'' associated with the averaged  purity \cite{jonay_coarse-grained_2018,zhou_entanglement_2020}, 
as discussed in Sec.~\ref{sec:Ising}.

At this point we have two independent functions, the line tension $r_R(v)$ for a single endpoint, and the line tension $r_B(v)$ for a bound state. 
A further simplification special to the 1+1D Haar-random brickwork circuit means that there $r_B(v)$ can \textit{also} be expressed in terms of $\mathcal{E}_2(v)$, so that in that particular case all of the functions introduced in this section can be expressed in terms of a single function.

Let us briefly relate the line tension functions $r_{L,R}(v)$ defined above to the OTOC, defined as
\begin{equation}
\operatorname{OTOC}({\vec x},t) = - \f{1}{2} \< [\mathcal{O}({\vec 0},t),\mathcal{O}({\vec x},0)]^2 \>.
\end{equation}
An effective lightcone may be defined using the OTOC: it includes those velocities ${\vec v}$ such that ${\operatorname{OTOC}({\vec v} t,t)}$ saturates to unity at large $t$.
In {1+1D} the lightcone is given by left  and right butterfly velocities ${(v_L, v_R)}$.
For rays outside this lightcone there is exponential decay \cite{khemani2018velocity,xu2020accessing,xu_locality_2019},
\begin{equation}
\operatorname{OTOC}({\vec x},t)  \asymp \exp ( - \seq \lambda(v) t ) 
\end{equation}
which defines another rate function $\lambda(v)$. This is simply related to $r_{R,L}(v)$.

In the random system, we must be more precise by specifying a type of average on the left-hand-side.
In the random circuit, $\overline{\operatorname{OTOC}({\vec x},t)}$
is proportional to the probability [in the Markov process for a string that starts at ${({\vec 0},0)}$]
that  ${({\vec x},t)}$ lies inside the string \cite{nahum_operator_2018,von_keyserlingk_operator_2018, khemani_operator_2018,rakovszky2018diffusive}.
Typically the right endpoint $x_R$ travels at velocity $v_R$, and the left endpoint at $v_L$, so that $\lambda(v)=0$ within the lightcone.
For ${v>v_R}$, $\lambda(v)$ is set by the probability that $x_R$  travels atypically far, so
\begin{align}
\label{eq:lambdaotoc}
\lambda(v) & = r_R(v) & &  \text{for $v>v_R$}, \\
\lambda(v) & = r_L(v) & &  \text{for $v<v_L$}.
\end{align}

\subsection{Fat and thin trajectories}

The trajectories contributing to $\overline{G(x,t)^2}$ have 
${x_L(0)=x_R(0)=0}$ and ${x_L(t)=x_R(t)=vt}$, where ${v=x/t}$.
Since the dominating trajectories can be either fat or thin, we have 
\begin{align}
\label{eq:Cdecayintro}
G_\text{rms}( x, t) &  \asymp \exp \lf - s_\text{eq}\, r( v)\, t \ri,
&
 v & =  x / t,
\end{align}
with
\begin{align}
\label{eq:rvalternative}
r(v) & = 
\f{1}{2} \times \left\{
\begin{array}{l}
\, r_R(v) + r_L(v)  \\
\, r_B(v),
\end{array}
\right.
\end{align}
with the second line on the RHS holding whenever the bound state exists, in which case the second line  is smaller than the first.
Using reflection symmetry, ${r_L(v)=r_R(-v)}$, and the identity Eq.~\ref{eq:detailedbalancerels} we can simplify the first line:
\begin{align}
\label{eq:rvalternative2}
r(v) & = 
 \left\{
\begin{array}{l}
\, r_R(v) + v  \\
\, \f{1}{2} \times r_B(v).
\end{array}
\right.
\end{align}
Eqs.~\ref{eq:Cdecayintro},~\ref{eq:rvalternative2} specify the exponential decay rate that gives the leading scaling of ${\ln \overline{G^2}}$ at large times. There are of course also subleading terms   (Sec.~\ref{sec:avversustypbrief}).

\cref{fig:rL_rR} in the previous subsection shows the type of nonanalyticity we can have in $r(v)$. In a range of $|v|$ at large $|v|$  we must choose the lower branch of the figure (the bound state tension).

\section{Binding transition in Haar circuit}
\label{sec:Haartransition}

\begin{figure}[h]
\centering
\includegraphics[width=\columnwidth]{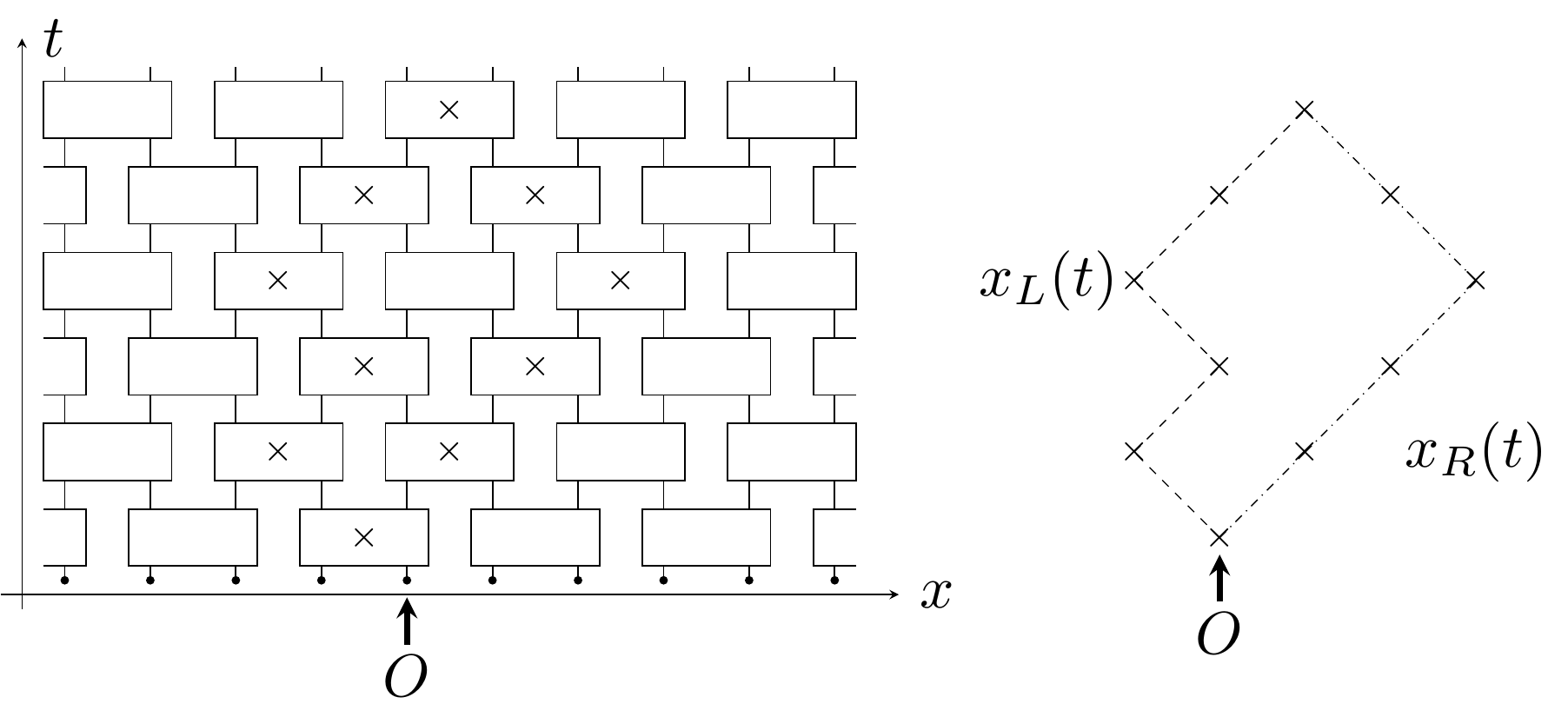}
\caption{The structure of the brickwork circuit. Crosses label the left and right boundaries of the Pauli strings in the time evolved operator $O(0,t)$.}
\label{fig:circuit_walk}
\end{figure}

\subsection{Calculation in cluster picture}

Here we compute $r(v)$ for the Haar-random brickwork circuit using the mapping to a stochastic process. This reveals an unbinding transition for the two paths $x_L(t')$ and $x_R(t')$ at a critical speed $v_c$.
In the Haar circuit, $v_c$ coincides with the butterfly velocity: $v_c=v_B$.

At the end of Sec.~\ref{sec:Haarsimplification} and in App.~\ref{app:transitionrates} we reviewed a mapping of the operator dynamics to a Markov process.  
This yields simple Markovian dynamics for the endpoints of the operator. 
Because of the brickwork structure of the circuit, 
it is better to think of the endpoints $x_{L,R}$ as living on \textit{bonds} of the 1D spatial lattice, rather than on sites \cite{nahum_operator_2018, von_keyserlingk_operator_2018}.
To do this, we simply associate a site of the spatial lattice, at a given time $t'$, with one of its adjacent bonds: 
namely the one which will receive a unitary in the next time step.
In this subsection we will label bonds by integers. 
Then $x_{L,R}(t')$ is an even or an odd integer depending on whether $t'$ is even or odd, 
but the transition probabilities for  $x_{L,R}$ are time-translation invariant.

In each time step $x_L$ changes be either $+1$ or $-1$, and similarly for $x_R$. The transition probabilities, denoted
\begin{equation}
W\big[ x_L(t'+1), x_R(t'+1); x_L(t'), x_R(t') \big],
\end{equation}
are as follows, in terms of a probability $p\equiv1/(q^2+1)$:

\begin{itemize}[leftmargin=3.5mm]
\ii If $x_L(t')\neq x_R(t')$, then $W$ factorizes into separate probabilities for each walk. 
The right-hand walker $x_R$ has probability $1-p$ for a step to the right and $p$ for a left step, 
and vice versa for the left walker. 
$x_R$ is biased towards right steps (since $p<1/2$), and vice versa for $x_L$, leading to   
 the nonzero butterfly speed ${v_B = 1-2p}$.
\ii If the two walkers coincide, $x_L(t')=x_R(t')$, then the probabilities are: 
$p$ if the two walkers remain together in the next time step 
(i.e. if $x_L$ and $x_R$ either both increase or both decrease);
and $1-2p$ if the walkers separate (i.e. $x_L$ decreases while $x_R$ increases).
We will see below that it is  convenient to separate out factors of $1-p$ and $p$ by writing these weights in the form
\ba
p&= p(1-p) \times V \times E, &
1-2p& = (1-p)^2 \times V,
\end{align}
with $V$ and $E$ defined in Eq.~\ref{eq:VE} below.
\eit

The correlator $\overline{G^2}$ of interest maps to a partition functions for two paths with the schematic form
\begin{equation}
Z(x) = \hspace{-2mm} \sum_{w_L, w_R} \prod_{t'} W\left[ x_L(t'+1), x_R(t'+1); x_L(t'), x_R(t') \right],
\end{equation}
where $w_L$ denotes the full trajectory $x_L(t')$ of the left walker and the boundary conditions involve both walks starting at $0$ and ending at $x$.

We can think of the walks as trajectories on a square lattice whose axes are rotated by 45 degrees with respect to the space/time axes, and whose sites are in correspondence with the unitary blocks --- see  Fig.~\ref{fig:circuit_walk}. 
When the walkers are separated, 
the weights $W$ consist of factors of $p$ and ${1-p}$.
The writing of the weights above shows that there is an additional weight $V$ when the two walkers meet at a vertex of the rotated square lattice, and an additional weight $E$ when they share an edge of the lattice, with 
\begin{align}
\label{eq:VE}
V&=\f{1-2p}{(1-p)^2},
&
E & = \f{1-p}{1-2p}.
\end{align}
Since the total number of right and left steps for each walker are fixed by the boundary conditions of the trajectories, we can factor out the $p$ and $1-p$ terms. Up to an unimportant boundary term from the final timestep,
\begin{equation}
\label{eq:Zpathensemble}
Z(x) = [p(1-p)]^t \sum_{w_L, w_R} V^{\lf \substack{\text{\# shared}\\ \text{vertices}}\ri} 
\, E^{\lf \substack{\text{\# shared}\\ \text{edges}}\ri}.
\end{equation}

The question is whether the interactions in Eq.~\ref{eq:Zpathensemble} lead the paths to bind.
Thanks to translation invariance of the weights, this reduces to a transfer matrix calculation involving only the \textit{relative} coordinate ${\Delta=(x_R-x_L)/2}$.
We find that the transfer matrix has a bound state when the net speed $|v|=|x/t|$ of the paths is larger than a threshold.\footnote{Note that as $|v|$ approaches the maximal value of 1, the entropy of the ensemble of paths is reduced. For example,  if $v$ is close to 1, then almost every step has to be a right step. Entropy, which promotes unbinding, becomes more important as $|v|$ is reduced.} 

In  more detail, define
\begin{align}
\Delta &= \f{x_R - x_L}{2},&
X & = \f{x_L + x_R}{2},
\end{align}
where $\Delta\in \{0,1,2,\ldots\}$.
We can  sum over the endpoint $x$ to  define the partition function with a fixed ``force'' on this endpoint:
\begin{equation}
\label{eq:Zlaplace}
Z(\mu) = \sum_x e^{-\mu x} Z(x). 
\end{equation}
We describe in Appendix~\ref{app:transfermatrixwalks} how this can be written using a transfer matrix  for the relative coordinate $\Delta$,
\begin{equation}
Z(\mu)  = \sum_{\{\Delta\}} \prod_{t'} T_{\Delta(t'+1), \Delta(t')},
\end{equation}
where the $\mu$ dependence is in the transfer matrix $T$.

The variable $\mu$ in Eq.~\ref{eq:Zlaplace}, which is conjugate to the total displacement $x=vt$, determines the
saddle-point velocity $v=v(\mu)$ of the trajectories that dominate the partition function. By the definition of $r(v)$,
\begin{equation}
\lim_{t\rightarrow\infty} t^{-1} {\ln Z(\mu)}  =  - \min_v \lf \mu v + 2\seq r(v) \ri.
\end{equation}
The minimization determines the relation between $v$ and~$\mu$.
The left hand side above is the logarithm of the leading eigenvalue of the transfer matrix, which is easily obtained (App.~\ref{app:transfermatrixwalks}).

We find that there is a bound state for speeds greater than $v_c$, with 
\begin{equation}
v_c = \f{q^2 - 1}{q^2 +1}.
\end{equation}
Remarkably, in this particular model this coincides with the butterfly speed.
In the present approach that looks like a coincidence: in Sec.~\ref{sec:Ising} we will explain it using the mapping of the Haar circuit to an Ising-like statistical mechanics model.

The rate functions may be written in terms of  a single symmetric function of $v$,
 \be\label{eq:E2annealed}
 \mathcal{E}_2(v) = \f{\ln \f{q^2+1}{q} + \f{1+v}{2} \ln  \f{1+v}{2} + \f{1-v}{2} \ln  \f{1-v}{2}   }{\ln q},
 \ee
whose interpretation is  reviewed in Sec.~\ref{sec:Ising} \cite{jonay_coarse-grained_2018,zhou_entanglement_2020}.
They have the remarkably simple forms:
\begin{align}
r_{R} (v) & =  \mathcal{E}_2(v) - v,  \\
r_L(v) & = \mathcal{E}_2(v) + v, \\
r_B(v) & = \mathcal{E}_2(v) + |v|.
\end{align}
From Eq.~\ref{eq:rvalternative}, the rate function for the two-point correlator $G_\text{rms}$ is 
\begin{align}
\label{eq:rvalternativeHaar}
r(v) & = 
  \left\{
\begin{array}{ll}
\, \mathcal{E}_2(v)   & \,\, \text{for $|v|<v_c$,}  \\
\, \lf \mathcal{E}_2(v)+|v| \ri/2  & \, \, \text{for $|v|>v_c$.}
\end{array}
\right.
\end{align}
This form is shown in Fig.~\ref{fig:rc_rv}
for local Hilbert space dimension $q=2$ and $q=3$.
The nonanalyticity at $v_c$ is quite weak, because $\mathcal{E}_2'(v_B)=1$ so that $r'(v)$ is continuous.
The above forms have the property $r(v_B)=v_B$, because of the general relation $\mathcal{E}_2(v_B)=v_B$.

Monte Carlo simulations of trajectories that  confirm the above predictions have been reported in Ref.~\cite{de2021rare}. They clearly show the bound/unbound phases at $v\lessgtr v_c$.

\begin{figure}[h]
\centering
\includegraphics[width=\columnwidth]{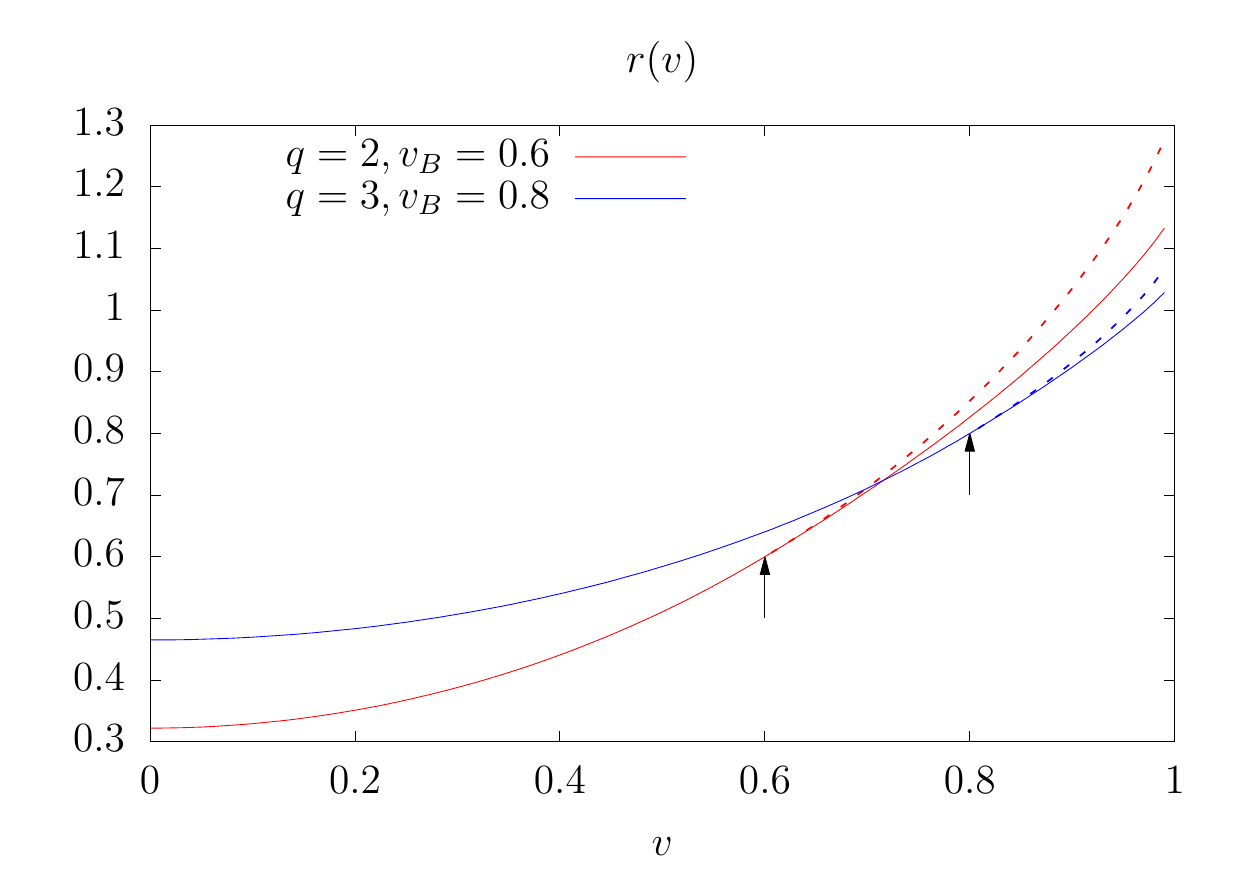}
\caption{The rate function $r(v)$ for the Haar circuit for $q = 2$ and $q = 3$ (solid lines). 
Arrows show the phase transition points.
The dashed lines are the analytic continuations of the $|v|<|v_B|$ forms.}
\label{fig:rc_rv}
\end{figure}

\subsection{Bound state size}

The typical  bound state size for ${|v|>v_c}$ is (App.~\ref{app:transfermatrixwalks})
 \be\label{eq:boundstatesizemaintext}
 \Delta_\text{typ}(v)= \f{2}{\ln \lf \f{1}{q^2} \times \f{1+|v|}{1-|v|}\ri}.
 \ee
 This size diverges (with a critical exponent equal to $-1$ \cite{fisher1984walks})  as the speed $|v|$ tends to $v_c$ from above.
 It \textit{vanishes} as the speed tends to the maximal speed (unity) allowed by the geometry of the circuit. This means that spacetime trajectories simplify in this limit. This phenomenon should occur in a larger class of systems, allowing a perturbative treatment of correlators at large speed.
 
The result for $r(v)$ can also be obtained from the Ising mapping, where the appearance of $\mathcal{E}_2(v)$ is much clearer: see Sec.~\ref{sec:Ising}.

\subsection{More general  models: $v_B$ and $v_c$}
\label{sec:vBvc}

A remarkable feature of the Haar circuit is that the unbinding speed $v_c$ coincides with the butterfly speed $v_B$, meaning that in principle the butterfly velocity could be determined from the two-point function.
However, this identity does not hold for all 1+1D systems or even for all 1+1D local random circuits. 
For example, we can induce $v_c < v_B$ by introducing unitaries that act on pairs of spins at separation larger than 1, as we will discuss in  Sec.~\ref{sec:higherdims} below.
In fact it is possible to  construct circuits in which the trajectories are bound for \textit{all} $v$.
We expect, but have not proved, that it is also possible to have the reverse situation $v_c>v_B$.\footnote{It should be possible to do an explicit calculation for a model that is a weak perturbation of the Haar circuit. See Sec.~\ref{sec:haarfinetuned}.} 

Nevertheless, the result in the previous section raises the question of whether there is a way to determine the butterfly velocity $v_B$ from a two-point function that works more generally. One possibility may be to use the relation ${r_R(v_B) = 0}$: if trajectories are unbound at $v_B$, then this relation together with the relations for $r_{L,R}$ from reflection and time reversal  symmetry (Sec.~\ref{sec:symmrels}) suffices to fix $r(v_B) = v_B$. 
In a large finite system with boundaries, the two-point function between an operator at the left boundary and one at the right boundary can also give a nonanalyticity at $v_B$ that could in principle also be used to detect $v_B$; this is discussed in Sec.~\ref{sec:Ising}.

\subsection{Average versus typical}
\label{sec:avversustypbrief}

So far we have discussed the average $G_\text{rms}$ in the random circuit
and established the basic picture
for the two phases.
A natural question is how this differs from $G$ in a specific realization of the circuit, 
or from the typical value $G_\text{typ}$ defined in Eq.~\ref{eq:Gtypdefn}.
This subsection  discusses these more subtle issues. Some readers may prefer to skip ahead.

The main claims in this subsection are:
(1) Going from $G_\text{rms}$ to $G_\text{typ}$ slightly modifies the quantitative values of the rate functions, but the basic picture relating them to line tensions for different kinds of trajectory remains intact.
(2) In a given realization, randomness of the unitaries leads to universal \textit{subleading} terms in ${\ln G}$ with Kardar-Parisi-Zhang/directed polymer exponents \cite{kardar1986dynamic,huse1985huse}.

As we have discussed, $\overline{G^2}$ is governed in the scaling limit by an effective classical partition function for either a single path 
(representing the bound state) or a pair of paths 
(representing the two endpoints). The weights in this effective partition function are translationally invariant.
In a given \textit{realization} of the random circuit,
we conjecture that in the scaling limit $G$ is governed by a similar classical partition function for paths whose weights are no longer translation invariant, but include quenched disorder
(including random signs).
The typical properties of these paths will be captured by $G_\text{typ}$ in Eq.~\ref{eq:Gtypdefn}. 

A slightly  more detailed picture for this, in the bound phase, is given in App.~\ref{app:signtransition}.
In the unbound phase a more quantitative analysis of the effect of disorder could be attempted using the approach of  Ref.~\cite{zhou_emergent_2019}, but we do not attempt this here.
Instead we summarize the consequences of this conjecture regarding the effects of disorder.

Disorder will ``dress'' the rate functions, so that the asymptotics of $G_\text{typ}$ involve rate functions $r^\text{typ}_{R,L,B}(v)$ that in general differ from the rate functions $r_{R,L,B}(v)$  calculated above for $G_\text{rms}$. We expect this dressing effect to be quantitatively small in the Haar circuit (see footnote below), so that the rate functions $r$ and $r^\text{typ}$  are numerically  close. (In Sec.~\ref{sec:numericalcasestudy} we study another random model, where the effect of dressing again seems to be  small.)

Spatiotemporal disorder also changes some universal properties of the paths, which will now be those associated with directed polymers in a random medium (a well-studied problem \cite{huse1985huse, kardar1986dynamic}). 
The distinction between bound and unbound phases still makes sense for paths in a random potential.\footnote{See Ref.~\cite{de2015crossing} for universal results for the unbound phase.} One effect of disorder will be on universal subleading terms in $\ln G$.

For simplicity, consider the bound regime, where after coarse-graining we have only a single path to consider.
For  $G_\text{rms}$, the mapping to a random walk means that
\begin{equation}
G_\text{rms}^2  \sim \f{c(v)}{\sqrt{t}} \, \exp \lf -  \seq r_B (v) t \ri,
\end{equation}
where $c(v)$ is a function of $v$ only.
For $G_\text{typ}$, or indeed $G$ in  a single realization of the circuit, there is a characteristic subleading term in the free energy  of order $t^{1/3}$ \cite{huse1985huse, kardar1986dynamic}:
\begin{equation}
G  \sim \exp \lf - \f{\seq}{2} r_B^\text{typ}(v) t +  t^{1/3} \chi + \ldots  \ri.
\end{equation}
Here $r_B^\text{typ}(v)$ is self-averaging, i.e. independent of the realization.
$\chi=\chi(x,t)$ is a random variable that depends on the realization, but which is of order 1; for $G_\text{typ}$, we use the average value of $\chi$, which is nonzero.

Disorder also changes the typical lengthscale for wandering of the paths (for example $t^\alpha$ in Fig.~\ref{fig:path_cartoon}) from the diffusive scale $t^{1/2}$ to $t^{2/3}$.

Finally let us note some more subtle points.
The first is to do with the sign of $G$.
In the bound phase, $G$
(in a realization) is essentially a partition function for a path with random weights of both signs \cite{kardar2007statistical,nguyen1985tunnel,medina1989interference,kim2011interfering}, see App.~\ref{app:signtransition}.
These random signs mean that $\overline{G}=0$ in the Haar circuit. In more general spatiotemporally random models 
$\overline{G}$ is not identically zero, but nevertheless $\overline{G/|G|}\ll 1$  for generic 1+1D models with spatiotemporal randomness (see App.~\ref{app:signtransition} for a comparison with higher dimensions).

Second, the quantities $r_{R,L}(v)$ computed for $G_\text{rms}$
obey identities relating them to a quantity $\lambda(v)$ defined using the OTOC 
(Eq.~\ref{eq:lambdaotoc})
and to 
a version of the entanglement line tension $\mathcal{E}_2(v)$
(see Sec.~\ref{sec:Ising}).
Disorder will also ``dress'' the values of $\lambda$ and $\mathcal{E}_2$,\footnote{The dressed version of the entanglement line tension $\mathcal{E}_2$ determines $\overline{S_2}$, while the version without dressing  determines ${\ln \overline{e^{-S_2}}}$ \cite{zhou_emergent_2019}. The dressed $\lambda$ determines the average of $\ln \operatorname{OTOC}$.} though this dressing effect is known to be small in the Haar circuit.\footnote{The effect of dressing on  $\mathcal{E}_2(v)$ was computed Ref.~\cite{zhou_emergent_2019}: it gives a correction that vanishes as  $1/[q^8\ln q]$ as $q\rightarrow\infty$, and also has a small numerical coefficient.}
It is natural to ask whether the identities between these quantities still hold when dressing is taken into account. We leave a proper investigation of this to the future. 
We do expect the identity $r_R^\text{typ}(v_B)=0$ to remain valid.
On the other hand the relation between $r_B(v)$ and 
$\mathcal{E}_2$
(the second line of Eq.~\ref{eq:rvalternativeHaar})
is special to $G_\text{rms}$ in the Haar circuit, as shown in Sec.~\ref{sec:Haarsimplification}.

For the most of this paper we will put the subtleties discussed in this section aside, for the following reason. 
If our ultimate aim is to model translationally invariant systems,
the simple translationally invariant effective model that arises from $\Grms$ may be a more useful guide. We will return to this in Sec.~\ref{sec:instances}.

 \section{Higher dimensions}
\label{sec:higherdims}

 For unitary dynamics in higher spatial dimensions $(d>1)$ composed of Haar-random, local unitary gates, the root-mean-square correlator $G_{\mathrm{rms}}(\boldsymbol{x},t)^{2} = \overline{G(\boldsymbol{x},t)^{2}}$
 (Eq.~\ref{eq:rmsdefn}; recall that  $\overline{\cdots}$  denotes an average over the choice of Haar-random unitary gates) is described by a classical Markov process for the growth of a $d$-dimensional cluster. 
 The support of the cluster at time $t$ is labeled by a binary occupation number $n_{{\bf y},t} \in \{0,1\}$ denoting an unoccupied or occupied site of the cluster, respectively.  The cluster trajectories in the description of $G_{\mathrm{rms}}(\boldsymbol{x},t)^{2}$ must obey the boundary conditions $n_{{\bf y},0} = \delta_{{\bf y},0}$ and $n_{{\bf y},t} = \delta_{{\bf y},{\bf x}}$ so that the cluster is exclusively supported at a single site at the initial and final times.  Since the cluster's typical behavior without this final-time condition would be to grow ballistically outwards, the cluster trajectories contributing to $G_{\mathrm{rms}}(\boldsymbol{x},t)^{2}$ describe rare events.  In this section, we argue that the cluster remains thin throughout these trajectories in dimensions $d \ge 2$, for any velocity, in stark contrast to the one-dimensional case. 
 We argue that this conclusion --- that the trajectories
 contributing to $G({\bf x},t)$ are thin for all $v$ ---
  holds generally for chaotic models in higher dimensions, even when there is no mapping to a classical Markov process.
 
 For concreteness, we specialize to spatial dimension $d=2$. 
 Consider a  ``fat'' spacetime trajectory which  grows to a large size with linear dimension $\sim t^{\alpha}$ $(\alpha > 0)$ 
 at intermediate times: we will argue that the contribution of such trajectories to 
 to $G_{\mathrm{rms}}(\boldsymbol{x},t)^{2}$ is suppressed.
 This contribution may be understood in
 terms of an effective coarse-grained Markov process for the one-dimensional boundary of the cluster.
 At a given intermediate time in the evolution, we may focus on a portion of the cluster boundary, of a linear extent $\ell$ that is much larger than the lattice spacing but much smaller than the average radius of curvature $\sim t^\alpha$ of the cluster. We assume that, after coarse-graining on scales $\lesssim\ell$, this portion is approximately flat, with outward normal vector  oriented in the direction $\theta$. 
 
In the absence of conditioning, the cluster boundary will on average advance in the direction of its normal at a butterfly speed ${v}_{B}(\theta)$, which in general depends on $\theta$ \cite{nahum_operator_2018}.
Advancing (or retreating) with any other speed is a rare event. 
By spatiotemporal locality, the probability that this portion of the cluster boundary subsequently evolves with a constant normal velocity of magnitude ${v}$ over time interval $\Delta t$ scales as 
\be
P \asymp \exp[-\ell\,\Delta t\,\gamma({v},\theta)],
\ee
where the \emph{surface tension} $\gamma({ v},\theta)$ vanishes when ${{{v} = {v}_{B}(\theta)}}$ and is negative for all other velocities.  Therefore, unless it grows outwards at the butterfly speed, this patch of the cluster boundary incurs a cost in probability which is exponentially small in the surface area of the spacetime trajectory.

\begin{figure}
$\begin{array}{cc}
\includegraphics[width=.5\columnwidth]{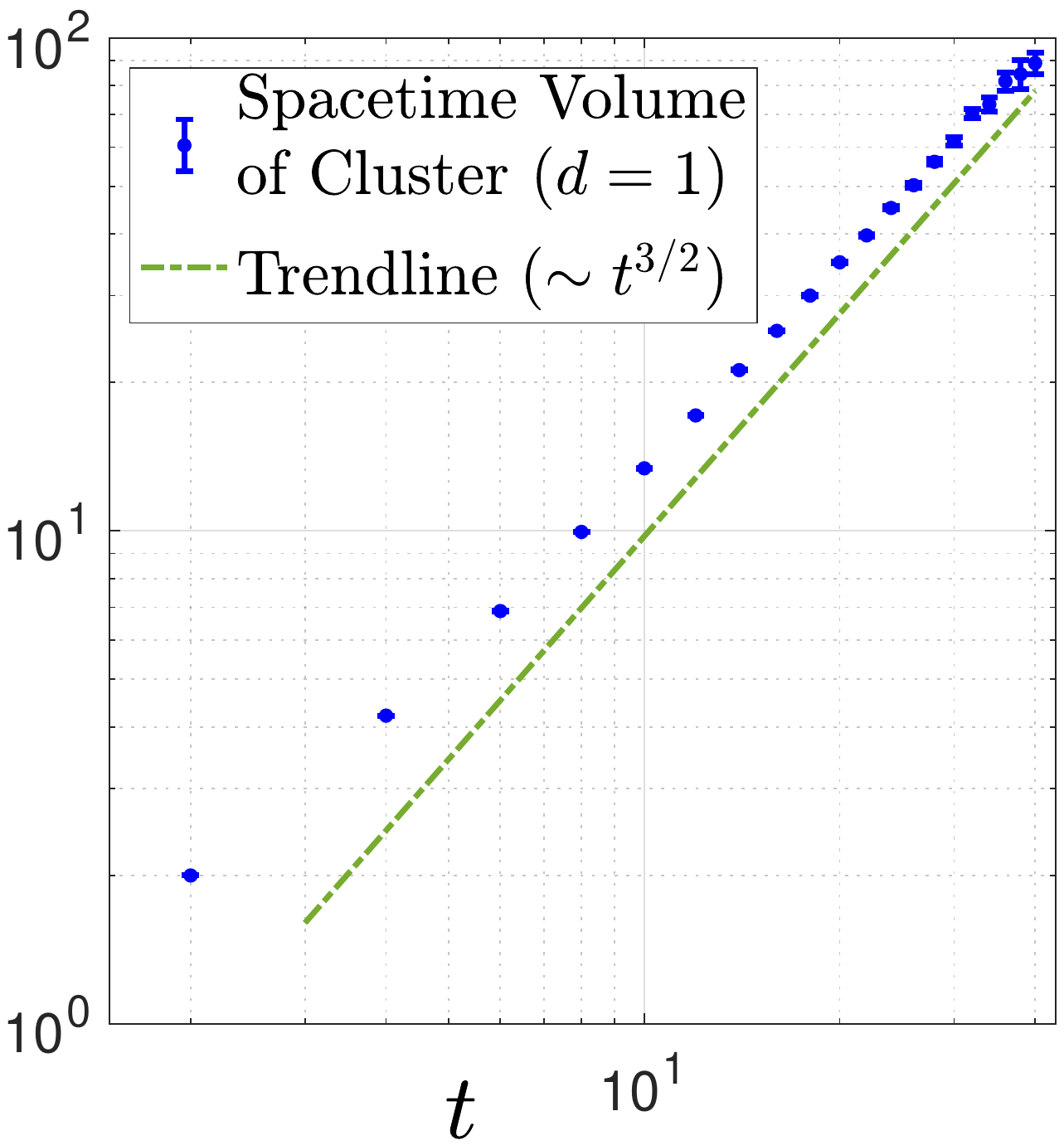} & 
\includegraphics[width=.48\columnwidth]{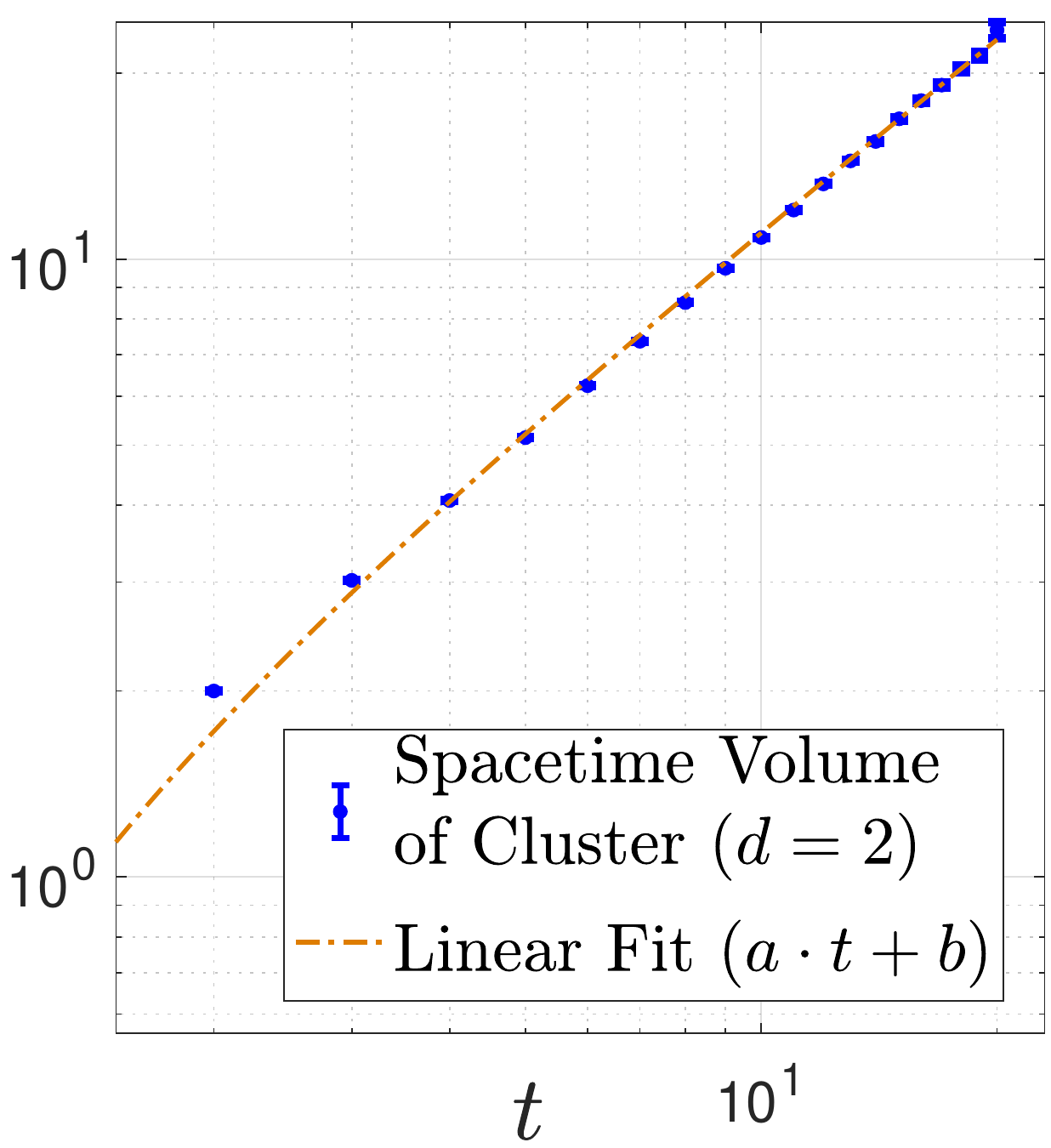}\\
\text{(a)} & \text{(b)}\\
\end{array}$
\caption{Scaling of the spacetime volume of clusters
in the classical Markov process, which contribute to $\overline{G(0,t)^{2}}$ in  dynamics with randomly-applied two-site unitary gates in (a) one and (b) two spatial dimensions.  The growth of the spacetime volume as $t^{3/2}$ in (a) is consistent with the growth of operators to a large size at intermediate times in one dimension, while the linear scaling in (b) suggests that operator histories remain an $O(1)$ size in two spatial dimensions.}
\label{fig:cluster_volume_scaling}
\end{figure}

In a ``fat" history of the cluster, the local growth rate of the cluster boundary must differ from the butterfly speed, at most points on its ``worldsurface'', on order to meet the condition that the cluster is of size 1 at the initial and final times.
It follows that the overall contribution of a ``fat" history of the cluster to $G_{\mathrm{rms}}(\boldsymbol{x},t)^{2}$ is exponentially small in the total surface area of the trajectory traced out by the boundary of the cluster. 
``Thin" trajectories of minimal surface area, where the cluster remains an $O(1)$ size at any point in the evolution, are thus always favored so that a binding/un-binding transition in the histories of operator trajectories is not expected to occur in dimensions $d\ge 2$.

We anticipate that this scaling argument also applies to the operator histories contributing to two-point correlations $G({\bf x},t)$ in generic, chaotic models even when there is no exact mapping to a classical Markov process (and even in the presence of conservation laws).  In this setting, a membrane tension may still be associated with fat operator trajectories  (Sec.~\ref{sec:instances}) which penalizes local growth rates that deviate from the butterfly velocity, so that thin  trajectories are always preferred.
 
Our result is confirmed in numerical simulations of the Markov process for the cluster in one and two spatial dimensions.  In contrast to quantum circuits with a regular brickwork array of unitary gates considered in previous sections, here we consider unitary dynamics in which two-site, Haar-random unitary gates are randomly applied to nearest-neighbor qubits (on-site Hilbert space dimension $q=2$) in both one and two spatial dimensions.  
In both cases, one timestep of the unitary evolution in an $N$-site system is defined by  the application of $N/2$ gates so that each site is acted upon by a single gate on average in a single timestep. 
Simulations of the Markov process for the 1+1D \textit{brickwork} circuit, using Monte Carlo on the ensemble of trajectories, were presented in Ref.~\cite{de2021rare}.

In one spatial dimension, we observe that the cluster histories contributing to $G_{\mathrm{rms}}(0,t)^{2}$ have a spacetime volume which scales with time as $t^{3/2}$, as shown in Fig. \ref{fig:cluster_volume_scaling}a.  This behavior is consistent with the dominant contribution being from  fat operator trajectories, which grow to a width $\sim t^{1/2}$ at intermediate times.  In contrast, the spacetime volume of trajectories contributing to $G_{\mathrm{rms}}(0,t)^{2}$ in two spatial dimensions only grows linearly in time, as shown in Fig. \ref{fig:cluster_volume_scaling}b, which is consistent with 
operator histories being thin.  Furthermore, an exponential decay of $G_{\mathrm{rms}}(0,t)^{2}\asymp \exp(-\Gamma t)$ in time is observed in Fig. \ref{fig:2d_autocorrelator} in two dimensions.
Here ${\Gamma=2 \seq r(0)=\seq r_B(0)}$ in our conventions.
An approximation to this decay rate is obtained by analytically summing over all cluster trajectories which remain \emph{maximally} thin during the evolution, so that the cluster is supported at a single site at any time.  In Appendix \ref{app:thin_cluster}, we demonstrate that this ``thin cluster approximation" gives an estimate $G_{\mathrm{rms}}(0,t)^{2} = \exp(-\Gamma_{\mathrm{thin}}(q)t)$ where 
\begin{align}
    \Gamma_{\mathrm{thin}}(q) = \frac{q^{2}-1}{q^{2}+1}
\end{align}
and $q$ is the on-site Hilbert space dimension.  A comparison of this approximation to the data can be seen in Fig.~\ref{fig:2d_autocorrelator}.

\begin{figure}
\includegraphics[width=.7\columnwidth]{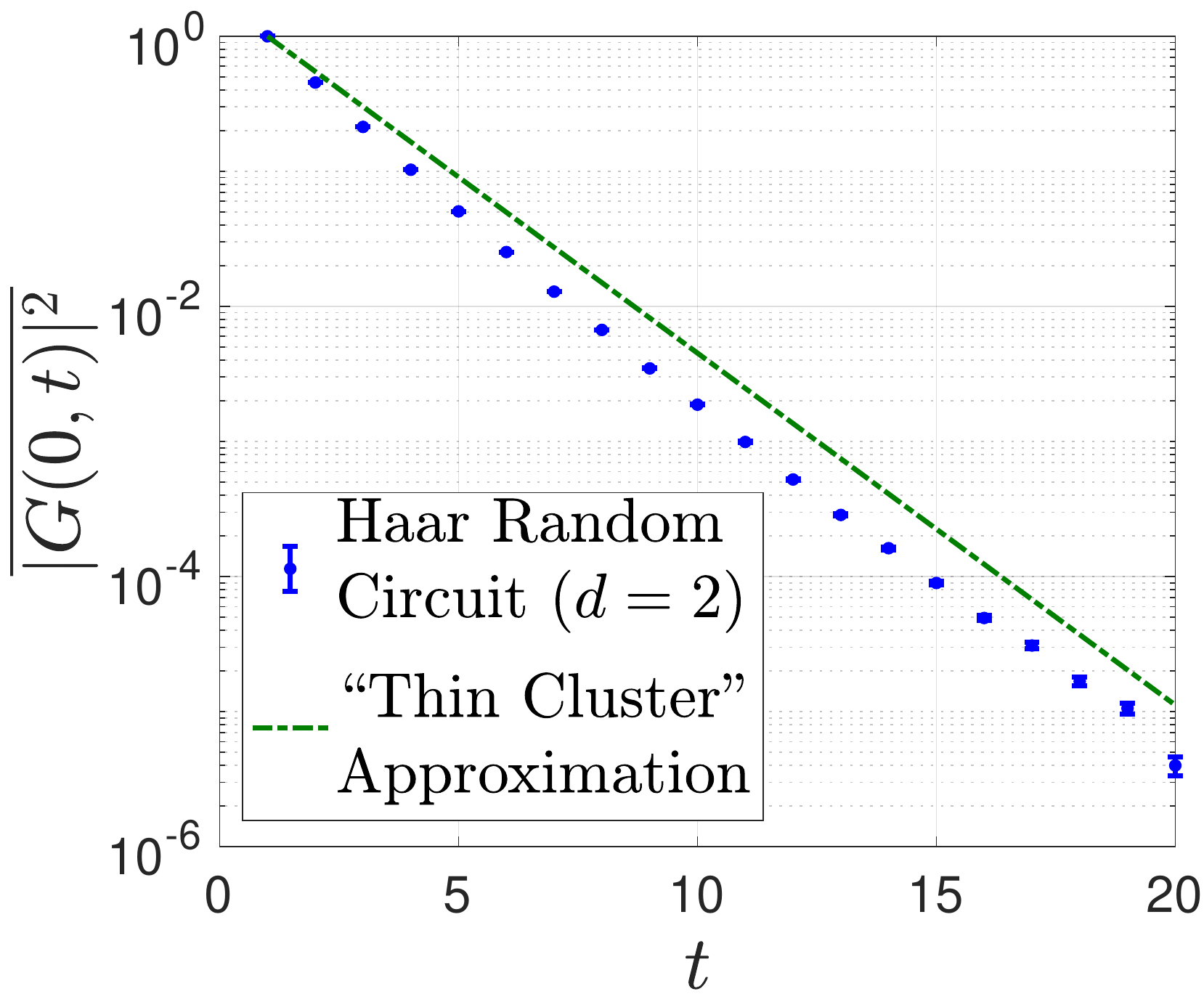}
\caption{The behavior of $\overline{G(0,t)^{2}}$ in Haar-random unitary dynamics with randomly-applied two-site unitary gates in two spatial dimensions is shown in blue.  An approximation of this quantity by analytically summing over ``maximally thin" trajectories of clusters in the classical Markov process is shown in green.}
\label{fig:2d_autocorrelator}
\end{figure}

These results for higher dimensions also suggest a way to modify a 1D model so as to remove the unbound phase.
Loosely speaking, a basic lesson of the foregoing is that spacetime operator trajectories with a large surface area are suppressed.
Therefore, consider a quasi-1D system
in the shape of a cylinder that is infinitely extended in the $x$ direction and of  circumference $\ell$ in the $y$ direction. The circumference $\ell$ will be taken finite, but sufficiently large.
In this setup, a bound trajectory can be constructed using  operator clusters of size $\ll \ell$, so that the surface area in spacetime does not grow with $\ell$.
However in an unbound trajectory,  the $x$-extent of the clusters becomes arbitrarily large (by definition), and therefore much larger than $\ell$.
In this situation the cluster fills up the periodic direction, and the surface area in spacetime scales with $\ell$, giving exponential supression in $\ell$.
Therefore if $\ell$ is large enough the bound trajectories are guaranteed to dominate.
We can also achieve a similar effect in a strictly 1+1D circuit by including gates with a longer range (this has the effect of increasing the effective surface area of the cluster).

\section{Ising picture for Haar circuit transition}
\label{sec:Ising}

\subsection{Setup and calculation}

In Sec.~\ref{sec:Haartransition}, we mapped  $\overline{G^2(x,t)}$ to a rare event probability in a  Markov process for the two endpoints of the operator string. 
In that picture, the transition in $r(v)$ was interpreted as a binding transition for these endpoints. 
We also found that the rate function $r(v)$ could be expressed in terms of the entanglement line tension function $\mathcal{E}_2(v)$. 

In this section we provide a complementary  perspective  by mapping  $\overline{G^2(x,t)}$ to an Ising-like lattice magnet \cite{nahum_operator_2018,zhou_emergent_2019,zhou_entanglement_2020,von_keyserlingk_operator_2018,chan_solution_2018,hunter2019unitary}, where the entanglement line tension function emerges naturally
\cite{jonay_coarse-grained_2018,zhou_entanglement_2020}. 
 We first give a  schematic review of this mapping.

When written as a tensor network, the dynamical quantity $\overline{G^2(x,t)}$ contains two forward-evolving copies of the circuit (i.e. of the evolution operator $U(t)$) and two backward-evolving copies (i.e. of $U(t)^*$, complex conjugated in a given basis). 
We can imagine these as stacked on top of each other so that they share the same $(x,t)$  coordinate system. Inside, each individual two-site gate $u$ is replicated (stacked) to give a tensor ${u\otimes u^* \otimes u \otimes u^*}$. 
 Altogether, this replicated tensor has  8 input legs and 8 output legs, each of dimension $q$, so is formally an operator acting on a $q^8$-dimensional Hilbert space.
However, the Haar average
over $u$
\cite{hamma2012quantum,harrow2013church,collins2021weingarten}
transforms
${u\otimes u^* \otimes u \otimes u^*}$ into a projection operator that projects down to a two-dimensional subspace of this large space, with basis states labelled $\{+, -\}$. These $\pm$ states are our Ising spins.

Formally these two states represent  two ways of pairing up the four layers $1,\bar 1, 2, \bar 2$ in the stacked circuit. The two pairing patterns are:
\begin{align}
\label{eq:pairing}
+:\quad &\contraction[1.5ex]{}{1}{}{\bar 1}
\contraction[1.5ex]{1\bar 1}{2}{}{\bar 2}
1\bar 1 2 \bar 2,
&
-: \quad&\contraction[2ex]{}{1}{\bar 1 2}{\bar 2}
\contraction{1}{\bar 1}{}{2}
1\bar 1 2 \bar 2.
\end{align}
Physically, pairings arise in a  way  analogous to the discussion in Sec.~\ref{sec:Haarsimplification}: most Feynman trajectories of the multilayer circuit will be killed by phase cancellation when the Haar average is performed. Those that survive have a locally paired structure that allows compensation of opposite phases from $u$ and $u^*$ layers.

\begin{figure}[t]
\centering
\includegraphics[width=0.9\columnwidth]{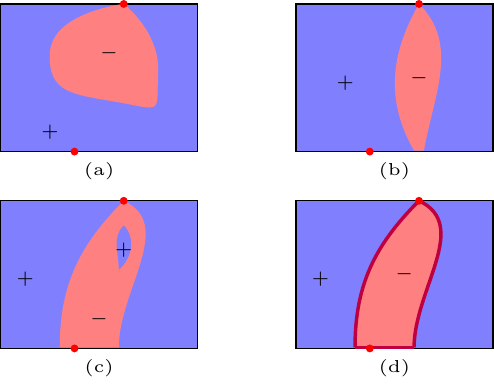}
\caption{Some disallowed (a-c) and allowed (d) configurations of Ising spins for $\overline{G^2}$.
Time runs upwards.
At the bottom the site $(0,0)$ 
where the initial operator is placed must lie in a domain of $-$. Thus (a) and (b) are not allowed. Domain wall annihilation events (note that time runs upwards) are forbidden in the bulk: for this reason configuration (c) is not allowed. 
The absence of annihilation events  also means that a $-$ domain must propagate all the way up to the top of the sample. 
At the top boundary, $-$ is allowed only at the point $(x,t)$ where the other operator is inserted, so the domain pinches off there as in (d), which shows an allowed configuration. Allowed configurations consist of two domain walls connecting the bottom boundary to the point $(x,t)$ at the top. Next, free energy minimization ensures that the domain walls are straight at the largest scales, with fixed velocities: see Fig.~\ref{fig:morph}.}
\label{fig:domain_config}
\end{figure}

\begin{figure}[t]
\centering
\includegraphics[width=0.9\columnwidth]{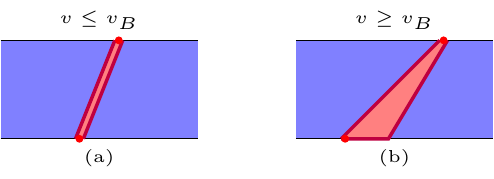}
\caption{Equilibrium configurations  for (a) $v \le v_B$ and (b) $v \ge v_B$. Here we have taken ${v > 0}$: the left domain wall then always has slope $x/t=v$. The wandering of the domain walls on scales ${\sqrt{t}\ll t}$ is not shown.}
\label{fig:morph}
\end{figure}

After averaging,  $\overline{G^2(x,t)}$
is a partition function  $Z$ for a 2D lattice model of Ising spins, with one spin for each ``block'' in the initial unitary circuit, and with boundary conditions that we specify below. These boundary conditions induce domain walls in the bulk. 
The rate function $r(v)$ is proportional to the free energy of this effective 2D model, defined here as $-\ln Z$.

The same Ising model, but with different boundary conditions, can also be used to compute other quantities, such as the purity of the time-evolved state. Its properties are best understood in terms of domain walls between $+$ and $-$.
The geometry of these domain walls is highly constrained.
Events in which a pair of domain walls annilates, as we proceed upwards in the time direction, are forbidden.\footnote{See Sec.~\ref{sec:haarfinetuned} for more detail. Unitary prevents the annihilation of an isolated pair of domain walls, but the effective model for the Haar circuit obeys further constraints.} 
Since an annihilation event is also 
an event in which a domain wall ``turns around'', this leads to a simple picture in terms of  domain walls that are \textit{directed} in the time direction.
At large scales these are  characterized only by a line tension, defined  as the free energy divided by the temporal extent.
Due to homogeneity after averaging, this line tension only depends on the local velocity $v = dx(s)/ds$. The explicit expression for a single isolated domain wall is in Eq.~\ref{eq:E2annealed}.
This is referred to as an ``entanglement line tension'' since it also determines entanglement generation  \cite{jonay_coarse-grained_2018,zhou_entanglement_2020,mezei_membrane_2018}.\footnote{More precisely, the quantity  denoted $\mathcal{E}_2$ here determines  ${-\ln\overline{\exp(-S_2)}}$, while $\overline{S_2}$ is determined by a dressed line tension.}

For left-right symmetric systems, the domain wall tension satisfies
\begin{align}
\label{eq:e2constraint}
  \mathcal{E}_2( v ) &\ge  |v|, & \mathcal{E}''_2( v ) &\ge 0,  \\
  \mathcal{E}_2( v_B ) &= v_B, & \mathcal{E}'_2( v_B )  &=1.
\end{align}
where $v_B$ is the butterfly velocity. These constraints indicate that the line tension function is a convex function whose graph is tangential with that of the function $|v|$ at $\pm v_B$.

Returning to the quantity $\overline{G^2(x,t)}$, we now determine the equilibrium domain wall configurations. Here we give a coarse-grained picture: details are similar to Refs.~\cite{von_keyserlingk_operator_2018, nahum_operator_2018,zhou_emergent_2019}. To begin with we consider a spatially infinite system.

If we  take the local operators whose correlator we are computing to be identity operators, then we  find that all the spins in the system are equal to $+$, and the partition function is equal to 1, giving a trivial correlator as expected. 
Here we take the local operators to be traceless. Then we find that at the top (final time) boundary  of the spacetime patch, all the spins, with the exception of  that at the location of the operator insertion $(x,t)$, are forced to be $+$.
At the bottom (initial time) boundary, the spin at the location of the operator insertion $(0,0)$ is forced to be $-$.
The other spins on the bottom boundary can be either $+$ or $-$, but each $-$ spin on the bottom boundary incurs a free energy cost $2\ln q=2\seq$, corresponding to a cost $\seq$ per unit  length of boundary (since each Ising spin is associated with a \textit{two-site} unitary gate).

These boundary conditions, together with the fact that domain walls cannot annihilate in the bulk,
imply that a pair of domain walls must span the system from the bottom to the top, where they meet at $(x,t)$. The dynamical quantity $\overline{G^2(x,t)}$ is therefore determined by the free energy of the two non-intersecting domain walls, see Fig.~\ref{fig:domain_config}.

We now find the configurations that minimize the total free energy $f_\text{tot}$. 
In general, in the scaling limit, the left and right domain walls will be straight lines with some velocities $v_1$ and $v_2$. The total free energy is a sum of the domain wall free energies $t\seq \mathcal{E}(v_{1,2})$ together with a possible  contribution from the bottom boundary. The boundary contribution is present if $v_1>v_2$  (as this leads to $-$ spins on the boundary) and is equal to $t\seq(v_1-v_2)$. Altogether,
\begin{equation}
f = t \seq \left[ \mathcal{E}_2(v_1) + \mathcal{E}_2(v_2)+ (v_1-v_2)\right],
\end{equation}
which must be minimized over $v_{1,2}$ within the allowed ranges.

It is convenient view the configuration as containing two paths from $(0,0)$ to $(x,t)$. One of these paths may contain a segment of the bottom boundary: see Fig.~\ref{fig:morph}(b).

Let the free energy of the left and right paths be $s_{\rm eq} f_L(v)$ and $s_{\rm eq} f_R(v)$ respectively (one of these may include a boundary contribution). 
Note that the parameter $v = x/t$ is the velocity of the ray connecting the operator insertion points (not necessarily equal to the velocity of a given  domain wall). 
Since the problem is symmetric, we have $f_L(v) = f_R(-v)$ and only need to work out $f_R(v)$. 

Let the right domain wall start at $(x_0, 0)$ with $x_0 \ge 0$. The boundary cost is $s_{\rm eq} x_0$. Hence at leading order, we compute the free energy as an optimization over domain wall slope:
\begin{equation}
\begin{aligned}
s_{\rm eq} f_R( v ) &=  s_{\rm eq} \min_{x_0 \ge 0}\left(\mathcal{E}\lf \frac{x - x_0}{t} \ri + \frac{x_0}{t} \right) \\
&=  s_{\rm eq} \min_{-1 \le v_2 \le v } \bigg( \mathcal{E}( v_2 ) - v_2 + v  \bigg)
\end{aligned}
\end{equation}
According to the properties of the line tension function in Eq.~\ref{eq:e2constraint}, the minimum is  
\begin{equation}
\label{eq:fR}
f_R(v) 
= 
\left\lbrace
\begin{aligned}
  & \mathcal{E}(v)  &&\text{if $v \le v_B$ (in which case $v_2 = v$)} \\
  & v  &&\text{if $v > v_B$ (in which case $v_2 = v_B$).} \\
\end{aligned} \right. 
\end{equation}
The domain wall configurations for these two cases are presented in Fig.~\ref{fig:morph}. We infer the free energy of the left path by symmetry: 
\begin{equation}
f_L(v) =
\left\lbrace
\begin{aligned}
  & \mathcal{E}(-v)  & \quad   v \ge -v_B \\
  & -v  & \quad v < - v_B  \\
\end{aligned} \right. 
\end{equation}
Summing them up, we have
\begin{equation}
\begin{aligned}
r(v) &= f_L(v) + f_R(v)  \\
& = \left\lbrace 
\begin{aligned}
  & \mathcal{E}_2(v) & \quad |v| \le v_B \\
  & (\mathcal{E}_2(v) + |v| ) / 2  & \quad |v| > v_B  \\
\end{aligned} 
\right. , 
\end{aligned}
\end{equation}
in agreement with (\ref{eq:rvalternativeHaar}).
The nonanalyticity is weak because
\begin{align}
\lim_{v \nearrow v_B } r(v) &= \lim_{v \searrow v_B } r(v) = v_B,\\
\lim_{v \nearrow v_B }  r'(v)   &= \lim_{v \searrow v_B}   r'(v)  = 1,
\end{align}
and the discontinuity occurs in the 2nd derivative at $v_B$. The rate functions for the random circuit with $q = 2$ and $3$ were shown in Fig.~\ref{fig:rc_rv}. 

The result above agrees with that from the  calculation in terms of the operator Markov process, and in fact by considering slightly more general boundary conditions (in order to separately control the velocities of the right and left string boundaries in the cluster growth picture)
it can be argued that ${r_R(v)=\mathcal{E}_2(v) -v}$ and
${r_L(v)=\mathcal{E}_2(v) +v}$, as  stated in Sec.~\ref{sec:Haartransition}.
However the spacetime trajectories that appear in the present formulation are somewhat different.\footnote{It should be noted that the mutually avoiding paths in Fig.~\ref{fig:morph}(a) are not bound together: they have the same velocity on scales of order $t$, but on scales of order $\sqrt{t}$ they  wander apart. So they are qualitatively similar to the operator endpoints in the unbound phase.}
The difference comes from a different choice of basis. The $\pm$ basis is non-orthogonal, but it is convenient because it exposes a $+\leftrightarrow -$ symmetry arising from the  possibility of permuting the layers.
The cluster picture used in previous sections arises from looking at the effective dynamics in a different (orthogonal) basis associated with the cluster occupation numbers $\{\emptystate,\fullstate\}$ discussed towards the end of Sec.~\ref{sec:Haarsimplification}. The different pictures are convenient for different calculations \cite{nahum_operator_2018,von_keyserlingk_operator_2018}.

\subsection{A special feature of the Haar circuit}\label{sec:haarfinetuned}

In this Section (which is not essential to the main development) we discuss a more subtle feature of the Ising picture.

We  noted above that the domain wall configurations are highly constrained by unitarity. However, the domain wall configurations for the Haar brickwork model in fact obey an additional restriction 
(beyond those imposed by unitarity)
which is generically relaxed when the distribution of the random unitaries is perturbed away from the Haar measure.  Such a perturbation leads to an additional allowed ``vertex'' in the effective Ising model which we discuss here.

It is convenient to use a transfer matrix language to specify the allowed configurations. For the purposes of this subsection, we will think of a transfer matrix that acts from the top of the spacetime patch to the bottom, i.e. we use the transfer matrix to go backwards by one time step.

Above we associated one Ising spin with each ``block'' in the unitary circuit, but in fact the Ising model can equivalently be formulated with one spin $\pm$ living at each spatial site. A local two-site transfer matrix is then associated with each ``block'' in the circuit. In the Haar case, the transition amplitudes for this transfer matrix $\hat T_\text{Haar}$ (see Eq.~\ref{eq:Thaar} in App.~\ref{app:bc} for a more formal definition) are
\begin{align} 
++ & \rightarrow ++  & &\text{with amplitude 1}  \label{eq:dw_rule_1} \\
++ & \not\rightarrow +- & &\text{(forbidden)} \label{eq:dw_rule_2} \\
+- & \rightarrow ++  & &\text{with amplitude $K$} \label{eq:dw_rule_3} \\
+- & \not\rightarrow -+ & &\text{(forbidden)}, \label{eq:dw_rule_4} \\
\end{align}
together with those related either by spatial reflection or Ising symmetry. Here $K=q/(q^2+1)$.

The prohibition $++ \not\rightarrow +-$ is a consequence of unitarity,\footnote{The mapping to an effective lattice magnet is also useful for non-unitary dynamics with measurement \cite{bao2020theory,jian2020measurement,nahum2021measurement,li2021statistical, li2021entanglement}.}  and can be given a meaning in any unitary circuit even without randomness \cite{zhou_entanglement_2020}. This rule means that, as we proceed downwards, it is \textit{not} possible to nucleate a pair of domain walls inside a uniform spin domain.
(Equivalently, as we proceed \textit{upwards}, it is not possible to annihilate an isolated pair of domain walls.)

In contrast the prohibition $+- \not\rightarrow -+ $ is special to the Haar distribution. 
This can easily be seen by considering a modified distribution in which the local unitary has some probability $(1-p)$ to be a Haar unitary, and the complementary probability $p$ to be a ``swap'' gate that exchanges the states at the two physical sites.
In this ensemble the action of the single-block transfer matrix is 
\begin{align}
++ & \rightarrow ++  & &\text{with amplitude 1} \\
++ & \not\rightarrow +- & &\text{(forbidden)}\\
+- & \rightarrow ++  & &\text{with amplitude $(1-p)K$} \\
+- & \rightarrow -+ & &\text{with amplitude $p$}
\end{align}
(together with the symmetry-related amplitudes).
The new ``vertex'' allows for one domain wall to split into three (and vice versa), for example ${++--\leftrightarrow +-+-}$.
In this Haar/swap ensemble the new vertex appears with a positive weight in the transfer matrix, but we can also define ensembles in which it has negative weight. The Brownian circuit is a notable example which is touched on in Sec.~\ref{sec:brownian}.

Fig.~\ref{fig:split_vertex}(b) shows an example of a new spacetime configuration that becomes possible when the new vertex is included.

We briefly consider the effect of this vertex on the coarse-grained Ising picture. 
\begin{figure}[t]
\centering
\includegraphics[width=\columnwidth]{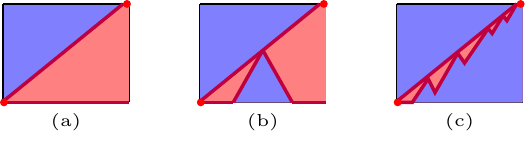}
\caption{Consequences of the domain wall splitting vertex. (a) One dominant configuration for $v > v_B $ in Haar circuit. (b) Configuration with one domain wall splitting event. (c) Configuration with multiple domain wall splitting event. Frequent splittings are favorable due to entropic considerations, leading to a "bound state" of left and right paths.}
\label{fig:split_vertex}
\end{figure}

First, this vertex will ``dress'' the local structure of a domain wall, in a way that changes the quantitative value of $\mathcal{E}(v)$ \cite{zhou_entanglement_2020}. 
For many applications of the Ising picture, including the calculation of the OTOC and the entanglement purity, this dressing is the only effect that needs to be taken into account, so that the structure of the  coarse-grained calculation is the same as for the Haar case.

This will also be true for the present correlator, $\overline{G^2}$, when $|v|<v_B$, if the two Ising domain  walls remain unbound.\footnote{We expect that if there is binding in the cluster picture for a given $v$ with $v<v_B$ (see Sec.~\ref{sec:vBvc})  there will also be binding in the Ising picture.}

However the boundary conditions for $\overline{G^2}$ are such that when $v>v_B$ the splitting events can also change the large scale structure of the configurations. This can be seen by considering the configurations in  Fig.~\ref{fig:split_vertex}. For simplicity, we consider the case where the vertex has a positive weight, so that the model has a simple classical interpretation.

Configuration (a) shows a domain wall with velocity ${v>v_B}$.
Configuration (b) is forbidden for the Haar model, but allowed in a more general model.
In this configuration we have used the ``new'' vertex to produce 
two additional domain wall segments, with velocities ${\pm v_B}$.
One may check that (a) and (b) have exactly the same free energy at leading order in $t$. The additional cost in (b) from the bulk domain walls is compensated by the additional $-$ region on the lower boundary. 
(This cancellation is special to the case where these additional domain walls have the optimal speed $v_B$; other choices of velocity would give a larger free energy.)

By similar reasoning, we can introduce further insertions of the new vertex without paying an extensive cost. 
Fig.~\ref{fig:split_vertex}~(c) shows a schematic. 
It follows that (in order to capitalise on the entropy of configurations with such insertions) the dominant configurations will have an \textit{extensive} number of insertions of the new vertex. This  produces a ``thin'' domain of $-$ spins (whose typical width remains finite when ${t\rightarrow\infty}$) whose coarse-grained speed is $v$. 
After coarse-graining, this thin domain can be assigned a line tension. 
This line tension is precisely $r_B(v)$ 
(defined in Sec.~\ref{sec:classesoftrajectory}), 
as can be seen by comparing with the expression for $\overline{G^2}$ obtained from the operator cluster picture in previous sections.
It should be noted that the line tension of this thin domain is no longer simply related to the line tension $\mathcal{E}_2$ of a single isolated Ising domain wall.\footnote{If there is binding for $v<v_B$ it will also be true there that the bound state tension is not simply related to $\mathcal{E}_2$.}

For the correlator, the main lesson from the above  is the following. 
We believe that in a general noisy model the quantities $r_{R,L}(v)$ 
are simply related to  the line tension 
$\mathcal{E}_2(v)$ of a single Ising domain wall.
(Here we are discussing  quantities involving  averages of ${U\otimes U^*\otimes U \otimes U^*}$, as relevant to $G_\text{rms}$: other kinds of averages will involve further dressing effects that we do not compute here, see Sec.~\ref{sec:avversustypbrief}.)
But for the \textit{bound state} rate function $r_B$, there is a simple relation with $\mathcal{E}_2$  only for the simplest averages in the Haar circuit, and not in more general models.

\section{Models without a bound state?}
\label{sec:numericalcasestudy}

We have  calculated the rate functions in the 1+1D Haar circuit using two different formalisms, and noted a special feature of this circuit.  To get insight into more general 1+1D models, we now present a numerical case study of  1D models in  which the data is consistent
with there being no bound state for any $v$. 
This is interesting because in the absence of a bound state there is a  specific relationship between  $G(\mathbf{x},t)$ and $\mathrm{OTOC}(\mathbf{x},t)$, at large $v$,  which is different from that of the Haar circuit.
 However, it is also possible that the models in this section do have a bound state at large $v$, just with very weak binding; this requires further examination.

We will discuss a noisy spin chain model and also the (qualitatively similar) ``Brownian circuit''~\cite{lashkari_towards_2013,shenker2015stringy,zhou_operator_2020,zhou_operator_2019,xu_locality_2019}.
We are restricted to modest sizes in numerics, so we maximise the spatial distance $x$ between the two operators by placing them at opposite ends of a system with open boundaries.
We denote this  edge-edge (EE) correlator by $G^\mathrm{EE}(x,t)$,
where ${x=L-1}$ for an $L$-site system.
We also define 
$\mathrm{OTOC}^\mathrm{EE}(x,t)$ as the OTOC with operators at the opposite ends. 
As before we define rate functions:
\begin{align}
G^\mathrm{EE}_\mathrm{rms}(vt,t)
&\asymp \exp[-s_\mathrm{eq}r^\mathrm{EE}(v)t], \\
\overline{\mathrm{OTOC}}^\mathrm{EE}(vt,t)& \asymp \exp[-s_\mathrm{eq}\lambda^\mathrm{EE}(v)t]
\label{eq:ee-corr}
\end{align}
Note that $v$ defines the aspect ratio of the space-time rectangle.

For the OTOC the rate function is in fact the same as in the case where the operators are not at the boundaries, 
\ie $\lambda^\mathrm{EE}=\lambda$, but we will retain the superscript to indicate how the numerical data was obtained.  
We will estimate asymptotic values of $r^\mathrm{EE}(v)$ by computing ${-\ln G^\mathrm{EE}_\mathrm{rms}(L-1,t)/s_\mathrm{eq}t}$ for $t=(L-1)/v$ and extrapolating the data to ${L\to\infty}$, and similarly for $\mathrm{OTOC}^\mathrm{EE}$.
See Appendix~\ref{app:noisy} for details.

\subsection{Morphologies for edge-edge correlators}
\label{sec:edgeedgemorphology}

Before discussing  specific models, let us classify possible  morphologies of the operator history for  edge-edge correlators.
Note that we are returning to the language of the operator string, which should not be confused with the Ising language in Sec.~\ref{sec:Ising}.

To simplify the discussion we assume the rate functions are parity-symmetric, so that the left and right butterfly speeds are equal: ${v_R=|v_L|=v_B}$.
In this case there are three possible morphologies, shown in Fig.~\ref{fig:edge-edge-morph}.
The new possibility in the finite system is the leftmost one, in which the operator string grows to span the entire system during part of the trajectory.
The others were already present in the bulk discussion: the thin (bound) and fat (unbound) trajectories respectively.

\begin{figure}[t]
\includegraphics[width=\linewidth]{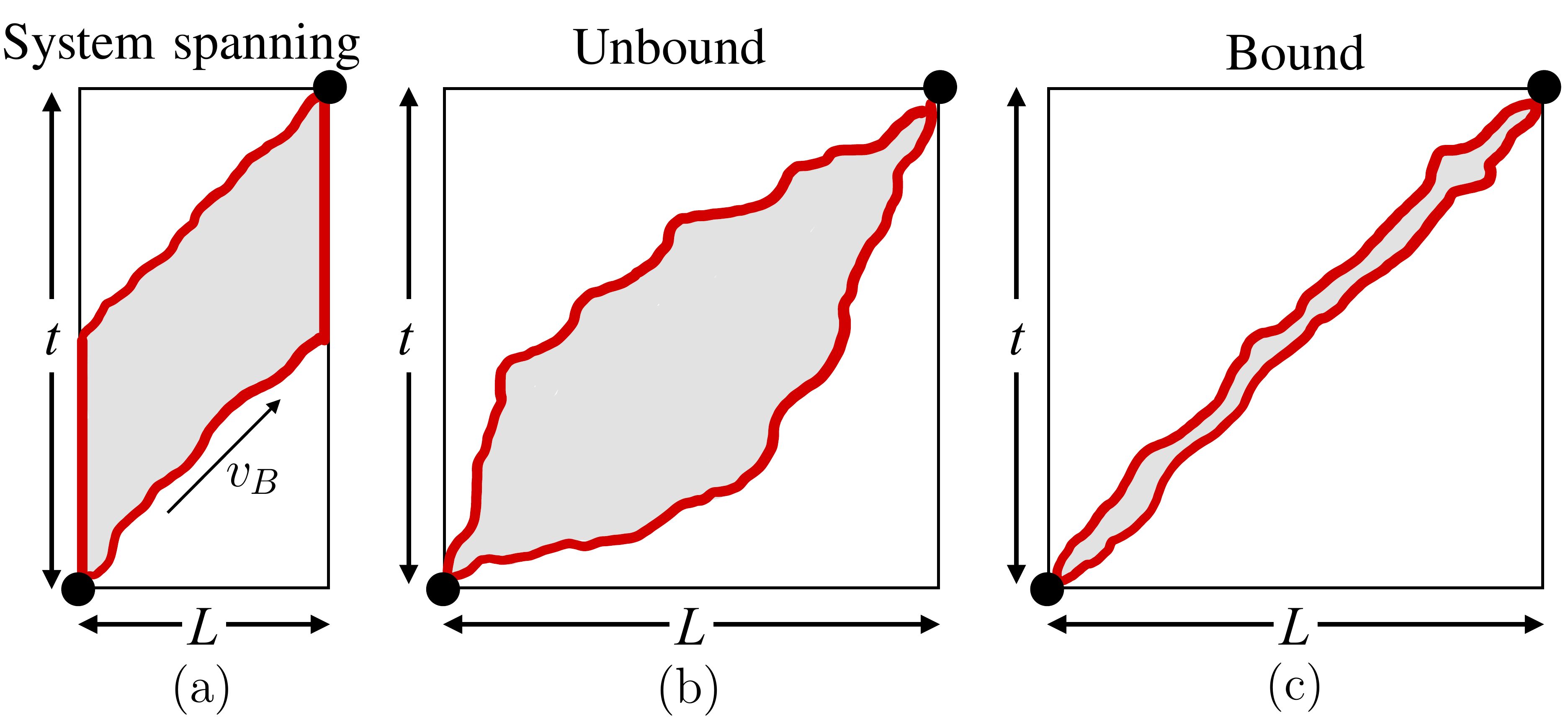}
\caption{The three possible morphologies of the operator history for the edge-edge correlators. (a) The system-spanning cluster which can appear only for  $v\le v_B$. (b) The cluster where the boundaries are unbound and which can appear only for $v>v_B$. (c) The case where the two strings form a stable bound state, which a priori can appear for both $v\ge v_B$ and $v \le v_B$. In the Haar case, the cluster shape switches directly from (a) to (c) at  $v=v_B$ whereas in a model with no bound state it transitions from (a) to (b) at $v_B$.}
\label{fig:edge-edge-morph}
\end{figure}

Let us now consider a situation where there is no bound state for any $v$. Trajectories of the type in Fig.~\ref{fig:edge-edge-morph}(c) are then never dominant.
Then, using the  properties of the functions $r_{R,L}$ in Sec.~\ref{sec:defininglinetensions}
we find that for $v\le v_B$, the system-spanning type  configurations are dominant
and the nontrivial boundary segments move at ${\pm v_B}$, as shown in Fig.~\ref{fig:edge-edge-morph}(a). On the other hand, for $v\ge v_B$, the morphologies with the unbound strings, Fig.~\ref{fig:edge-edge-morph}(b) dominate. We therefore have
\begin{align}
\label{eq:rEEunbound}
r^\mathrm{EE}(v) = 
\begin{cases}
v;\quad &v\le v_B\\
r_R(v)+v;\quad & v\ge v_B
\end{cases}\,.
\end{align}
The second line is the expression for the bulk rate function, in the case where trajectories are unbound.
In models where there is a bound state for some range of velocities,
$r^\mathrm{EE}(v)$ is the minimum of Eq.~\ref{eq:rEEunbound} and $r_B(v)/2$. 
The Haar circuit is a special case where there is a switch from the system-spanning trajectories to bound trajectories exactly at $v_B$ for the edge-edge correlators (\Cref{app:bd_eff_haar}).

Here, however, we focus on the case without a bound state, where 
Eq.~\ref{eq:rEEunbound} applies.
In this case, from Eq.~\ref{eq:lambdaotoc} and Eq.~\ref{eq:rEEunbound}, we see that $r^\mathrm{EE}$ has a very simple relationship with the rate function for the OTOC,
\begin{equation}
r^\mathrm{EE}(v)-v = \lambda^\mathrm{EE}(v)\,.
\label{eq:rEE-lamEE}
\end{equation}
We now describe models in which this identity appears to be obeyed to good precision, which is consistent with the absence of a bound state (or perhaps a bound state with only very weak binding at large $v$).

\subsection{Noisy spin-1/2 chain}
\label{sec:simulatedIsingmodel}

In this section and the next we  study Hamiltonian models where the couplings fluctuate randomly in time. 
Operator trajectories exist for arbitrarily large speeds in these models, unlike the brickwork circuit where the circuit geometry imposes a strict causal cone at speed 1.

The first such model has a stroboscopic-like protocol where the  Hamiltonian changes whenever time $\Delta t$ has elapsed.
The evolution operator is
\eq{U(t)=\prod_{n=1}^T\exp[-i \Delta t H^{(n)}]\,,
    \label{eq:U-noisy}}
where $t=T\times \Delta t$ and
\eq{
H^{(n)} = J\sum_{l=1}^{L-1}Z_{l}Z_{l+1}+\sum_{l=1}^L[h_l^{(n)}Z_l+g_l^{(n)}X_l].
\label{eq:H-noisy}
}
Here $\{X_l,Y_l,Z_l\}$ are the Pauli matrices at site $l$, and $\{h_l^{(n)}\}$ and $\{g_l^{(n)}\}$ are uncorrelated random numbers drawn from the  uniform distribution on  ${[h-W,h+W]}$. We use ${\Delta t=0.2}$, ${J=0.5}$, ${h=1}$ and  ${W=3}$. 

\begin{figure}
\includegraphics[width=\linewidth]{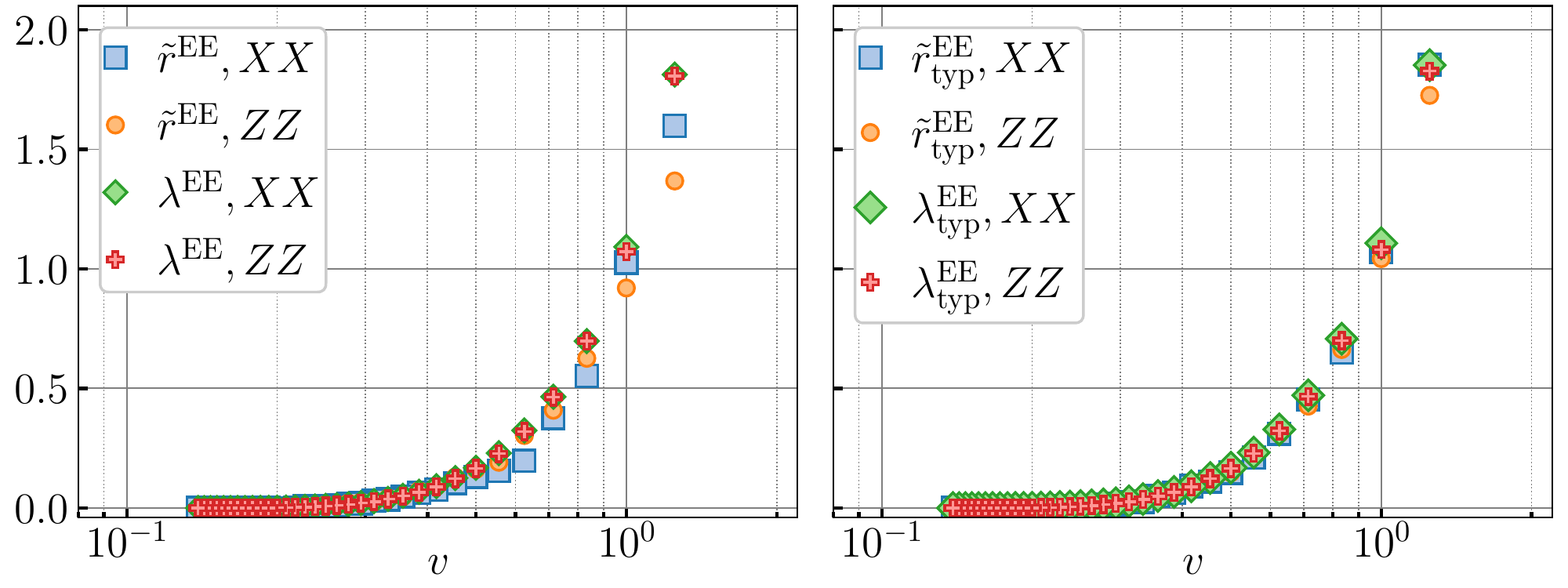}
\caption{Left: $\tilde{r}^\mathrm{EE}(v)\equiv r^\mathrm{EE}(v)-v$ and ${\lambda}^{EE}(v)$ defined in Eq.~\ref{eq:ee-corr} as a function of $v$ for the model defined via Eqs.~\ref{eq:U-noisy} and \ref{eq:H-noisy}, suggesting that $\tilde{r}^\mathrm{EE}(v)={\lambda}^\mathrm{EE}(v)$ in this case, consistent with the absence of a bound state. Right: The rates $r^\mathrm{EE}_\mathrm{typ}$ and $\lambda^\mathrm{EE}_\mathrm{typ}$ obtained from the typical values of $(G^\mathrm{EE})^2$ and $\mathrm{OTOC}^\mathrm{EE}$ are close to the those in the left panel (obtained from the mean).}
\label{fig:nsc-r-lam}
\end{figure}

We show the results in Fig.~\ref{fig:nsc-r-lam}(left) where $XX$ denotes the case where the two operators are $X_1(0)$ and $X_L(t)$, and similarly for $ZZ$. 
(See App.~\ref{app:noisy} for details.)
The data are consistent with Eq.~\ref{eq:rEE-lamEE}.

The right panel of Fig.~\ref{fig:nsc-r-lam} shows similar results, but for rate functions defined using the \textit{typical} values of $G^2$ and $\mathrm{OTOC}$, rather than the mean values. The typical and the mean values are close here, indicating that the ``dressing'' effects discussed in Sec.~\ref{sec:avversustypbrief} are weak.

\subsection{Brownian circuit}
\label{sec:brownian}

Since the time step $\Delta t=0.2$, after which the couplings in the Hamiltonian \eqref{eq:H-noisy} change randomly, is quite small, the model is reminiscent of the ``Brownian circuit'' --- a continuous time model with fields and spin-spin couplings that fluctuate like white noise~\cite{lashkari_towards_2013,shenker2015stringy,zhou_operator_2020,zhou_operator_2019,xu_locality_2019} and which  allows for a partial analytical simplification. 
We show here that data for the Brownian circuit is qualitatively similar to that in the previous section. 

Averaging over the randomness gives an effective ``classical'' description for the Brownian circuit, in continuous time. As in the Haar circuit, this can be formulated either as a Markov process in the basis of strings, or, using a different basis, as an effective Ising model; see Appendix~\ref{app:bc} for details.

First we note that the trajectories of the Markov process can be shown to simplify if we take the limit ${v\rightarrow\infty}$ at fixed large $x$. 
The  left and right operator endpoints $x_{L,R}$ then become unbound, biased random walkers, i.e. there is no bound state (we will discuss this elsewhere).
This limit $v\rightarrow\infty$ at fixed $x$ is distinct from the limit ${t,x\rightarrow \infty}$ at fixed $v$, however, so it is possible there is a bound phase at large $v$, with the strength of binding vanishing only as $v\rightarrow\infty$.

\begin{figure}
\includegraphics[width=\linewidth]{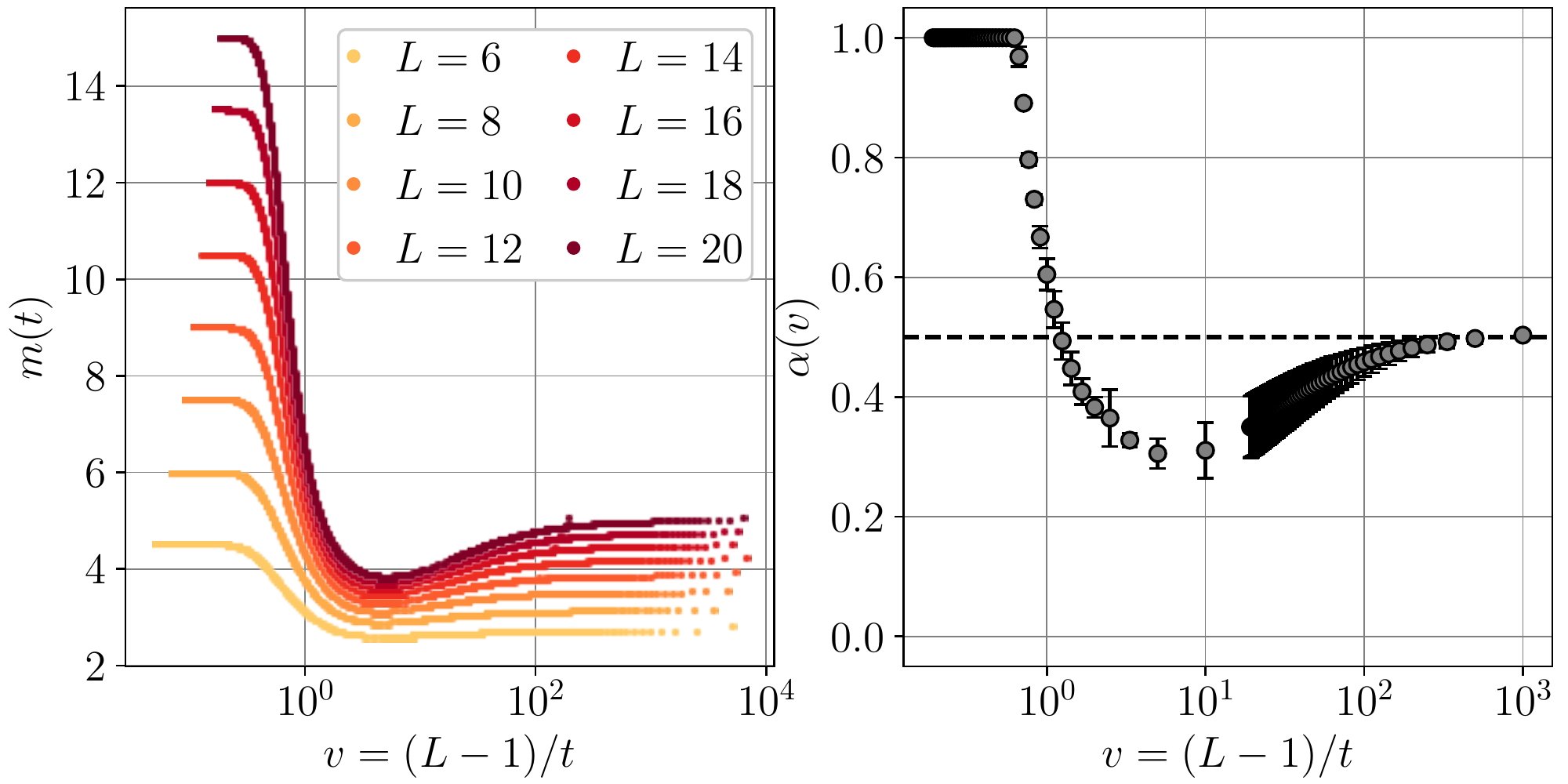}
\caption{Left: The cluster mass $m$, defined as the average number of occupied sites at  time $t/2$ halfway through  the trajectory,
for the  edge-edge correlator in the Brownian circuit. Data is shown as a function of velocity for several $L$. 
Right: Effective scaling exponent from a numerical fit  ${m\sim a + b L^\alpha}$ for the data at a given $v$.}
\label{fig:bc-cluster}
\end{figure}

We have studied the correlator numerically for general $v$ using the classical mappings  (treating the state as a $2^L$-dimensional vector).  $G^\mathrm{EE}_\mathrm{rms}$ and $\overline{\mathrm{OTOC}}^\mathrm{EE}$ are shown in Fig.~\ref{fig:r-lam-bc} (Left). 
Extracting the rate functions as in  \cref{sec:simulatedIsingmodel}  gives excellent agreement with the identity in  Eq.~\ref{eq:rEE-lamEE}, consistent with $G$ being in an unbound phase.

\begin{figure}
\includegraphics[width=\linewidth]{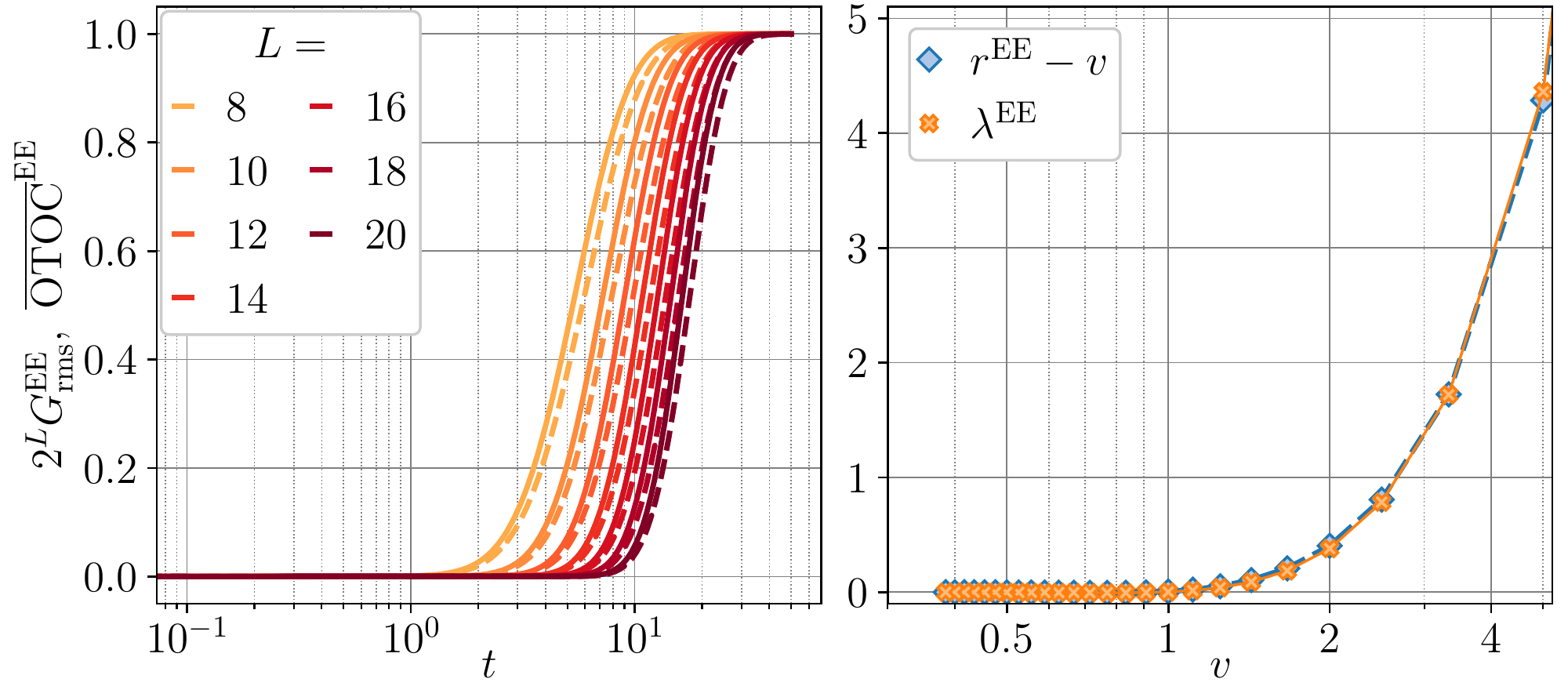}
\caption{Left: The edge-edge correlator, $G_\mathrm{rms}^\mathrm{EE}$ (dashed), and the edge-edge OTOC for the Brownian circuit for different $L$. Right: The corresponding rates $r^\mathrm{EE}$ and $\lambda^{EE}$ as a function of $v$ follow Eq.~\ref{eq:rEE-lamEE}  to a very good precision.}
\label{fig:r-lam-bc}
\end{figure}

This suggests that there is no bound state, for any $v$, in the Brownian circuit.
However, there is an  alternative possibility, which is a very weakly bound state (\ie with a large typical size and a small binding energy).

For an initial exploration we have  plotted the average mass $m$ of the cluster at the time $t/2$ midway through the trajectory (Fig.~\ref{fig:bc-cluster}).
For configurations of type (a), (b), and (c) in Fig.~\ref{fig:edge-edge-morph}, this average mass should be of order $L$, of order $L^{1/2}$, and of order $1$ respectively.
We find the expected $L$ scaling for ${v\lesssim v_B}$, and we find the expected  $L^{1/2}$ scaling 
in the limit  ${v\rightarrow\infty}$ (at fixed $L$). But for intermediate $v$ the cluster mass is growing  with $L$ but more slowly than $L^{1/2}$
(see Fig.~\ref{fig:bc-cluster}, Right)
i.e. finite size effects prevent a  conclusion as to which phase we are in.
This could be  resolved with  Monte Carlo 
studies of the Markov trajectories for large $t$ \cite{de2021rare}.

\section{Non-random Floquet and Hamiltonian systems}
\label{sec:instances}

Random circuits are a convenient laboratory for exploring the structures that we have discussed, but these structures are also relevant to more conventional many-body Hamiltonians, which need not have any randomness (and need not have a circuit structure).

In Sec.~\ref{sec:evolutionstring} we began by formulating the sum over operator trajectories (Eq.~\ref{eq:Gasamplitude})  in a given system (i.e. without any averaging):
\begin{equation}
\label{eq:Gsumreprise}
G =  \sum_{\mathcal{S}_{t-1}, \ldots \mathcal{S}_{1}} 
\UU^{(t)}_{\mathcal{S}_F, \mathcal{S}_{t-1}}
\ldots
\UU^{(2)}_{\mathcal{S}_2, \mathcal{S}_{1}}
\UU^{(1)}_{\mathcal{S}_1, \mathcal{S}_{I}}. 
\end{equation}
We  discussed the case of a particular instance  of a Haar circuit in Sec.~\ref{sec:avversustypbrief}, but we could also consider a Floquet system  with discrete space and time translation invariance 
(i.e. without randomness), or indeed a system with a fixed Hamiltonian and therefore with continuous time-translation invariance. (In the latter case there is an additional feature, energy conservation, which we comment on below.)
For simplicity we continue to use the language of discrete time evolution, but this is not crucial.

In a particular circuit we no longer have a simple microscopic mapping to a Markov process.
We argue in this section that the basic structures that we have discussed survive, 
in particular the distinction between bound and unbound phases in 1+1D. 
We restrict here to a qualitative discussion.

\subsection{Bound phase}
\label{sec:instancebound}

The universal physics of the bound phase is the simplest, and also the most generic once we go beyond 1+1D, so we start with this. We describe the simplest scenario (additional features are possible which are mentioned briefly below).

By definition,  the dominant operator histories in the bound phase involve operator strings with a finite typical length $\xi$.
The simplest  scenario is then that  coarse-graining to scales beyond $\xi$ gives an effective theory for the path of a pointlike ``particle'', i.e. for the position $x_\text{cm}$ of the operator string. 
Heuristically,
this  particle is then   characterised by an amplitude $K_{t;\Delta t}( x'; x)$ for the particle to propagate from $x$ to $x'$ in a coarse-grained   time interval $\Delta t\gg 1$.  In the case with both space and time translation symmetry  we can write this as ${K_{\Delta t}(x'-x)}$.
$K$ is real, but can be either positive or negative. At first sight it looks similar to the propagator for  Schrodinger evolution of a particle,  but a key difference is that $K_{\Delta t}(x'-x)$ is not unitary.\footnote{Unless the dynamics is fine-tuned and does not cause strings to grow (e.g. a circuit of SWAP operators, or a free fermion Hamiltonian acting on fermion creation operators).} 
This is because we have effectively projected the full unitary dynamics of the operator string to a restricted subspace of small strings. This leads to exponential decay.\footnote{In general the decay could be accompanied by oscillations \cite{bertini_exact_2018}.}
Below we discuss one way to make this more precise in the microscopic model.

In the bound phase it is possible to compute the rate function ${r(v)=r_B(v)}$ perturbatively, taking into account larger and larger strings at higher order. As discussed below,  this picture of  a simple bound phase connects with recent work in Refs.~\cite{kos_correlations_2021} and \cite{von2021operator} that
appeared after the results in this paper for the random circuit were obtained.
The perturbation theory will simplify in any limit where the typical cluster size ($\xi$ above) becomes small: for example in the Haar circuit, this happens when $v$ approaches its maximal value, as can be seen in Eq.~\ref{eq:boundstatesizemaintext}.

Above we discussed a propagator for an effective  ``particle'' heuristically. For a practical computation we should work with objects that are well-defined on the lattice scale. 
One possibility, which is practical in the strongly bound regime \cite{zhou_entanglement_2020}, is to
distinguish ``short'' operator strings
(taken to be single Pauli operators $\sigma^\alpha$)
from ``long'' strings,
and define amplitudes for excursions of any given duration $\Delta t$ 
outside the space of short strings.
(See also Ref.~\cite{claeys_absence_2021}.)
In the strongly bound phase these amplitudes decay rapidly with $\Delta t$ and working with them eliminates most finite size effects.\footnote{For concreteness, consider a translation-invariant spin-1/2 Floquet model with unit time period. Let 
$Z^{\alpha\beta}( x,  t)=\langle \sigma^{\alpha}_{x}( t) \sigma^{\beta}_{0}\rangle$
be the amplitude for a string to propagate from Pauli operator $\sigma^\beta_0$ at a site $0$ to Pauli operator  $\sigma^\alpha_{x}$ at site $x$ in a time ${t\geq 1}$.
Now we define the elementary amplitudes $W^{\alpha\beta}( x,  t)$. For $t=1$ we have simply 
 $W^{\alpha\beta}( x,  1)= Z^{\alpha\beta}( x,  1)$, but for $t>1$ we define $W$ to be the amplitude to propagate from 
$\sigma^\beta_0$ to $\sigma^\alpha_{x}$ without being a ``short'' string at any intermediate time. $W$ may be obtained  recursively from $Z$:
\be\notag
W^{\alpha\beta}(x,t) = 
Z^{\alpha\beta}(x,t) 
- \sum_{\substack{t'<t; y}}
W^{\alpha\gamma}(x-y,t-t')Z^{\gamma\beta}(y,t').
\ee
Numerically, $Z$ and therefore $W$ can be obtained up to some $t_\text{max}$. The  rate function $r_B(v)$ may then be extracted from the Laplace transform of $W$ with respect to $x$ and $t$, see \cite{zhou_entanglement_2020} for an implementation of a similar procedure for the entanglement membrane. }

In Ref.~\cite{kos_correlations_2021}, a  perturbation theory for correlation functions in a quantum circuit close to the ``dual-unitary'' \cite{bertini_exact_2018,bertini_entanglement_2018,akila2016particle,bertini_exact_2019-1,bertini_operator_2020,bertini_scrambling_2020,gopalakrishnan_unitary_2019,prosen_many_2021,piroli_exact_2020,claeys2021ergodic} limit was developed.
In the dual unitary circuit, correlations are confined to a lightcone, so that the only contributing operator trajectories consist of a cluster of size 1 travelling along the light cone.
The perturbation theory about this limit can be developed in terms of segments of such paths connected by vertices \cite{kos_correlations_2021}.
The authors found that there was a regime in which this perturbation theory was convergent. In the present language, this regime lies within the bound phase.

The fact that the operator trajectories that dominate 2-point functions involve much smaller strings than the trajectories that dominate the OTOC has practical consequences for numerical simulations \cite{rakovszky2022dissipation,von2021operator}. Refs.~\cite{rakovszky2022dissipation,von2021operator} develop a matrix-product-operator based approach to calculating two-point functions in the Heisenberg picture (for a model with a conserved density) which exploits the possibility of discarding long strings. 

\subsection{Conserved densities}

If the model has a conserved density this will affect the nature of the trajectories and the rate functions. If for example the two-point function is sensitive to a diffusive mode, then for ${|x|\ll t}$ we have ${G(x,t)\sim c(v) t^{-1/2} \exp(- x^2/2Dt)}$
(assuming parity symmetry),
which means that the rate function close to $v=0$ obeys 
\be
r(v) \simeq \f{v^2}{2 \seq D}  + \mathcal{O}(v^4)
\ee
(in the  convention of  Eq.~\ref{eq:Gratefn}).
The dominating trajectories in this regime of small $v$ are thin ones, that can be thought of heuristically  as being dominated by the short local operators  that represent the conserved density
\cite{khemani2018operator,rakovszky2018diffusive,von2021operator}: 
for example the Pauli $Z_x$ operators, in a random circuit with conserved $\sum_x Z_x$, or local energy densities in a model with conserved energy.

Once $v$ is of order 1 
(i.e. when $x$ and $t$ are taken to be large and of the same order) 
it is no longer guaranteed that the conserved densities will dominate the correlator: in principle we might imagine a 1+1D model with a bound phase at small $v$ (in which the strings had significant overlap with the conserved density) which gave way to an unbound phase, dominated by longer strings,
at larger $v$. 
However, in spatial dimensions above 1 we expect thin trajectories for all $v$, regardless of the presence or absence of a conserved density, for the reasons discussed in Sec.~\ref{sec:higherdims}.

\subsection{Unbound phase}

Above we have discussed how to think about the bound phase in a particular circuit, without any averaging. The unbound phase is perhaps more interesting, since the sum over diagrams for $G$ in Eq.~\ref{eq:Gsumreprise} is more nontrivial, involving spactime histories, or Feynman diagrams, of spatial width much larger than the lattice spacing. Here we suggest that a picture similar to the one we developed in Sec.~\ref{sec:classesoftrajectory} survives.

We continue to think of $G$ as a partition function for spacetime diagrams (cluster histories).
A difference from a standard partition function for a classical model is that the local weights defining the partition function (determined by $\UU_{S,S'}$) are can be negative.
Nevertheless, we may try to define free energy densities associated with different types of local structure. 
Given the minus signs, this is only a conjecture.

Consider a cluster whose typical spatial size is much larger than microscopic scales.  
The free energy density of the ``vacuum'' outside the operator cluster is manifestly zero ($\UU$ acting on the identity gives the identity). There will be a nontrivial free energy associated with cluster boundary, precisely as in earlier sections. 
What remains to consider is the bulk of the cluster. At first sight we should associate a free energy density $f_\text{bulk}$ with this interior. However, we expect that $f_\text{bulk}=0$, in order to be consistent with operator spreading (for example when we modify the final-time boundary condition so as to pick out the weight $a_\mathcal{S}$ of a long string).\footnote{In the Markov process of earlier sections,
$f_\text{bulk}=0$ follows from the fact that the probability-conserving dynamics has an equilibrium state that corresponds to the interior of the cluster \cite{merolle2005space,liu2021void}.}
If these assumptions are correct, then we recover a picture like the one discussed in Sec.~\ref{sec:defininglinetensions}, with the asymptotics of $G(x,t)$ in the unbound phase set by coarse-grained line tensions $r_{L,R}$. 
However, as we have discussed, we expect the unbound phase to be special to 1+1D models (absent fine-tuning).

\section{Outlook}
\label{sec:outlook}

We list some unresolved questions and directions for the future.

We have argued that it is useful generally to classify spacetime Feynman diagrams 
according to their coarse-grained geometry, and to characterize them by line tensions for various types of paths,
and that in 1+1D there can be phase transitions between distinct classes as a function of velocity or model parameters. 
It would be interesting to explore this phenomenology in other contexts. 

First, it would be interesting to explore realistic models numerically and potentially  in experiment.

Numerical exploration of the unbinding transition as a function of $v$ would benefit from optimized  methods for calculating two-point functions. Directly computing the two-point function and fitting its asymptotic form is unlikely to be the optimal approach, because this involves boundary effects from the initial and final time.

There is also a numerically more tractable version of the unbinding transition that could be explored.
In 1+1D (if there are no conservation laws) we can  have an unbinding transition for a correlator at a \textit{single} position, ${\< \mathcal{O}(x,t)\mathcal{O}(x,0)\>}$, as a function of a parameter in the model. 
Numerically, the simplest case is where $x$ lies at a spatial boundary of the system. 
The parameter driving the transition could be the strength of a boundary coupling. 
The trajectories can either  ``stick'' to the spatial boundary or unbind from it. Since the correlator has $v=0$ this case is less demanding numerically.

In the bound phase with a thin cluster, $r(v)$  is (in principle) efficiently computable on a classical computer, while at first sight it is much more challenging to compute $r(v)$ in the unbound  phase  due to the large Hilbert space of long operator strings.
However, our results suggest that there should exist efficient numerical techniques that isolate the contributions from near the boundary of the cluster which determine the rate function. 
More generally, it is interesting to ask about the computational complexity of evaluating correlators in various settings and phases. A transition in complexity as a function of circuit depth in a dual unitary circuit was found in 
Ref.~\cite{suzuki_computational_2022}.

Second, it will be interesting to attempt analytical calculations in settings where we cannot exploit simplifications from random averaging.

The bound phase can be handled by a direct perturbation theory, at least in principle, 
which should make it possible to examine a wide range of models quasi-analytically if they are in this phase. This could include simple quantum field theories. Spacetime Feynman diagrams may simplify at large $v$ in some cases, in analogy to the random circuits (both Haar when ${v\rightarrow 1}$ and Brownian when ${v\rightarrow \infty}$).

The  calculations here were for models with a simple infinite temperature equilibrium state, so that equal-time correlations vanish. More generally they will be present. How do they modify the spacetime picture?

In the unbound phase, where we have to  handle the interior of the operator string, 
it may be enlightening to explore further the simpler setting of  lattice models at infinite temperature.
The OTOC can be handled even in the translation-invariant case by locally separating the multi-copy Hilbert space into different sectors \cite{zhou_entanglement_2020, mcculloch2021operator}, but the treatment of the two-point function may be more subtle.

There are also outstanding questions even within the  realm of random circuits.
For example, it would be technically interesting to study fluctuations of G in the Haar circuit using replicas \cite{zhou_emergent_2019}. 
This  would shed light on which identities (Sec.~\ref{sec:symmrels}) are special to $G_\text{rms}$ and which  hold even at the level of a single realization of the circuit (this may also shed light on the status of these identities in translationally invariant models).

It may be interesting to explore the consequences of the vanishing vertex weight discussed in Sec.~\ref{sec:Haarsimplification} for more general (e.g. higher-point) correlation functions in the Haar circuit. 

Perturbative calculations may shed light on the phase diagram in various limits.
The numerics in Sec.~\ref{sec:brownian} did not show signs of a bound state in the Brownian circuit; this could be examined further, perhaps using the fact that trajectories simplify when ${v\rightarrow\infty}$. 
We have argued that the identity ${v_c=v_B}$ is special to the 1+1D Haar circuit; it would be interesting to calculate $v_c(\lambda)$ and $v_B(\lambda)$ as a function of a small parameter $\lambda$ that tuned the random circuit ensemble  away from the Haar ensemble.

\begin{acknowledgments}
We thank Fabian Essler, Juan P. Garrahan,  David Huse, Sergio de  Queiroz, Curt von Keyserlingk, Tibor Rakovszky, Andrea De Luca, and Pierre Le Doussal for discussions. We have surely had other useful feedback on talks based on these results since 2018. 
AN acknowledges support from EPSRC Grant No.~EP/N028678/1 and from a Royal Society University Research Fellowship for support during parts of this work.  SR acknowledges an ICTS-Simons Early Career Faculty Fellowship. SR also acknowledges EPSRC Grant No. EP/S020527/1 for support during his stay at the University of Oxford where a part of the work was carried out. TZ was supported by a postdoctoral fellowship from the Gordon and Betty Moore Foundation, under the EPiQS initiative, Grant GBMF4304, at the Kavli Institute for Theoretical Physics. TZ is currently supported as a postdoctoral researcher from NTT Research Award AGMT DTD 9.24.20 and the Massachusetts Institute of Technology.

\end{acknowledgments}

\appendix

\section{Review of transition probabilities}
\label{app:transitionrates}

The transition probabilities for the process on Pauli strings \cite{oliveira_generic_2007,dahlsten_emergence_2007,znidaric_exact_2008} or clusters \cite{nahum_operator_2018, von_keyserlingk_operator_2018} associated with the brickwork Haar circuit  (see \Cref{sec:Haarsimplification,sec:Haartransition}) are as follows.

Consider a time step where a pair of sites $(x, x+1)$ receive a unitary.
The operator string on these two sites (which is part of the possibly larger string $\mathcal{S}'$) 
is updated probabilistically.
The trivial string $\mathbb{1}\otimes \mathbb{1}$ is left unchanged by the update.
If the string is nontrivial then the probabilities for the outcomes are independent of which nontrivial initial string we have to start with:
this string is  randomly replaced with any  of the other nontrivial basis operators on these two sites, with equal probability for each of the $q^4-1$ possibilities. 
For the a qubit chain, the 15 possible nontrivial strings are of the forms 
 $\mathbb{1}\otimes \sigma^{j}$, or  $\mathbb{1}\otimes \sigma^j$, or $\sigma^j\otimes \sigma^k$, for for Pauli labels $j,k=x,y,z$. 
 
As mentioned in Sec.~\ref{sec:Haarsimplification}, the string $\mathcal{S}'$ defines an occupation number $n_{\bf x}$ for each site which is equal to 1 (represented  $\begin{tikzpicture}
\fill[black] (0,0) circle (0.1cm); \end{tikzpicture}$) if  the site is in the support of $\mathcal{S}'$, i.e. if the site hosts a nontrivial basis operator,   and to zero (represented  $\begin{tikzpicture}
\draw (0,0) circle (0.1cm); \end{tikzpicture}$) if it does not.
The stochastic process above defines a simple stochastic process for these occupation numbers. When a unitary is applied to a pair of unoccupied sites we have $\emem\rightarrow\emem$, i.e. they remain unoccupied.
When a unitary is applied to a pair of sites whose total occupancy is nonzero, then the outcome is either $\emfu$ with probability $p$, $\fuem$ with probability $p$, or $\fufu$ with probability $1-2p$, 
where these probabilities are determined simply by the fraction of nontrivial strings that are of each type, giving ${p=1/(q^2+1)}$, i.e. $p=1/5$ for the case of qubits.

The simple, symmetric structure of these rules means that the endpoints of the operator string $\mathcal{S}'$ satisfy autonomous random walk dynamics. The transition probabilities were reviewed in Sec.~\ref{sec:Haartransition}. 

The stochastic process above satisfies detailed balance with respect to a simple equilibrium measure in which sites are uncorrelated, and a given site is
equally likely to be any of the basis operators. Since these include the $q^2-1$ nontrivial operators, together with the identity,  this means that in the equilibrium state a given site is occupied with probability $(q^2-1)/q^2$, i.e. $3/4$ for a qubit chain.

In Sec.~\ref{sec:classesoftrajectory},  detailed balance with respect to this distribution was used to relate $r_R(v)$ to $r_R(-v)$, and similarly for $r_L(v)$. 
There we needed the fact that the state 
 $x_R+d$ is more likely than $x_R$ by a factor $\exp ( 2 \seq d)$, 
 which in the case of a qubit chain is $4^d$.
This is just the number of possible states for the extra length of  string  in the region $(x_R,x_R+d]$.

\section{Transfer matrix for walks}
\label{app:transfermatrixwalks}

The Laplace-transformed partition function is:
\begin{equation}
Z(\mu) = \sum_x e^{-\mu x} Z(x).
\end{equation}
In this partition function, the final time boundary condition for the centre of mass coordinate ${X=(x_L+x_R)/2}$ is free (but there is a ``force'' $\mu$ on this point). 
The final time boundary condition for the relative coordinate $\Delta = (x_R-x_L)/2$ is $\Delta(t)=0$.
The initial boundary conditions are $X=\Delta=0$.

The factors of $e^\mu$ can be absorbed into the weights. Schematically,
\begin{equation}
Z(\mu) = 
\sum_{\{X(t')\}}
\sum_{\substack{
\{ \Delta (t') \} \\
\Delta (t)=0
}}
 \prod_{t'} \left( 
 W \times
 e^{- \f{\mu}{2} (\delta x_R  + \delta x_L) } 
  \right),
\end{equation}
where $\delta x_{L,R}$ is the change of $x_{L,R}$ in a given time step.
We can sum over the possibilities for the centre of mass coordinate $X(t')$, with the trajectory $\Delta(t')$ fixed.
If $\delta \Delta=\pm 1$  in a given timestep that means that the walkers move in opposite directions, so that $X(t')$ does not change.
If $\delta \Delta=0$ in a given timestep, then there are two possibilities to sum over, one where the two walks move to the right, with an additional weight $e^{-\mu}$, and one where they move to the left, with an additional weight $e^\mu$. 
We obtain
\begin{equation}
\label{eq:zmuappendix}
Z(\mu)  = [p(1-p)]^t  \, \sum_{\{\Delta\}} \prod_{t'} T_{\Delta(t'+1), \Delta(t')}.
\end{equation}
The sum is now over $\Delta(1),\ldots, \Delta(t-1)$, with ${\Delta(t'+1) - \Delta(t') = \pm 1}$ and $\Delta \geq 0$. $T$ is a semi-infinite transfer matrix
\begin{equation}
T = T_1 T_2,
\end{equation}
with 
\begin{align}
T_1 & = \left(
\begin{array}{ccccc}
E M & 1 & 0 & 0 & \ldots \\
1 & M & 1 & 0 & \ldots \\
0 & 1 & M & 1 & \ldots \\
\ldots &&&&
   \end{array}
\right),
& \hspace{-2mm}
T_2 & = \left(
\begin{array}{ccccc}
V &  0 & 0  &  0 &  \ldots \\
0 & 1 & 0 & 0  &  \ldots  \\
0 & 0 & 1 & 0  &  \ldots  \\
  \ldots  &&&&
   \end{array}
\right)
\end{align}
where 
\be 
M = e^\mu + e^{-\mu}.
\end{equation}
Up to boundary terms that we are not considering here, we can replace $T$ with the symmetrized transfer matrix $T' = T_2^{1/2}T_1 T_2^{1/2}$:
\begin{equation}
T' = \left(
\begin{array}{ccccc}
V E M & V^{1/2} & 0 & 0 & \ldots \\
V^{1/2} & M & 1 & 0 & \ldots \\
0 & 1 & M & 1 & \ldots \\
\ldots &&&&
   \end{array}
\right).
\end{equation}
We would like to consider when $T'$ has a bound state. If a bound state is present its form will be
\begin{equation}
\label{eq:boundstateansatz}
\psi = (1, A, Ak, Ak^2,\ldots)
\end{equation}
with $T' \psi = \lambda \psi$, so that 
\begin{align}
\label{eq:transfereigeq1}
A V^{1/2}+ VEM    & = \lambda, \\ \label{eq:transfereigeq2}
A(k+M) + V^{1/2} & = A \lambda\\ \label{eq:transfereigeq3}
1+M k + k^2 & = k \lambda.
\end{align}
To find the point at which the bound state appears, we fix $k=1$: this gives 
\begin{equation}
\label{eq:muc}
M_c = q^2+q^{-2}
\quad \Longrightarrow \quad
e^{\mu_c} =q^2. 
\end{equation}
The bound state exists when $|\mu|$ is \textit{larger} than $\mu_c$. When $|\mu|$ is smaller than $\mu_c$ there is no solution with $k<1$.

Next we ask what critical velocity this corresponds to. The velocity is a function of $\mu$, 
which will be continuous but nonanalytic at $\mu_c$. 
This is easy to determine for $|\mu|<\mu_c$ (no bound state).
Since the walks then wander far apart we can neglect the contact interactions, and the  factors of $p$ and $(1-p)$ are independent of $v$, 
so the the only weights we need take into account are factors of  $e^{\mu/2}$ or $e^{-\mu/2}$ each time one of the walks takes a step to the left or right respectively. 
Averaging over these possibilities gives the velocity in the unbound regime:
\begin{equation}
v(\mu) = \f{e^{\mu/2} - e^{-{\mu/2}}}{e^{\mu/2} + e^{-\mu/2}} \qquad \text{for $|\mu|<\mu_c$},
\end{equation}
so from (\ref{eq:muc})  the critical speed (above which there is a bound state) is
\begin{equation}
v_c = \f{q^2 - 1}{q^2 +1}. 
\end{equation}
This critical velocity coincides with the butterfly speed $v_B$ in the circuit.

In the unbound regime, where the two walks are well-separated, the scaling of the partition function is given by the rate functions for isolated walks, so it is enough to consider the unconstrained dynamics of an isolated $x_R$ walker. The probability to travel a distance $x$ is 
\begin{equation}
P_R(x) = (1-p)^{(t+x)/2} p^{(t-x)/2}
\left(
\begin{array}{ccc}
t  \\
  (t+x)/2
\end{array}
\right).
\end{equation}
Setting $P_R(x)\sim \exp( - \seq r_R(v) t)$, with $\seq = \ln q$ and $v=x/t$, this gives
 \be
 r_R(v) =  \mathcal{E}_2(v) - v
 \ee
 where $\mathcal{E}_2(v)$ is symmetric in $v$,
 \be
 \mathcal{E}_2(v) = \f{\ln \f{q^2+1}{q} + \f{1+v}{2} \ln  \f{1+v}{2} + \f{1-v}{2} \ln  \f{1-v}{2}   }{\ln q}.
 \ee
 
 In the bound state regime $\mu > \mu_c$
 (where $v<-v_c$; results for $v>v_c$ are analogous) 
 the transfer matrix  has the leading eigenvalue (from Eqs.~\ref{eq:transfereigeq1}-\ref{eq:transfereigeq3}) 
 \ba
 \lambda(\mu) = {(q+1/q)} \lf q e^{-\mu} + q^{-1} e^{\mu} \ri
\end{align}
 and typical size $\Delta_\text{typ}$:
 \be\label{eq:boundstatesizeapp1}
 1/\Delta_\text{typ} \equiv - \ln k =  \ln (e^{\mu}/q^2).
 \ee
The free energy $- \ln Z(\mu) $ at large $t$ may be written either in terms of 
the line tension for the bound state  or in terms of the transfer matrix eigenvalue
$\lambda(\mu)$: 
 \be\label{eq:legendreappendix}
\min_v \lf \mu v + \seq r_B (v) \ri = - \ln \lambda(\mu) - \ln p(1-p)
 \ee
(the final term is from the prefactor in Eq.~\ref{eq:zmuappendix}).
The Laplace transform is inverted by:
\begin{equation}
r_B(v) = \mathcal{E}_2(v) + |v|.
\end{equation}
The velocity appearing in (\ref{eq:legendreappendix}) is $v(\mu)={(q^2-e^{2\mu})}/{(q^2+e^{2\mu})}$, so we can write the bound state size (\ref{eq:boundstatesizeapp1}) as 
 \be\label{eq:boundstatesizeapp2}
 \Delta_\text{typ}(v)= \f{2}{\ln \lf \f{1}{q^2} \times \f{1+|v|}{1-|v|}\ri}.
 \ee
The size of the bound state diverges as $|v|\rightarrow v_c$:
\begin{equation}
 \Delta_\text{typ}(v) = \f{4 q^2}{(q^2+1)^2} \lf \f{1}{|v|-v_c}\ri + \ldots 
\end{equation}
and it vanishes as $v$ approaches the maximal possible speed $|v|=1$ allowed by the geometry of the brickwork circuit.

\section{Boundary effects in Haar circuit}
\label{app:bd_eff_haar}

The presence of a boundary can change the scaling of $G$. We analyze the boundary effect in the Ising domain wall picture. 

For simplicity, we take $v > 0$ and place $(0,0)$ at the left boundary. If $v$ is small, the domain wall has an alternative path to exit through the left boundary as shown in Fig.~\ref{fig:morph_fs}(a)
(the microscopic mechanism is shown in Fig.~15 of \cite{zhou_emergent_2019}).
There is no contribution to the free energy cost from the spatial boundary of the system.

Let us do a quantitative comparison. 
\begin{figure}[h]
\centering
\includegraphics[width=0.8\columnwidth]{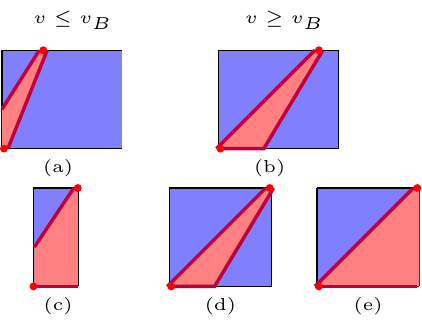}
\caption{The Ising domain wall configurations when one (top row) or both of the operators (bottom row) are at the spatial boundaries. When $v < v_B$, the left domain wall has the option of exiting from the boundary ((a),(c)). Configurations (d) and (e) are degenerate and have the same free energy. In (b), (d), the right-hand domain wall is of speed $v_B$.}
\label{fig:morph_fs}
\end{figure}
When $(x,t)$ is far away from the right boundary, the line tension of the right domain wall is the same as Eq.~\ref{eq:fR}. For $v > 0$, we have
\begin{equation}
f_R(v) 
= 
\left\lbrace
\begin{aligned}
  & \mathcal{E}(v) & \quad  0 \le  v \le v_B   \\
  & v  & \quad  v_B <  v  
\end{aligned} \right. 
\end{equation}
The left domain wall costs
\begin{equation}
f_L(v) = \min_{v'\le v } \mathcal{E}(v') \frac{x}{v'} \frac{1}{t} = v \min_{v'\ge v } \frac{\mathcal{E}(v') }{v'}
\end{equation}
By the properties of the line tension function, it is
\begin{equation}
\label{eq:0L_fL}
 f_L(v) =
\left\lbrace
\begin{aligned}
  & v  & \quad   0 \le v \le v_B \quad \implies v' = v_B  \\
  & \mathcal{E}(v)   & v_B < v \quad \implies v' = v  \\
\end{aligned} \right. 
\end{equation}
Therefore the rate function becomes
\begin{equation}
r(v) = \frac{1}{2} (f_L(v) + f_R(v) ) = \frac{1}{2} ( \mathcal{E}(v) + v ) 
\end{equation}
and the transition in $r(v)$ disappears. 

When we also place $(x,t)$ on the right boundary, a transition emerges again. The analysis the of the left domain wall is the same as Eq.~\ref{eq:0L_fL}. However, in the presence of the right boundary, the right domain wall can legitimately exit the system as in Fig.~\ref{fig:morph_fs} (b) (d). In that case, the  cost of the right-hand path from $(0,0)$ to $(x,t)$ comes from the segment on the lower boundary and is equal to $v t$. Since this is never greater than the cost in Eq.~\ref{eq:fR}, we have
\begin{equation}
f_R(v ) = v. 
\end{equation}
Therefore, the rate function is
\begin{equation}
r(v) =
\left\lbrace
\begin{aligned}
  & v & \quad 0 \le v \le v_B \\
  & \frac{1}{2}(v + \mathcal{E}(v) )  & \quad v_B < v \\
\end{aligned} \right. 
\end{equation}
This phenomenon is useful in eliminating the finite size effect in small system numerics.

\section{Thin cluster approximation in $d=2$}
\label{app:thin_cluster}

Consider $q$-dimensional qudits on the sites of an $L\times L$ square lattice. We consider dynamics of this system which consist of choosing a random bond on the lattice, and applying a two-site unitary gate to the qudits at the ends of the bond.  The unitary gate is chosen from the Haar measure over the unitary group U$(q^{2})$.  A single timestep of the dynamics is defined by applying $L^{2}/2$ such gates, so that each qudit is acted upon by a single unitary operator in each timestep, on average.

The evolution of $\overline{G(0,t)^{2}} \equiv \overline{\langle \mathcal{O}(0,t)\mathcal{O}(0,0)\rangle^{2}}$ where $\overline{\cdots}$ denotes an average over the Haar-random unitary gates is given by a Markov process for a two-dimensional cluster, whose support at the initial and final times is exclusively at the origin $\boldsymbol{x} = (0,0)$.  While many trajectories contribute to this averaged correlator, we may focus our attention on completely thin trajectories where the cluster does not grow, and remains supported exclusively at a single site at every timestep of the evolution. 

To determine the contribution of these trajectories to the correlator, we first observe that when a unitary gate is applied, the probability that a particular site is not acted upon by that gate is $p_{0}(L) = (2L^{2}-4)/(2L^{2})$ since each site is attached to four distinct bonds, and there are $2L^{2}$ bonds in the system.  On the other hand, the probability that the same site is acted upon by unitary gates in a single timestep, and afterwards remains exclusively supported at a single site is 
\begin{align}
    p_{1}(L) = \Big[1-p_{0}(L)\Big] \frac{2(q^{2}-1)}{q^{4}-1} 
\end{align}
The total weight $w(L)$ of cluster histories which ($i$) start at the origin, ($ii$) are maximally thin, and ($iii$) are allowed to end at any point at the final time is then given by
\begin{align}
    w(L) = \Big[p_{0}(L) + p_{1}(L)\Big]^{(L^{2}t/2)}
\end{align}
In the thermodynamic limit $L\rightarrow\infty$, this simplifies to
\begin{align}
    &\lim_{L\rightarrow\infty}w(L) = e^{-\Gamma(q)t}
\end{align}
with the $q$-dependent decay rate
\begin{align}
    \Gamma(q) = \frac{q^{2}-1}{q^{2}+1}.
\end{align}

\section{Brownian circuit: cluster and Ising pictures}
\label{app:bc}

In \cref{sec:Ising}, we derived the Ising picture for the correlator transition in the Haar random circuit. The $\pm$ Ising spins in Eq.~\ref{eq:pairing} and their evolution rules (Eqs.~\ref{eq:dw_rule_1} to \ref{eq:dw_rule_4}) come from  averaging of the tensor product $u \otimes u^* \otimes u \otimes u^*$. 

In this Appendix, we review a continuous time noisy spin chain model, the Brownian circuit, and derive the evolution rules in the $+/-$ basis. The  average of ${U \otimes U^* \otimes U \otimes U^*}$ with $H$ given by the Brownian circuit generates a slightly different rate matrix compared to the Haar-random gate. We also write the rate matrix in the  $\emptystate$/$\fullstate$ basis. 

A Brownian circuit \cite{lashkari_towards_2013,zhou_operator_2020,zhou_operator_2019,xu_locality_2019,shenker2015stringy}  with general two-body interactions has the  infinitesimal Hamiltonian increment $dH$ (playing the role of ``$Hdt$'')
\begin{equation}
\begin{aligned}
dH &= \sum_{i< j} J_{ij} dh_{ij}, \\ 
dh_{ij} &=  \sum_{\mu_i= 0}^{q^2-1}\sum_{\mu_i= 0}^{q^2-1}  \sigma_i^{\mu_i} \sigma_j^{\mu_j}  dB(t)_{ij}^{\mu_i \mu_j }.
\end{aligned}
\end{equation}
The $J_{ij}$ are fixed coupling strengths that can be chosen arbitrarily. 
Each two-body term $dh_{ij}$ contains a collection of random interaction terms with fluctuating strengths given by Brownian motions. They are statistically independent. In the It\^{o} formalism,
\begin{equation}
\label{eq:brownian_ito}
dB(t)_{ij}^{\mu_i \mu_j} dB(t)_{kl}^{\mu_k \mu_l} = \delta_{ik}\delta_{jl} \delta_{\mu_i \mu_k} \delta_{\mu_j \mu_l} dt. 
\end{equation}
Here $\sigma^{\mu_i}_i$ is a set of Hermitian basis on site $i$ that generalizes the Pauli matrices to a local Hilbert space of dimension $q$. We assign them to be (using the convention in \cite{zhou_operator_2020})
\begin{equation}
\sigma^\mu = \left\lbrace
             \begin{aligned}
               & \I_q & \quad  \mu = 0 \\
               & \sqrt{ 2q} T_a & \quad \mu = a > 0  \\
             \end{aligned} \right. 
\end{equation}
where $T_a$ are the standard $\rm SU(q)$ generators with the normalization convention
\begin{equation}
T_a T_b = \frac{1}{2q} \delta_{ab} \I_q + \frac{1}{2} \sum_{c=1}^{q^2 - 1} (d_{ab}{}^c + i f_{ab}{}^c ) T_c .
\end{equation}
We then have the inner product
\begin{equation}
\tr( \sigma^\mu \sigma^\nu ) = q \delta_{\mu\nu} .
\end{equation}

Define $U(t+ dt) = G \times U(t)$, where $G$ is the infinitesimal evolution operator. For evolution with Brownian motions, we expand to the second order:
 \begin{equation}
G = 1 - i dH - \frac{1}{2} dH \, dH . 
\end{equation}
The infinitesimal (multiplicative)  change of ${U \otimes U^* \otimes U \otimes U^*}$ is then $G \otimes G^* \otimes G \otimes G^*$. We take the average, and will interpret the result in terms of an operator $\hat W$ acting in the replicated (tensor product) space:
\be
\overline{G \otimes G^* \otimes G \otimes G^*} 
=
e^{\hat W d t}.
\ee   
There are simplifications resulting from the It\^{o} calculus in Eq.~\ref{eq:brownian_ito}: 
\begin{itemize}[leftmargin=1.28em,labelindent=16pt]
\item The change factorizes into separate contributions from each interaction term $(ij)$, i.e. 
\begin{equation}
\begin{aligned}
 &\overline{G \otimes G^* \otimes G \otimes G^*} \\
& \qquad =\prod_{i<j} \overline{g_{ij}  \otimes g_{ij}^* \otimes g_{ij} \otimes g_{ij}^*} \\
& \qquad =1 + \sum_{i<j} (\overline{g_{ij}  \otimes g_{ij}^* \otimes g_{ij} \otimes g_{ij}^*}  - 1 )
\end{aligned}
\end{equation}
where
\begin{equation}
g_{ij} = 1 - i J_{ij} dh_{ij} - \frac{1}{2} J^2_{ij} dh_{ij}^2
\text{ (no summation)}.
\end{equation}
\item The contribution from the  interaction ($ij$) that survives the  average is proportional to $J^2_{ij}$. 
\end{itemize}

Therefore, it is sufficient to  work out the result for a fixed $i$ and $j$ 
(a two-site example), 
and to set ${J_{ij} = 1}$.
With this simplification, we suppress the $i,j$ indices in $g$ and $h$ for clarity. The infinitesimal change is
\begin{equation}
\begin{aligned}
  dg &= 1 - i dh- \frac{1}{2} \sum_{\mu =0}^{q^2-1}\sum_{\nu=0}^{q^2-1}  \sigma_i^{\mu } \sigma_j^{\nu}  \sigma_i^{\mu } \sigma_j^{\nu}  dt \\
  &= 1 - i dh -\frac{1}{2}  q^4 \I dt, 
\end{aligned}
\end{equation}
where we have used an identity for a complete set of Hermitian orthonormal operators $A^{\Delta}$
\begin{equation}
\label{eq:cmpt_rel}
\sum_\Delta A^{\Delta}_{\alpha \beta} A^{\Delta} _{\sigma \gamma} = \delta_{\alpha \gamma}\delta_{ \beta \sigma }
\end{equation}
with ${A^{\Delta=(\mu,\nu)}_{\alpha \beta} =q^{-1}  ({\sigma_i^{\mu }} \otimes  {\sigma_j^{\nu}})_{\alpha \beta}}$ (where $\alpha,\beta$ run over $q^2$ values).
Hence
\begin{equation}
\begin{aligned}
  &\overline{g \otimes g^* \otimes g \otimes g^*}  =  - 4 \frac{1}{2} q^4 dt  \I \otimes \I \otimes \I \otimes \I \\
  & \qquad + \overline{(1 -i dh ) \otimes (1+i dh^{\top} ) \otimes (1 -i dh ) \otimes (1+i dh^{\top} )}.
\end{aligned}
\end{equation}
We use Eq.~\ref{eq:cmpt_rel} to compute the possible contractions in the last term, and denote it diagrammatically as
\begin{equation}
\begin{aligned}
  &\sum_{\mu,\nu} \sigma_i^\mu \sigma_j^\nu \otimes \sigma_i^\mu \sigma_j^\nu = \sum \sumbasis = q^2 \bcswitch \\
  &\sum_{\mu,\nu} \sigma_i^\mu \sigma_j^\nu \otimes (\sigma_i^\mu)^{\top} (\sigma_j^\nu)^{\top} = \sum \sumbasis[0][0][0][1] = q^2\bcwick.
\end{aligned}
\end{equation}
In the middle, a line with an up arrow represents a product of Paulis, $\sigma^\mu_i \sigma^\nu_j$, and 
a line with a down arrow represents the transposed version, 
$(\sigma_i^\mu)^{\top} (\sigma_j^\nu)^{\top}$.
Note that each line represents {\it two} spatial sites, and the two lines stand for the two  terms in the tensor product (the two ``replicas'').
On the right-hand side, the lines \textit{without} arrows denote the identity maps  between the corresponding Hilbert spaces,
i.e. a $\delta_{ab}$, where $a,b$ run over $q^2$ values because of the two spatial sites.
If we label the state on each site explicitly, then
$\delta_{ab}=\delta_{\alpha,\beta}\delta_{\alpha',\beta'}$, where
$a$ is the two-site multi-index $(\alpha, \alpha')$ and similarly for $b$.

With this setup, we have the following Wick contractions: 
\begin{align}\notag
&\overline{(1 -i dh ) \otimes (1+i dh^{\top} ) \otimes ( 1-i dh ) \otimes (1+i dh^{\top} )} -1\\ \notag
& \quad = q^2 dt  \left( \bcwickfouru[0] + \bcwickfouru[1] + \bcwickfouru[2] + \bcwickfouru[3]  \right)\\
& \quad - q^2 dt \left( \bcswitchfouru[0] + \bcswitchfouru[1]  \right). 
\end{align}
Writing
\begin{equation}
\begin{aligned}
\overline{g \otimes g^* \otimes g \otimes g^*}= 
\mathbb{I} \otimes \mathbb{I} \otimes \mathbb{I} \otimes \mathbb{I}  +\hat{W} { dt}  ,
\end{aligned}
\end{equation}
the  average of the  infinitesimal evolution for a two-site Brownian interaction gives
\begin{equation}
\begin{aligned}
\hat{W}  =
&  - 2 q^4   \bcidfouru  \\
 &+ q^2   \left( \bcwickfouru[0] + \bcwickfouru[1] + \bcwickfouru[2] + \bcwickfouru[3]  \right) \\
 &-q^2  \left( \bcswitchfouru[0] + \bcswitchfouru[1]  \right). \\
\end{aligned}
\end{equation}

We then work out the evolution in the space of $+$/$-$ states. 
For a single site we have 
${|+ \rangle=\idket}$,
${|- \rangle=\swapket}$, so that for two sites we have the identification
\begin{equation}
\begin{aligned}
|++ \rangle &= \idket \idket & 
|+ - \rangle & = \idket \swapket \\
| - + \rangle &= \swapket \idket &
|--\rangle & =  \swapket \swapket.
\end{aligned}
\end{equation}
The states such as $\idket$ represent the identity maps between the Hilbert spaces (they are the definitions of the boundary $\pm$ states, see notations in \cite{zhou_emergent_2019}). 

In this basis, we have 
\begin{equation}
\begin{aligned}
\hat{W} | ++ \rangle &= 0\\
\hat{W} | +- \rangle &= -2 q^4 | +- \rangle  + 2q^3 ( | ++ \rangle + | -- \rangle ) - 2q^2 | -+ \rangle \\ 
\hat{W} | -+ \rangle &= -2 q^4 | -+ \rangle  + 2q^3 ( | ++ \rangle + | -- \rangle ) - 2q^2 | +- \rangle \\ 
\hat{W} | -- \rangle &= 0.
\end{aligned}
\end{equation}
In general the above transition operator is proportional to the squared coupling strength $J^2$, which we have so far set to unity. To simplify the factors let us now take ${J^2 = (2q^3)^{-1}}$: this gives the transition operator
\begin{equation}
\label{eq:W_brownian}
\hat{W}_{\rm Brownian} 
\begin{bmatrix}
|++\rangle \\
|+-\rangle \\
|-+\rangle \\
|--\rangle 
\end{bmatrix}
= \underbrace{
\begin{bmatrix}
  & & & \\
1 & -q & - \frac{1}{q} & 1\\
1 & -\frac{1}{q}& -q  & 1\\
 &  &  &  \\
\end{bmatrix}
}_{W_{\rm Brownian}^{\top}} 
\begin{bmatrix}
|++\rangle \\
|+-\rangle \\
|-+\rangle \\
|--\rangle 
\end{bmatrix}.
\end{equation}
(Here $\hat{W}_{\rm Brownian}$ is viewed as an operator on the Hilbert space spanned by the $\pm$ states, while 
$W_{\rm Brownian}^{\top}$
is a matrix defined componentwise.)

An infinite time evolution with a Brownian two-body interaction will lead to a uniformly (Haar) random unitary evolution operator. 
Therefore we expect 
\be
\hat {T}_{\rm Haar} = \lim_{t \rightarrow \infty}\exp( \hat{W}_{\rm Brownian} t)
\ee
to agree with  the  transition matrix corresponding to a Haar gate, whose action was specified in Eqs.~\ref{eq:dw_rule_1}-\ref{eq:dw_rule_4}.

The result
\begin{equation}
\hat{T}_{\rm Haar} 
\begin{bmatrix}
|++\rangle \\
|+-\rangle \\
|-+\rangle \\
|--\rangle 
\end{bmatrix}
= \underbrace{
\begin{bmatrix}
1 & & & \\
\frac{q}{q^2+1} & 0& 0& \frac{q}{q^2+1}\\
\frac{q}{q^2+1} & 0& 0& \frac{q}{q^2+1}\\
  & & & 1\\
\end{bmatrix}}_{T_{\rm Haar}^{\top}} 
\begin{bmatrix}
|++\rangle \\
|+-\rangle \\
|-+\rangle \\
|--\rangle 
\end{bmatrix}
\end{equation}
is consistent with the Haar average Eqs.~\ref{eq:dw_rule_1}--\ref{eq:dw_rule_4}. 

We can also define a ``continuous time'' Haar circuit  dynamics in which, in a time interval $dt$, each bond has a probability $\gamma dt$ of receiving a Haar random unitary. 
This is equivalent (in the thermodynamic limit) to the dynamics in Sec.~\ref{sec:higherdims} and App.~\ref{app:thin_cluster}.
Then the  evolution operator for infinitesimal time is ${e^{ \hat W_\text{c-Haar}^{(\gamma)} dt}
= (1- \gamma dt) + \gamma dt  \hat T_\text{Haar}}$:
\be
\hat W_\text{c-Haar}^{(\gamma)} =
\gamma 
(\hat T_\text{Haar} - 1).
\ee
Define $\hat W_\text{c-Haar}$ (without a superscript) as the case where $\gamma=\frac{q^2+1}{q}$, such that the rate for the transition ${\ket{+-}\rightarrow\ket{++}}$ is unity, as in Eq.~\ref{eq:W_brownian}:
\begin{equation}
\label{eq:c_haar}
\hat{W}_{\rm c-Haar} 
\begin{bmatrix}
|++\rangle \\
|+-\rangle \\
|-+\rangle \\
|--\rangle 
\end{bmatrix}
= \underbrace{
\begin{bmatrix}
 0 & & & \\
1 & -\frac{(q^2 +1)}{q} & & 1\\
1 & & -\frac{(q^2 + 1)}{q} & 1\\
 &  &  &  0 \\
\end{bmatrix}}_{W_{\rm c-Haar}^{\top}} 
\begin{bmatrix}
|++\rangle \\
|+-\rangle \\
|-+\rangle \\
|--\rangle 
\end{bmatrix}.
\end{equation}

We can also consider mixed dynamics, in which the spins are acted on by the Brownian circuit with coupling strength $J^2 =\frac{(2q^3)^{-1}}{1 + \kappa}$ 
\textit{and} Haar unitaries applied at rate ${\gamma=\frac{q^2+1}{q}
\frac{\kappa}{1 + \kappa}}$.
Then we have the continuous time evolution operator:
\be
\hat W_\kappa= \frac{1}{1 + \kappa} \hat{W}_\text{Brownian} + \frac{\kappa}{1 + \kappa} \hat{W}_\text{c-Haar}
\ee
which interpolates between Brownian circuit and continuous time Haar circuit dynamics.

For completeness, we list the basis transformations from the $+$/$-$ to the  $\emptystate$/$\fullstate$ basis. 

$\hket[0]$ represents an identity operator and $\hket[1]$ represents a non-identity operator, therefore
\begin{equation}
\hket[0] \rightarrow \frac{\I \otimes \I }{q}, \qquad \hket[1] \rightarrow \frac{{F}}{q} 
\end{equation}
where
\begin{equation}
    {F} \equiv \frac{1}{\sqrt{q^2 - 1}} \sum_{a=1}^{q^2-1} \sqrt{2q} T_a\otimes \sqrt{2q} T_a.
\end{equation}
The normalization ensures $\tr( F^2 ) = 1$. 

From their inner products with the $+$/$-$ basis, we have the transformation: 
\begin{equation}
\begin{aligned}
\idket &= q \hket[0] \\
\swapket  &= ( q^2 - 1) \hket[1] + \hket[0]. 
\end{aligned}
\end{equation}

Therefore 
\begin{equation}
\hat{W}_{\rm Brownian} 
\begin{bmatrix}
\hket[00] \\
\hket[01] \\
\hket[10] \\
\hket[11] \\
\end{bmatrix}
=
\begin{bmatrix}
 0 & & & \\
 & -q + \frac{1}{q} &   & q - \frac{1}{q} \\
 &  & -q + \frac{1}{q}  & q - \frac{1}{q} \\
 & \frac{1}{q} & \frac{1}{q} & \frac{2}{q} \\
\end{bmatrix}
\begin{bmatrix}
\hket[00] \\
\hket[01] \\
\hket[10] \\
\hket[11] \\
\end{bmatrix}
\end{equation}
and 
\begin{equation}\label{eq:Thaar}
\hat{T}_{\rm Haar} 
\begin{bmatrix}
\hket[00] \\
\hket[01] \\
\hket[10] \\
\hket[11] \\
\end{bmatrix}
=
\begin{bmatrix}
 1 & & & \\
   & \frac{1}{q^2+1} & \frac{1}{q^2+1} & \frac{q^2-1}{q^2+1}  \\
   & \frac{1}{q^2+1} & \frac{1}{q^2+1} & \frac{q^2-1}{q^2+1}  \\
   & \frac{1}{q^2+1} & \frac{1}{q^2+1} & \frac{q^2-1}{q^2+1}  \\
\end{bmatrix}
\begin{bmatrix}
\hket[00] \\
\hket[01] \\
\hket[10] \\
\hket[11] \\
\end{bmatrix}
\end{equation}
\begin{equation}
\hat{W}_{\rm c-Haar} 
\begin{bmatrix}
\hket[00] \\
\hket[01] \\
\hket[10] \\
\hket[11] \\
\end{bmatrix}
=
\begin{bmatrix}
 0& & & \\
 & -q & \frac{1}{q}  & q - \frac{1}{q} \\
 & \frac{1}{q} & -q   & q - \frac{1}{q} \\
 & \frac{1}{q} & \frac{1}{q} & \frac{2}{q} \\
\end{bmatrix}
\begin{bmatrix}
\hket[00] \\
\hket[01] \\
\hket[10] \\
\hket[11] \\
\end{bmatrix}
\end{equation}


\section{Data for edge-edge correlations \label{app:noisy}}
In this Appendix we provide some numerical details pertinent to the extrapolation of $r^\mathrm{EE}$ and $\lambda^\mathrm{EE}$ to the ${L\to\infty}$ limit. A given velocity $v$  corresponds to ${t=(L-1)/v}$ for the edge-edge correlator/OTOC for a system of size $L$. Therefore
\eq{
-v\ln G^\mathrm{EE}_\mathrm{rms}(vt,t)/s_\mathrm{eq} =
(L-1)r^\mathrm{EE}(v) \,,
}
so that plotting the LHS of the above equation against $L$ should yield a straight line whose slope is  asymptotically equal to  $r^\mathrm{EE}(v)$. This is shown in the top left panel in Fig.~\ref{fig:nsc-extrap} for the the $ZZ$ correlator in the noisy spin chain (Sec.~\ref{sec:simulatedIsingmodel}). The results are similar for the $XX$ correlator. The asymptotic $r^\mathrm{EE}(v)$ so obtained is superposed on the finite-sized $r^\mathrm{EE}_L(v)\equiv -\ln G^\mathrm{EE}_\mathrm{rms}(vt,t)/s_\mathrm{eq}t$ in the top right panel of Fig.~\ref{fig:nsc-extrap}. The lower row corresponds to an identical analysis for the $\overline{\mathrm{OTOC}}^\mathrm{EE}$ and $\lambda^\mathrm{EE}$.

\begin{figure}[t]
\includegraphics[width=\linewidth]{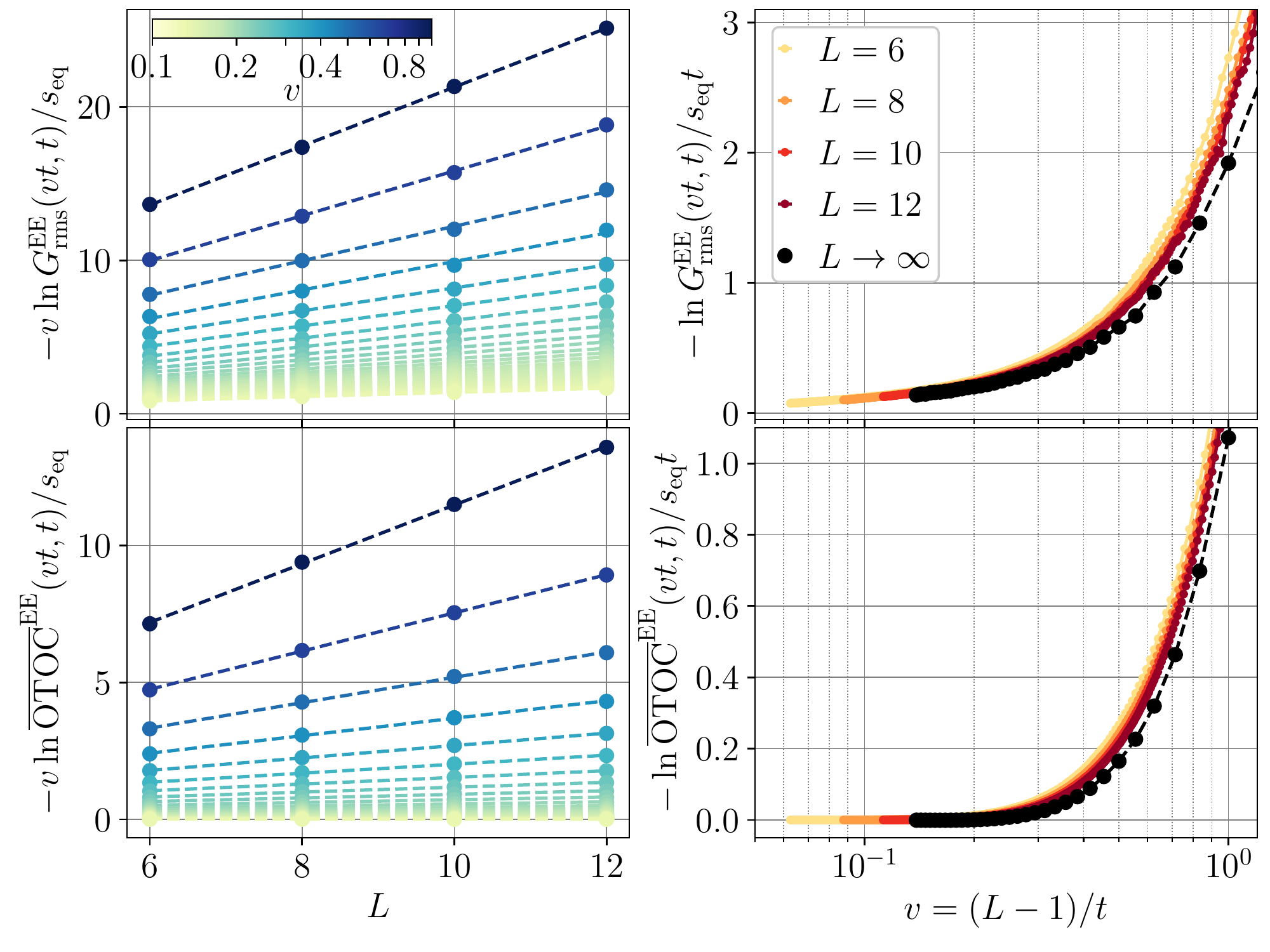}
\caption{Extrapolation of $r^\mathrm{EE}$ and $\lambda^\mathrm{EE}$ to $L = \infty$ for the $ZZ$ correlation function and the $ZZ$ OTOC for the noisy spin chain described in Sec.~\ref{sec:simulatedIsingmodel}. The top and bottom rows correspond to $G^\mathrm{EE}_\mathrm{rms}$ and $\overline{\mathrm{OTOC}}^\mathrm{EE}$ respectively. The top left panel shows $-v\ln G^\mathrm{EE}_\mathrm{rms}(vt,t)/s_\mathrm{eq}$ as a function of $L$ for several $v$ (colourbar) such that its asymptotic slope corresponds to the asymptotic value $r^\mathrm{EE}$. The top right shows the finite-size approximations $r_L^\mathrm{EE}(v)$ for different $L$, with the black dots denoting the extrapolated result. Similar analysis for the $\overline{\mathrm{OTOC}}^\mathrm{EE}$ in the bottom row.}
\label{fig:nsc-extrap}
\end{figure}

\section{Boundary conditions in Ising mapping}

In this section, we record details of the boundary conditions required for the calculation of  $\overline{G^2(x,t)}$ (and $\overline{\text{OTOC}}$ \cite{nahum_operator_2018}) in the Ising language and discuss numerical implementation.

From the definitions of the correlator $\overline{G^2(x,t)}$, the bottom $(t=0)$ boundary of the stacked circuit has kets $\ket{+}=\idket$ attached at all of the  sites except at the site where the initial operator $O$ is placed.  The site with the operator has the state $|\fineq[-.4ex][0.3][0.3]{\oidst[0][0][$O$][$O$][r]}\rangle$ attached, which we  denote  $|O_{+} \rangle$, or call it a $O_+$ spin. Similarly on the top boundary, we have $\bra{+}$ bras attached at every site except the site with the operator $O'$, which has $\langle O'_{+} |$. That is, 
\begin{equation}
\begin{aligned}
    &\overline{G^2(x,t)} = \frac{1}{q^{2L}} \times \\
    &\langle + \cdots O'_{+} \cdots + | \overline{U\otimes U^* \otimes U \otimes U^*}| + \cdots O_{+} \cdots + \rangle .
\end{aligned}
\end{equation}
We will also average $G^2$ over the choice of operators, $O\rightarrow u O u^\dag$ where $u$ is a single-site Haar unitary.
In the Haar-random circuit this does not change the result at all, since the 
state $|O_{+} \rangle $ is contracted with (copies of)  a random gate: 
the unitary invariance of the Haar measure means that further single-site unitary averaging does not change anything.
For the Brownian circuit this is not an identity at the microscopic level, but we are free to perform the additional operator averaging without changing the asymptotics.

Thus $|O_{+} \rangle$ can be replaced with $\overline{u\otimes u^* \otimes u \otimes u^* }|O_{+} \rangle $ without affecting $\overline{G^2(x,t)}$. For a traceless operator, this  average is
\begin{equation}
    \overline{u\otimes u^* \otimes u \otimes u^* }|O_{+} \rangle = \frac{\tr(O^2)}{q^2 - 1}\left( |- \rangle - \frac{1}{q} | + \rangle  \right),
\end{equation}
where $\tr(O^2)$ is restricted to the local Hilbert space of $O^2$ (thus $\tr(\sigma_a^2) = q$ in contrast to $\Tr( \sigma_a^2 ) = q^L$). 
We notice that $\frac{1}{q^2 - 1} \left(  |- \rangle - \frac{1}{q} | + \rangle \right)$ is the dual basis state $|-^* \rangle$, which is orthogonal to $+$, and has inner product $1$ with $|- \rangle$. Thus the boundary condition at the site with $O$ is $-^*$. This forces the  spin associated with the gate directly above this boundary point to be $-$. 
We then replace $\langle O'_+|$ by the averaged value
\be\label{eq:braav}
\overline{\langle O'_+|}=
\frac{\tr(O'^2)}{q^2 - 1}\left( \langle - | - \frac{1}{q} \langle + | \right).
\ee
The contribution from the second term gives zero: this is because it gives a term of the form 
\ba\notag
& \langle + \cdots + \cdots + | 
\overline{U\otimes U^* \otimes U \otimes U^*}| + \cdots -^* \cdots + \rangle \\
& = \langle + \cdots + \cdots + |  + \cdots -^* \cdots + \rangle
= 0 ,
\end{align}
where we used the invariance of the all $+$ state under unitary evolution.
Therefore in (\ref{eq:braav}) we keep only the  $\langle - |$ term. This gives
\begin{equation}
\begin{aligned}
    &\overline{G^2(x,t)} = \tr(O^2) \tr(O'^2) \frac{1}{q^{2L}}  \frac{1}{q^2 - 1} \times \\
    &\langle + \cdots - \cdots + | \overline{U\otimes U^* \otimes U \otimes U^*}| + \cdots -^* \cdots + \rangle .
\end{aligned}
\end{equation}

Next we consider the boundary condition of the OTOC. 
\begin{equation}
    \overline{\text{OTOC}} = - \frac{1}{2} \frac{1}{q^L} \text{Tr}( [ O(x,t), O'(0,0)]^2 ).
\end{equation}
Except for the locations with operator insertions, there are always $+$ spins on the top boundary and $-$ spins at the bottom boundary. At the site with $O'$, we  have $\langle O'_+ |$, while at the site with $O$, we have  $|\fineq[-0.4ex][0.3][0.3]{\oswapst[0][0][][$O^2$][r]}\rangle -|\fineq[-0.4ex][0.3][0.3]{\oswapst[0][0][$O$][$O$][r]}\rangle $. Again we perform an additional averaging of the operators: 
\begin{align}
    |\fineq[-0.4ex][0.3][0.3]{\oswapst[0][0][][$O^2$][r]}\rangle -|\fineq[-0.4ex][0.3][0.3]{\oswapst[0][0][$O$][$O$][r]}\rangle &\rightarrow q\tr(O^2) | -^* \rangle \\
    \langle O'_+ | &\rightarrow  \tr( O'^2 ) \langle -^* | 
\end{align}
We expand $\langle -^* |$ into the $\langle + |$ and $\langle - |$. The $\langle + |$ branch vanishes for the same reason as in the correlator. Therefore we have
\begin{equation}
\begin{aligned}
    &\overline{\text{OTOC}}(x,t) = \tr(O^2) \tr(O'^2) \frac{1}{q^L}  \frac{q}{(q^2 - 1)} \times \\
    &\langle + \cdots - \cdots + | \overline{U\otimes U^* \otimes U \otimes U^*}| - \cdots -^* \cdots - \rangle .
\end{aligned}
\end{equation}

Now we discuss how to implement the calculation numerically, in the case where the operators are placed at the two spatial boundaries as in Sec.~\ref{sec:numericalcasestudy}.

The $|+ \rangle$ and $|- \rangle$ basis states are not orthogonal; they have inner product
\begin{align}
    \langle + |+ \rangle = \langle - |- \rangle = q^2, \\
    \langle + |- \rangle = \langle + |- \rangle = q. 
\end{align}
However this does not significantly change a transfer matrix calculation.  We give one way to do it in coordinates.
In the case of continuous time dynamics we have
\be
\overline{U\otimes U^* \otimes U \otimes U^*} = e^{\hat W t},
\ee
where the operator $\hat W$ was specified in App.~\ref{app:bc}.  Let the coordinate of the initial state $|- + \cdots + \rangle$ in terms of the $|+ \rangle$ and $| - \rangle$ basis on each site to be $\vec{v}_{1}$, then the coordinate of $\overline{U\otimes U^* \otimes U \otimes U^*} | - + \cdots + \rangle$ is $e^{ Wt} \vec{v}_1$ (note that $W$ transpose without hat has been worked out in App.~\ref{app:bc}). When evaluating the inner product in the expression $\langle + \cdots - | \overline{U\otimes U^* \otimes U \otimes U^*} | - + \cdots + \rangle / q^{2(L-1)}$, each $+$ in the expansion basis expansion of $\overline{U\otimes U^* \otimes U \otimes U^*} | - + \cdots + \rangle$ contributes a factor of $1$, and each $-$ contributes a $1/q$. Hence
\begin{equation}
\label{eq:G2_num}
    \overline{G^2(x,t)} =  \frac{\tr(O^2) \tr(O'^2) }{q(q^2 - 1)}  \sum_{\vec{v}, \sigma_L = -}
    \frac{ \vec{v}^\dagger \cdot  e^{ Wt}  \vec{v}_1 }{q^{N_{-}}},
\end{equation}
where $\vec{v}$ enumerates the coordinates of all the $+$/$-$ basis on $L$ sites with the right most one to be $-$. $N_{-}$ is the number of $-$ in each basis. Similarly the OTOC can be evaluated as 
\begin{equation}\notag
\overline{\text{OTOC}(x,t)} = \frac{q^L \tr(O^2) \tr(O'^2) }{q(q^2 - 1)} 
  \hspace{-2mm}
  \sum_{\vec{v}, \sigma_L = -}
   \hspace{-2mm}
    \frac{\vec{v}^\dagger \cdot  e^{ Wt}  \vec{v}_1 }{q^{N_{+}}}.
\end{equation}
In a random circuit, their saturation values are
\begin{align}
    \overline{G^2(x, t\rightarrow \infty)} \rightarrow \frac{\tr(O^2) \tr( O'^2) }{q^{2L+2}} \\
    \overline{\text{OTOC}}(x, t \rightarrow \infty) \rightarrow \frac{\tr( O^2 ) \tr( O'^2 )}{q^2}
\end{align}
When taking the local operators $O = \sigma_a$ and $O' = \sigma_b$, we have $\tr(O^2) = \tr( O'^2) = q$ and
\begin{align}
    \overline{G^2(x, t\rightarrow \infty)} \rightarrow \frac{1}{q^{2L}} \\
    \overline{\text{OTOC}}(x, t \rightarrow \infty) \rightarrow 1. 
\end{align}
Both are consistent with the cluster picture calculation.

\section{Sign phase transition for $G(x,t)$ in $d>1$}
\label{app:signtransition}

We discuss the sign of the correlator $G(x,t)$ in the random circuit, adding to the discussion in Sec.~\ref{sec:avversustypbrief}. We restrict here to the bound phase.

In the Haar circuit the  amplitudes $K_{t,\Delta t}(x'; x)$ (see Sec.~\ref{sec:instancebound}) have a random sign (because $\UU^{(t)}_{S,S'}$ can be  positive or negative). Therefore we expect the path sum for the bound state coordinate to be in the universality class of the partition function for a directed path with random signs. This is a well-studied problem that appears for path expansions of correlation functions in various disordered media
\cite{kardar2007statistical,nguyen1985tunnel,medina1989interference,kim2011interfering}. The amplitude $\overline{G^2}$ has the scaling forms discussed in Sec.~\ref{sec:avversustypbrief}.
A similar picture will apply for the Haar circuit in higher dimensions, for any value of $v$ (since in higher dimensions we have binding for all~$v$).

The random signs in the weights mean that the average \textit{sign} of the correlator, i.e. the average of ${\operatorname{sgn} G= G/|G|}$, vanishes in the Haar case. 
In 1+1D, the average sign is believed to vanish as $t\rightarrow \infty$  for any (non-fine-tuned) random path model where negative weights are allowed  \cite{kim2011interfering, baldwin2018sign}.
In higher dimensions a phase where  the sign does \textit{not} average to zero at large $t$ is also possible (the two phases are separated by a ``sign phase transition'' \cite{kardar2007statistical}). 
It is likely that this other phase can be accessed in a higher-dimensional circuit that is more weakly random (such as a  Floquet circuit perturbed by randomness that weakly breaks space and time translation symmetry). 
Varying the strength of disorder would then give a transition between a phase with $\overline{G/|G|}\ll 1$ at large $t$  and a phase where 
$\overline{G/|G|}$ is of order 1.

\bibliography{physical_corr}

\end{document}